# An Interfacial Balance Rule Governs Binder–Electrolyte Coupling in Lead-Free Perovskite Energy Storage

Arun Kumar[†, 1], Ayush Kumar Pandey[†, 1], Ankur Yadav[†], Vishnu Saraswat[^], Shiladitya Sengupta[†], Abhishek Tewari[#, $, *] and Monojit Bag[†, ‡, *]

[†]*Department of Physics, Indian Institute of Technology Roorkee, Roorkee 247667, Uttarakhand, India*

‡*Centre for Nanotechnology, Indian Institute of Technology Roorkee, Roorkee 247667, Uttarakhand, India*

^*Department of Electronics and Communication Engineering, SR University, Warangal 506371, Telangana, India*

[#]*Department of Metallurgical and Materials Engineering, Indian Institute of Technology Roorkee, Roorkee 247667, Uttarakhand, India*

[$]*Mehta Family School of Data Science and Artificial Intelligence, Indian Institute of Technology Roorkee, Roorkee 247667, Uttarakhand, India*

[1]*Equal contribution*

*Corresponding author: (Monojit Bag): monojit.bag@ph.iitr.ac.in,

(Abhishek Tewari): abhishek@mt.iitr.ac.in

**Abstract**

Electrode binders are conventionally regarded as inert structural components. Here we show that in lead-free perovskite supercapacitors the binder defines the optimal electrolyte composition. Across a factorial matrix of poly(vinylidene fluoride) (PVDF) loadings and LiTFSI concentrations in $CsSnCl_3$ electrodes, the capacitance optimum shifts systematically with binder content along a single linear relationship, described by the Interfacial Balance Rule ($\lambda+\theta=1$), where $\lambda$ and $\theta$ are the normalized lithium-supply and polymer contributions at the optimized interfacial state. The same relationship holds in hybrid $MASnCl_3$, showing that the optimum is governed by the polymer–electrolyte interface rather than the perovskite lattice chemistry. Simulations with a pre-trained MACE machine-learned interatomic potential show PVDF adopting a planar configuration on $CsSnCl_3$, interacting simultaneously with cationic and anionic surface sites. This configuration homogenizes lithium adsorption energetics, introduces fluorine-mediated coordination, and confines lithium to a two-dimensional interfacial region while preserving lateral mobility. Tuning polymer coverage through surface density and chain length reveals a finite interfacial lithium accommodation capacity that marks the onset of out-of-plane aggregation. The Interfacial Balance Rule provides a macroscopic descriptor of this finite interfacial resource, balancing polymer-mediated lithium stabilization against limited accommodation space. Binder loading is therefore an active design parameter for polymer-regulated energy-storage interfaces.

## 1. Introduction

The increasing global need for efficient and quick storage of energy is pushing research efforts towards developing electrode materials that combine high energy density provided by lithium-ion batteries and high-power density typical of dielectric capacitors. Among the recently emerged candidates for such applications are metal-halide perovskites (MHPs). These materials are already widely used in optoelectronics and have been investigated as electrode materials for electrochemical energy storage.[1–3] The reason for their ability to act as energy storage devices lies in their highly adjustable band gaps, defect tolerance and mixed ionic-electronic conductivity,[4–7] allowing for fast multidirectional charge transfer.[8–10] Moreover, the shift from lead-containing perovskites towards lead-free, namely the inorganic $CsSnCl_3$ and hybrid organic-inorganic $MASnCl_3$ systems, opens the door for developing environment friendly energy storage devices.[11,12] However, while the aforementioned materials show favourable bulk transport properties, the practical implementation of perovskite nanocrystals in liquid-electrolyte supercapacitors is limited by interface instabilities. In the traditional approach to electrochemical energy storage, solid-electrolyte interphase (SEI) serves as the key component of the system. SEI, being the product of electrolyte electrochemical decomposition on electrode surfaces, acts as an essential mediator of ion transport and structural protection for active material. However, the fragile ionic lattice of MHPs is particularly sensitive to both polar solvents and ionic impurities. Thus, electrochemical reactions at the perovskite-liquid interface are often plagued with phase instabilities, quick dissolution, and high rates of capacity fading. While a lot of attention has been paid to composition control and morphology manipulation of the bulk phase of perovskites, chemical processes at the binder-electrolyte-perovskite interface have received comparatively little attention.–[13–17] An important, but systematically governing process at the interface is the dynamic interaction of the polymer binder with the electrolyte salt. In typical electrode fabrication, poly(vinylidene fluoride) (PVDF) is extensively used as a structural binder due to its high affinity and electrochemical window, at an arbitrarily chosen mass fraction of about 10 to 15%.-[18–20] On the other hand, for optimization of capacitive behavior, it is commonly assumed that increasing concentration of electrolyte salt (such as Li-ions) will always decrease solution resistance and consequently improve double-layer or pseudo capacitance kinetics. Yet, PVDF is a strongly fluorinated polymer, while lithium salts are highly reactive.[21–23] In such chemically active environment as the perovskite interface, we propose that the binder and the solvated ions are not independent spectator materials, but actively participate in a coupled interfacial interaction that determines the composition of the interfacial layer.[24–27] So far, the stoichiometry of this process and its effect on capacitive behavior have not been comprehensively explored. The projected area of an electrode imposes a finite interfacial space in which both polymer chains and solvated ions participate. Therefore, binder loading and electrolyte concentration may not act as independent optimization parameters; instead, changes in polymer coverage could modify the available environment for ion accommodation and interfacial charge storage. Understanding this coupling is essential for establishing rational design principles for polymer–electrolyte interfaces in perovskite-based energy-storage systems. Here, we investigate how the interplay between polymer binder chemistry and electrolyte lithium concentration governs charge storage at lead-free perovskite–electrolyte interfaces. Using $CsSnCl_3$ nanocrystals as a model system, complemented by studies on the organic–inorganic analogue $MASnCl_3$, we combine electrochemical measurements, spectroscopic analysis, and atomistic simulations to resolve the molecular origin of the experimentally observed composition-dependent behavior. We demonstrate that PVDF is not an inert structural component, but an active interfacial regulator that controls lithium coordination, spatial distribution, and the dimensionality of interfacial storage. This work establishes a general framework in which polymer–ion interactions and finite interfacial accommodation capacity determine the optimal balance between lithium availability and reversible charge storage in mixed ionic–electronic materials. To elucidate the molecular origin of this behavior, machine-learned interatomic potential calculations are employed to investigate the polymer adsorption configuration, interfacial lithium coordination environment, and structural evolution under increasing lithium loading.

## 2. Results and Discussion

### 2.1. Synthesis and Structural Elucidation of Lead-Free $CsSnCl_3$ Nanocrystals

To establish a phase-pure baseline for interfacial electrochemical studies, crystalline and lead-free $CsSnCl_3$ perovskite nanocrystals (NCs) were synthesized and characterized. As illustrated in Figure 1, the $CsSnCl_3$ NCs were prepared using the ligand-assisted reprecipitation (LARP) method and subsequently used for electrode fabrication. The perovskite active material was homogeneously mixed with conductive carbon and PVDF binder and coated onto a graphite electrode. This configuration enables direct evaluation of material–electrolyte interfacial interactions. The structural quality and phase purity of the as-prepared $CsSnCl_3$ NCs were examined using X-ray diffraction (XRD). According to figure 2a, the diffraction pattern exhibits well-defined diffraction peaks that match with the simulated standard diffraction pattern of cubic perovskite phase (Pm3m space group). In particular, the main diffraction peaks with 2θ angles of around 22.5°, 31.0°, 39.5°, 45.9°, 49.5°, and 54.7° are assigned to (110), (200), (211), (220), (301), and (222) crystallographic planes, respectively.[11,28] No additional diffraction peaks corresponding to oxidized $SnO_2$ or the $Cs_2SnCl_6$ phase are observed, confirming the phase purity of the $CsSnCl_3$ nanocrystals. Additional information on morphology and microstructure of the perovskite nanocrystals was obtained through the transmission electron microscopy (TEM) technique. Bright field TEM micrograph (figure 2 (b)) shows a uniform distribution of nanocrystals with predominantly cubic-to-rectangular morphology, consistent with a crystalline perovskite structure. Statistical analysis of TEM data (inset) indicates a narrow-size distribution; the nanocrystals are found to have an average diameter of 3.4 ± 0.9 nm. This nanoscale dimension provides a high surface-area-to-volume ratio, which increases the availability of electrochemically accessible sites for charge storage. To further evaluate atomic-scale crystallinity, SAED and HRTEM techniques were used. The SAED pattern (figure 2 (c)) shows discrete, bright concentric rings of diffraction corresponding to (110), (200), (220) and (222) planes consistent with the XRD analysis, further confirming the polycrystalline nature of the nanocrystals. Moreover, the HRTEM image (figure 2 (d)) reveals continuous and uniform lattice fringes across the nanocrystal. The measured interplanar distance of 0.27 nm corresponds to the (200) crystallographic plane of $CsSnCl_3$ cubic structure. Together, these results demonstrate that, the highly crystalline, pure $CsSnCl_3$ electrodes have been synthesized successfully, providing a suitable platform for studying the interface interactions between PVDF and the electrolyte.

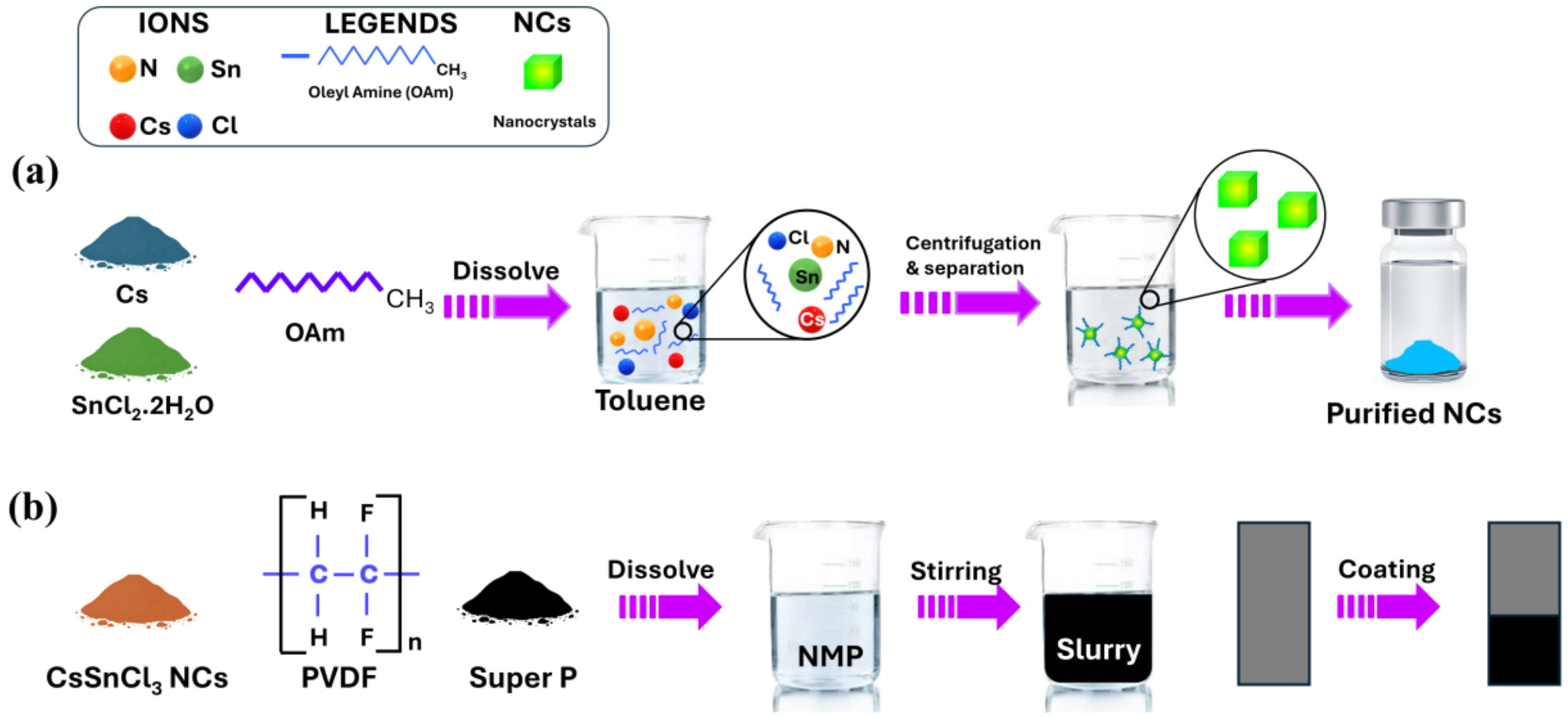

***Figure 1.*** *Schematic of the (a) colloidal synthesis and (b) subsequent PVDF-integrated electrode fabrication.*

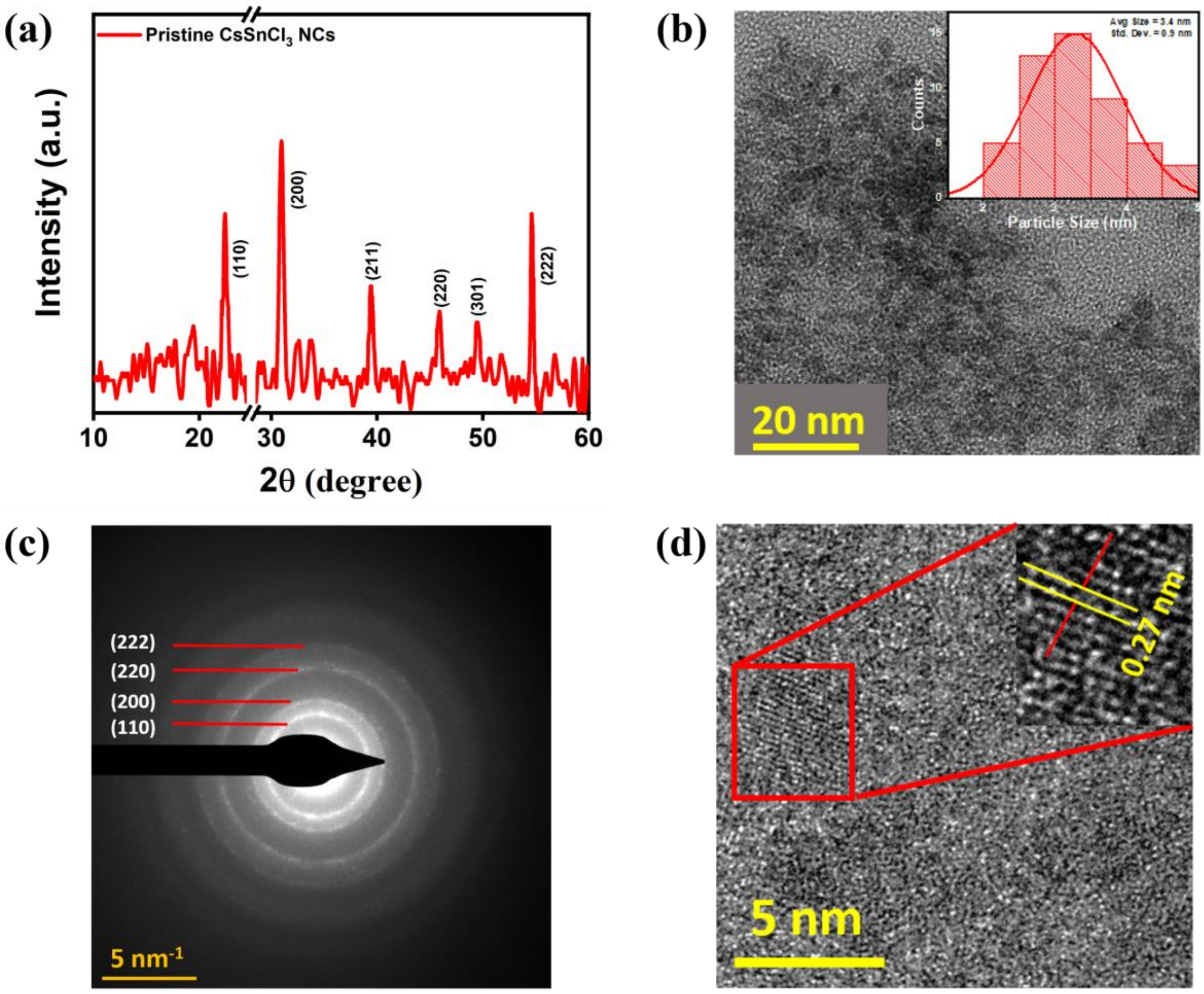

***Figure 2.*** *Structural and morphological characterization of pristine $CsSnCl_3$ nanocrystals. (a) XRD pattern matching the cubic perovskite phase. (b) Bright-field TEM image (inset: particle size distribution histogram indicating an average size of 3.4 ± 0.9 nm). (c) Corresponding SAED pattern confirming polycrystalline attributes. (d) HRTEM image highlighting the atomic lattice, with a magnified inset showing a measured d-spacing of 0.27 nm for the (200) facets.*

### 2.2. Electrochemical Performance: Mapping the Shifting Interfacial Optimum

To systematically evaluate the charge-storage behavior and PVDF–electrolyte interfacial effects, the fabricated $CsSnCl_3$ perovskite electrodes were investigated in a systematic multi-dimensional matrix electrochemically. Rather than evaluating a single binder composition, we altered the PVDF mass loading (5, 10, 15, and 20 wt%) and tested each binder content against a variety of LiTFSI salt concentrations (0.05 M to 0.20 M). Mapping capacitance across this compositional space revealed a coupled dependence on binder and electrolyte content, showing that ionic concentration alone is insufficient to determine capacitive performance. This behavior was attributed to a competition between polymer-mediated interfacial stabilization and excess lithium accumulation beyond the optimized interfacial regime, a mechanism investigated through atomistic simulations in Section 2.7. As shown in figure 3 (a), the specific capacitance determined from the galvanostatic charge-discharge cycle at the current density of 0.2 A/g, as marked by the patterned bars, reveals a strong dependence on the binder-to-salt ratio. Performance parameters have been calculated using equations S1-S6. At the lowest PVDF loading (5 wt%), the specific capacitance increases progressively with salt concentration, achieving its maximum of about 80 F/g at 0.20 M. Conversely, as the amount of highly fluorinated PVDF is gradually increased in the electrode composition, the LiTFSI concentration required for maximum capacitance progressively decreases. In the case of 10 wt% PVDF electrodes, the performance shows its maximum value at 0.15 M (124 F/g) exhibiting a bell-shaped dependence with further addition of the electrolyte resulting in a sharp drop in electrochemical performance. This inverse shift becomes more pronounced in the case of higher amounts of binder loading. In the case of 15 wt% PVDF, the optimum interfacial state occurs at 0.10 M giving a peak specific capacitance value of 126 F/g. Finally, in the case of the highest binder concentration, namely 20 wt% PVDF, the trend is reversed – the maximum capacity value (120 F/g) is obtained at the lowest concentration of 0.05 M LiTFSI, and further addition of the electrolyte results in a significant reduction in charge storage performance. To confirm that this trend reflects an intrinsic interfacial effect rather than a measurement artifact of galvanostatic measurements, areal capacitances were extracted from CV curves at 5 mV/s. As seen from figure 3 (b), the areal capacitance trends closely reproduce those observed for specific capacitance. From 5 wt% PVDF to 20 wt% PVDF, the concentration at which areal capacitance is maximised decreases linearly from 0.20 M at 5 wt% PVDF to 0.05 M at 20 wt% PVDF. This systematic electrochemical analysis demonstrates that the reactive perovskite electrode interface is dictated by the coupled effects of polymer matrix and electrolyte ions in solution. The progressive shift in the maximum of capacitance indicates that optimization cannot be achieved by varying either parameter independently.[27] However, this macroscopic observation requires a comprehensive, predictive model, the formation of which is developed in the following section, since the chemical basis of this optimized interfacial state needs to be established. Complete GCD and CV profiles for all binder/electrolyte combinations are provided in Figures S1–S12, which further support the observed trends in the electrochemical behavior of those materials. Calculations of specific and areal capacitance have been done using S1-S6.

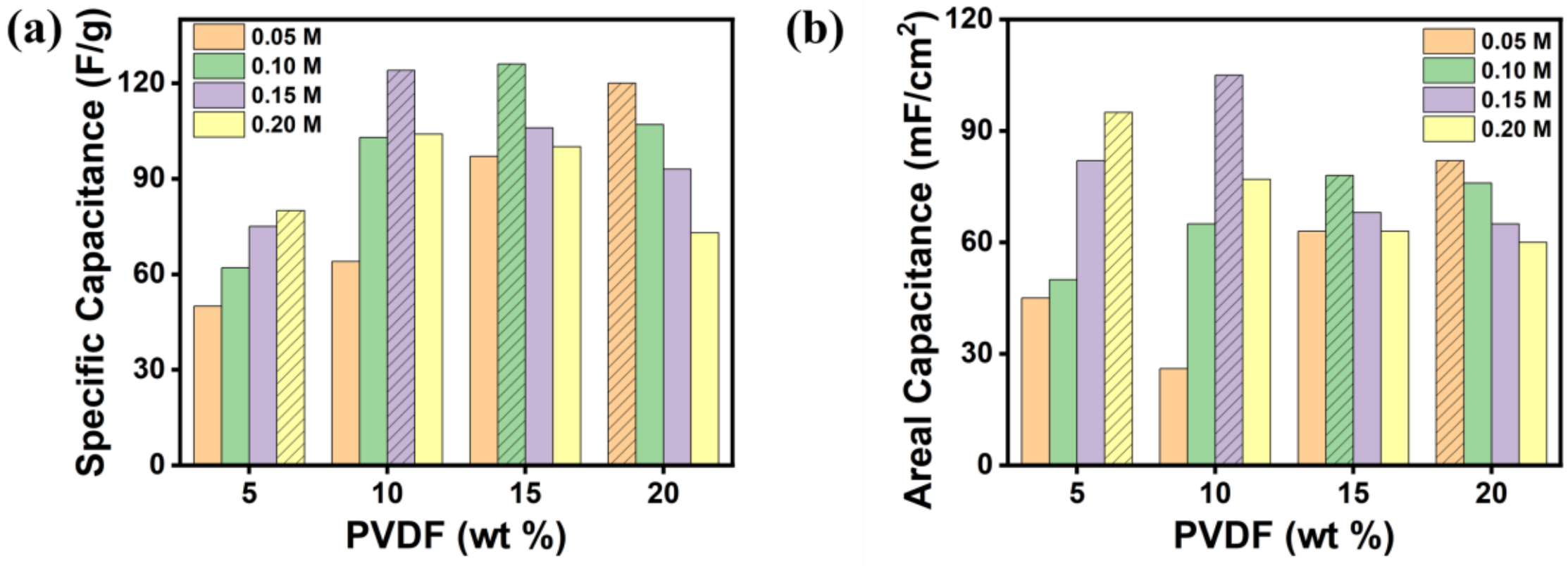

***Figure 3.*** *Electrochemical mapping of the shifting interfacial optimum. (a) Specific capacitance and (b) corresponding areal capacitance of the $CsSnCl_3$ electrodes as a function of LiTFSI concentration (0.05-0.20 M) across varying PVDF mass loadings (5-20 wt%). The patterned bars explicitly highlight the optimal electrolyte concentration required to maximize charge storage for each specific binder formulation, demonstrating a continuous inverse relationship.*

### 2.3. Mechanistic Elucidation of the Interfacial Balance Rule and Structural Preservation

To elucidate the mathematical and chemical underpinnings of the optima shift in figure 3, the multi-dimensional experimental matrix was converted into a unified thermodynamic map for the $CsSnCl_3$ perovskite system. As illustrated in the 2D contour heatmap in Figure 4 (a), the specific-capacitance landscape reflects the coupled influence of PVDF loading and LiTFSI concentration. Rather than exhibiting isolated optima, the map displays a smooth ridge of maximum performance. With an increasing weight percentage of the highly fluorinated PVDF binder in the electrode matrix, there is a highly correlated decrease in the molarity of the LiTFSI electrolyte to achieve the optimal capacitance. Formulations outside this ridge exhibit substantially lower capacitance owing to either insufficient polymer-mediated stabilization or excessive binder coverage, leading to active-site passivation and mass-transport limitations. To quantify the relationship governing this region, the capacitance optimum was determined for each PVDF loading. The experimentally determined optimum lithium concentration ($[Li^+]_{opt}$) decreases linearly with increasing PVDF loading ($PVDF_{wt\%}$) (Figure 4 (b)). Across the complete 4 × 4 composition matrix, each polymer loading corresponds to a single electrolyte concentration that maximizes capacitance: 0.20, 0.15, 0.10, and 0.05 M for 5, 10, 15, and 20 wt% PVDF, respectively. Each optimum satisfies the same linear constraint:

$$100 \times [Li^+]_{opt} + PVDF_{wt\%} = \xi_{Int}$$

where $\xi_{Int} = 25$ for the present experimental grid. Considering the 0.05 M concentration resolution used to identify the optimum, the uncertainty in this constant is ±2.5, giving $\xi_{Int} = 25 \pm 2.5$. This relationship indicates that the electrochemical optimum is governed not by bulk electrolyte availability or polymer passivation independently, but by a finite interfacial accommodation capacity shared between lithium supply and polymer-defined structural regulation. Whether this relationship represents a general feature of polymer–electrolyte interfaces or is specific to the $CsSnCl_3$ lattice is examined through extension to the $MASnCl_3$ system in Section 2.4.

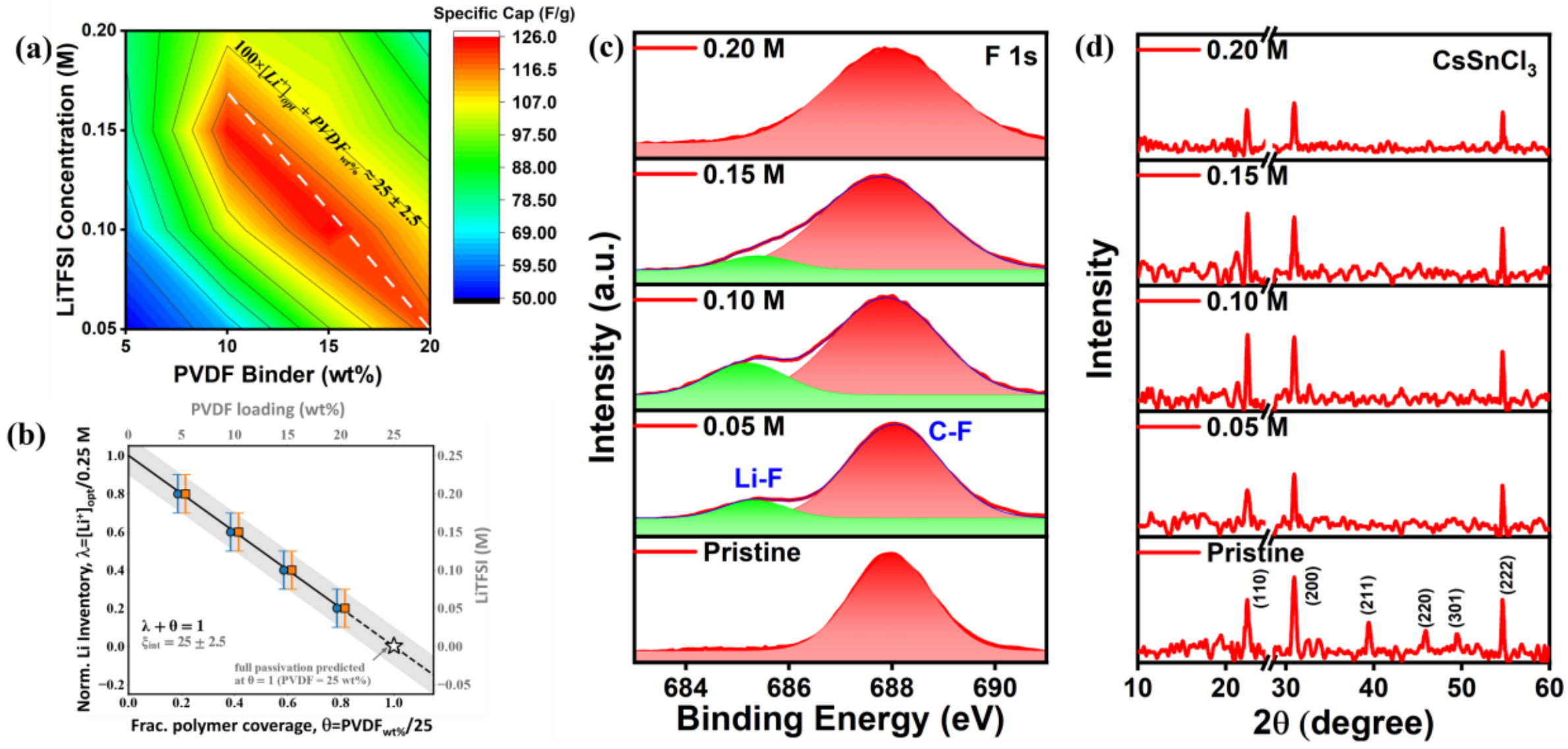

***Figure 4.*** *Interfacial balance rule and structural validation. (a) 2D specific capacitance heatmap across the binder-electrolyte matrix highlighting the optimal performance trajectory. (b) Linear regression of optimal LiTFSI concentration versus PVDF weight fraction. (c) High-resolution F 1s XPS spectra tracking the evolution of C-F and Li-F bonds as a function of electrolyte molarity. (d) XRD patterns confirming the structural stability of* $CsSnCl_3$ *across different electrolyte concentrations.*

The dimensional relationship can be expressed in a normalized form by defining the lithium and polymer contributions relative to their limiting values:

$$\lambda = \frac{[Li^{+}]_{opt}}{0.25\ M},\ \theta = \frac{PVDF_{wt\%}}{25}$$

yielding:

$$\lambda + \theta = 1$$

The experimental concentration resolution introduces an uncertainty of approximately ±0.10 in λ, arising from the 0.05 M spacing between adjacent electrolyte concentrations used to identify the optimum. This dimensionless relationship indicates that polymer loading and lithium availability represent complementary contributions to the optimized interfacial state. To probe the chemical origin of this balance and determine how electrolyte concentration influences the PVDF-mediated interfacial environment, high-resolution X-ray photoelectron spectroscopy (XPS) was performed. Core level F 1s spectra as a function of electrolyte concentration at a fixed 15wt% PVDF loading are shown in Figure 4 (c). Complementary C 1s, Sn 3d, and Cl 2p spectra (Figures S14–S15) confirm that the Sn–Cl lattice environment remains unchanged, excluding ligand loss as the origin of the electrochemical behavior.

**The Pristine/Fresh State:** The entire range exhibits a single symmetric peak centered at ~688.5 eV, which corresponds to the covalent carbon-fluorine (C-F) bonds of pristine poly(vinylidene fluoride) backbone.

- **The Low Concentration Regime (0.050 M):** With the addition of LiTFSI electrolyte, an additional peak emerges at the lower binding energy of ~685.5 eV. It is consistent with Li–F-associated interfacial species arising from fluorine coordination with lithium ions.

However, because of the limited lithium concentration, the Li–F-associated population remains low, and the measured capacitance is correspondingly lower.

- **The Optimal Regime (0.100 M):** At 0.10 M, the concentration predicted by the interfacial balance relationship for a 15 wt% PVDF loading, the Li–F component reaches its maximum fraction of the F 1s signal while the polymeric C–F component near 688.5 eV is retained, so the two coexist rather than the polymer being converted entirely into Li–F species. This composition coincides with the highest specific capacitance measured in the series, 126 F/g at 0.2 A/g.
- **The Over-Saturated Regime (0.150 M to 0.200 M):** Unlike the expectation that high salt molarity would continue to promote Li–F coordination, the increase in concentration above the stoichiometric limit leads to a clear and gradual reduction in the Li-F coordination peak. This indicates that the interface possesses a finite accommodation capacity for Li–F coordination. Once the available fluorinated coordination sites become saturated, any further arrival of lithium to the interface would not be coordinated through the PVDF film, promoting lithium aggregation outside the polymer-regulated interfacial environment. This reduces the fraction of Li–F-associated species observed by XPS, thus explaining the weakening of the XPS intensity. At higher electrolyte concentrations, the interface progressively deviates from the optimized coordination environment, with the emergence of less-coordinated lithium environments and a reduced contribution from Li–F-associated species. This loss of interfacial regulation leads to less uniform lithium accommodation and a corresponding decrease in capacitance (Figure 3). The atomistic origin of this capacity limit—specifically the transition from a confined planar lithium arrangement to out-of-plane aggregation—is investigated through computational modeling in the subsequent sections.

To investigate whether this optimized interfacial composition is effective in protecting the sensitive perovskite bulk against chemical dissolution or breakdown in the liquid medium, ex-situ X-ray diffraction (XRD) was done after conducting an extensive electrochemical test at all electrolyte concentrations. As seen in Figure 4d, the crystal structure of the $CsSnCl_3$ nanocrystals remains unchanged. The characteristic X-ray diffraction peaks from the crystallographic planes (100), (200), (211), (220), (301), and (222) are sharp and intense without any positional shift as compared to the pristine perovskite crystal structure. No traces of any secondary phases such as degraded tin chloride or complexed salt are observed in the analyzed spectrum.-[13,29] This structural stability provides direct macroscopic evidence of our interfacial engineering approach: under conditions satisfying the Interfacial Balance Rule ($100 \times [Li^+]_{opt} + PVDF_{wt\%} \approx 25$), the in-situ formed Li–F-associated interfacial environment maintains the structural integrity of lead-free perovskite during electrochemical operation. These observations demonstrate that optimized polymer–electrolyte interfacial balancing preserves the structural integrity of $CsSnCl_3$ while maintaining electrochemical activity. The distinction between bulk structural stability and interfacial optimization motivates the investigation of whether this balance extends beyond the $CsSnCl_3$ lattice, which is addressed using $MASnCl_3$ in the following section. Ex situ SEM (Figure S13) likewise shows no detectable morphological degradation after electrochemical testing.

### 2.4. Generalization of the Interfacial Balance Rule in the $MASnCl_3$ System

To determine whether the interfacial balance relationship identified in $CsSnCl_3$ is specific to the inorganic lattice or reflects a more general polymer–electrolyte interfacial phenomenon, we extended the analysis to the organic-inorganic hybrid perovskite $MASnCl_3$. Whereas $CsSnCl_3$ forms a rigid all-inorganic lattice, $MASnCl_3$ contains bulky and highly polar methylammonium ($MA^+$) cation that considerably changes the dielectric properties of the bulk.[13,28] However, because the dominant charge-storage processes occur at the polymer-modified two-dimensional interface, we hypothesized that this balance between polymer loading and lithium availability should remain independent of the A-site cation chemistry. As evident from figure

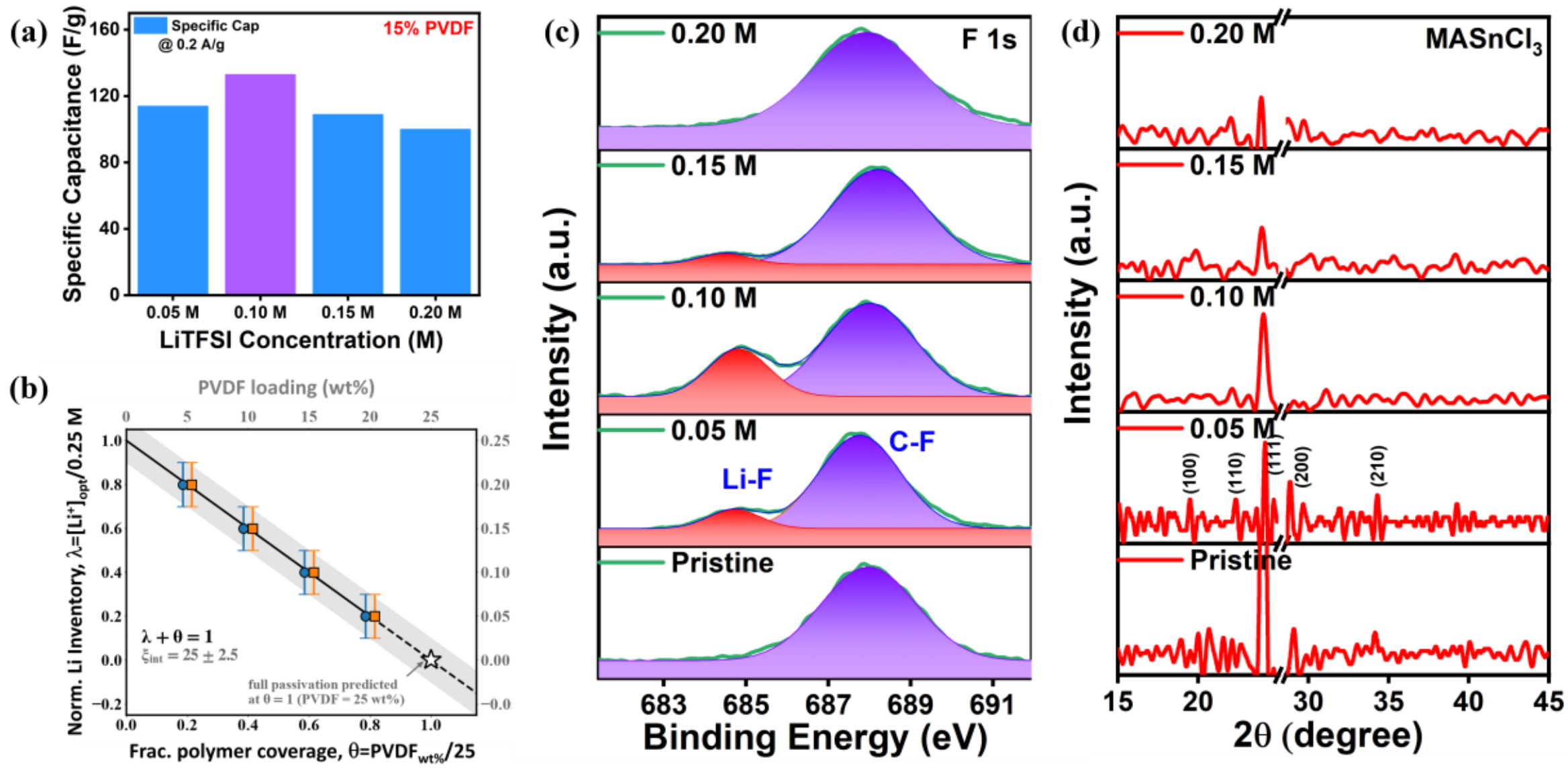

***Figure 5.*** *Validation of the Interfacial Balance Rule and structural stability. (a) Specific capacitance profile evaluated at 0.2 A/g for the 15 wt% PVDF electrode across varying LiTFSI concentrations. (b) Linear correlation of the optimal LiTFSI concentration versus PVDF mass loading. (c) Concentration-dependent evolution of the high-resolution F 1s XPS spectra. (d) Ex-situ XRD patterns confirming the strict preservation of the bulk* $MASnCl_3$ *crystal structure across all electrolyte concentrations.*

5 (a), the electrochemical behavior of $MASnCl_3$ at different $Li^+$ concentration closely reproduces the bell-shaped dependency obtained earlier for the inorganic perovskite material. Transferability of this interfacial balance relationship to the hybrid perovskite is supported by the corresponding GCD and CV data and the two-dimensional contour maps (Figures S16–S25). In case of 15 wt% PVDF loading, the specific capacitance is maximized at 0.10 M reaching the values comparable to those achieved by optimized $CsSnCl_3$ devices. Plotting the optimum electrolyte concentration against PVDF loading for $MASnCl_3$ reproduces the same inverse relationship observed in $CsSnCl_3$. The corresponding interfacial constant remains unchanged:

$$100 \times [Li^{+}]_{opt} + PVDF_{wt\%} = \xi_{Int} \approx 25$$

within the experimental uncertainty of the concentration grid. In normalized form, both perovskite systems collapse onto the same relationship, $\lambda + \theta = 1$, demonstrating that the governing parameter is the polymer–electrolyte interfacial balance rather than the specific A-site chemistry of the perovskite lattice.

The F 1s spectra of the $MASnCl_3$ electrodes (Figure 5(c)) show the same concentration dependence. Considering the 15 wt% PVDF matrix, the optimal 0.10 M solution produces the maximum population of Li-F-associated interfacial species while maintaining the fluorinated polymer framework. The coexistence of the polymeric C-F component (~688.0 eV) and the Li-F-associated component (~685.0 eV) indicates an optimized fluorine-mediated interfacial environment rather than complete conversion of the polymer into Li-F species. Deviation from the optimal 0.10 M concentration prevents establishment of this balanced coordination environment. Below 0.05 M concentration, there is an underutilization of reactive ions leading to development of an incomplete conductive network. Above the optimum (0.10 M), increasing

LiTFSI concentration promotes ion pairing and steric crowding associated with the bulky $TFSI^-$ anions. In the same way seen in the inorganic system, the electrostatic crowding effects reduce the effective availability of interfacial $Li^+$ for productive polymer-regulated accommodation, thereby limiting further growth of the fluorine-associated interfacial population. Consequently, additional lithium is no longer incorporated efficiently into the optimized polymer-defined interfacial environment, resulting in reduced Li-F-associated spectral intensity and deterioration of charge-transfer characteristics. Elemental cross-XPS analysis (figures S27-S28) further confirms that the Li-F-associated interfacial signature tracks the optimized polymer-electrolyte state in both inorganic and hybrid systems. Further support for this interpretation is provided by the XRD patterns (Figure 5 (d)). Despite the huge variations in electrochemical performance at various concentrations, the bulk $MASnCl_3$ crystal structure remains preserved. In addition to this structural evidence, ex-situ SEM analysis of the $MASnCl_3$ electrodes following the electrochemical analysis (Figure S26) confirms complete morphological preservation and the absence of significant particle aggregation or active-phase dissolution at compositions satisfying the interfacial balance condition. These results decouple interfacial failure from bulk structural degradation, showing that performance losses arise from exceeding or underutilizing the finite accommodation capacity of the polymer–electrolyte interface rather than from degradation of the perovskite lattice.

### 2.5. Computational Investigation of $CsSnCl_3$ – PVDF Interfacial Energetics

Having established the applicability of the Interfacial Balance Rule across both inorganic and organic–inorganic perovskites, we next investigated the atomic-scale origin of this interfacial balance. The objective was to identify the structural and energetic factors governing PVDF-mediated surface passivation and lithium accommodation. All atomistic simulations were performed using the MACE machine-learned interatomic potential pre-trained against the meta-GGA $r^2$SCAN reference dataset within the Atomic Simulation Environment (ASE).[30–33] The potential accurately describes the structural characteristics of the $CsSnCl_3$ lattice, β-PVDF[34,35] backbone, and lithium environment while enabling nanosecond-scale simulations of the 906-atom interfacial systems considered here. Its transferability was assessed for each component of the composite interface, including $CsSnCl_3$, β-PVDF, and metallic lithium (Table S1). Structural properties, including lattice parameters and bond geometries, are reproduced within 3% of reference experimental or first-principles values. The potential also captures bulk energetic properties with good quantitative agreement, while preserving the correct structural and bonding characteristics across all constituent materials. In particular, the C–C and C–F bond lengths of β-PVDF, which define the alternating fluorine arrangement responsible for interfacial interactions, agree closely with reported crystallographic values (Table S1). Since the mechanistic analysis is based primarily on relative energetic differences among chemically related interfacial configurations, the identified trends are less sensitive to systematic errors in absolute energy prediction. A $CsSnCl_3$ (100) slab was constructed from the cubic $Pm\bar{3}m$ structure. Cleavage along the [100] direction produces inequivalent $SnCl_2$- and CsCl-terminated surfaces, resulting in an intrinsically asymmetric slab. Because these two terminations possess distinct chemical environments, their surface energies cannot be obtained by assigning half of the total slab energy to each surface. The two terminations were therefore treated independently. The individual surface energies were obtained using the Simultaneous Equations Method[36]which separates the contributions of the two inequivalent terminations. The calculated minimum surface energies are 0.058 J $m^{-2}$ for the relaxed $SnCl_2$ termination and 0.080 J $m^{-2}$ for the CsCl termination. These values agree well with previous PBEsol calculations, which reported termination-dependent surface energies ranging from 0.055–0.148 J $m^{-2}$ for CsCl and 0.055–0.135 J $m^{-2}$ for $SnCl_2$ under different chemical-potential conditions. [37] Because the machine-learned potential does not explicitly sample chemical-potential variation, comparison is made against the minimum reported surface energy for each termination. Although direct comparison between different exchange-correlation functionals should be interpreted cautiously because of differing reference states, the machine-learned potential successfully reproduces the essential physical trends. The calculated $SnCl_2$ surface energy of 0.058 J $m^{-2}$ agrees within 0.001 J $m^{-2}$ with the PBEsol value (0.057 J $m^{-2}$, 500 eV cutoff) under $SnCl_2$-rich conditions. Both approaches predict low surface energies (<0.1 J $m^{-2}$) at their respective minima, consistent with the intrinsically soft and weakly bound nature of halide perovskite lattices.[37–40] The $SnCl_2$ termination is lower in energy than CsCl (0.058 vs. 0.080 J $m^{-2}$) and was therefore selected as the reference surface for subsequent interfacial calculations. To investigate PVDF adsorption on the perovskite surface, a hydrogen-capped all-*trans* oligomer ($C_8H_{10}F_8$) was placed on both the $SnCl_2$- and CsCl-terminated $CsSnCl_3$ surfaces in five distinct initial orientations (Figures 6(a–d) and S29): side-on, F-down, H-down, perpendicular, and tilted (45°). Each orientation was optimized from multiple initial surface separations. The lowest-energy relaxed structure for each orientation was retained for analysis. On the $SnCl_2$-terminated surface, the side-on configuration (Figure 6(d)) is the global minimum with an adsorption energy of −4.615 eV. In this geometry, the polymer backbone lies parallel to the surface, enabling simultaneous interaction of the –$CF_2$– groups with surface $Sn^{2+}$ cations and the –$CH_2$– groups with surface $Cl^-$ anions. The remaining orientations are substantially less favorable, with adsorption energies of −0.418 eV (H-down), −0.247 eV (F-down), −0.184 eV (tilted), and −0.150 eV (perpendicular), corresponding to an orientational energy spread of 4.47 eV. The same energetic ordering is preserved on the CsCl-terminated surface (Figure S30). The side-on configuration again represents the global minimum (−4.612 eV), followed by F-down (−4.213 eV) and perpendicular (−3.692 eV), whereas H-down (−0.112 eV) and tilted (−0.084 eV) remain only weakly adsorbed, spanning an energy range of 4.53 eV. The smaller energetic separation between the side-on and F-down configurations reflects the weaker steric constraint imposed by the CsCl termination, allowing fluorine-rich configurations to approach the surface more closely than on the $SnCl_2$ termination. The consistent energetic ordering across both surface terminations indicates that the side-on geometry is the thermodynamically preferred adsorption mode of PVDF, while the magnitude of the stabilization depends on the surface termination. This orientational preference originates from simultaneous interaction with both ionic sublattices of the perovskite surface (Table S2). Among the examined configurations, the side-on geometry is the only orientation that simultaneously establishes multiple F···cation and H···anion contacts, yielding four F···Sn and eleven H···Cl contacts on the $SnCl_2$ termination (Table S2). In contrast, the H-down and F-down configurations interact predominantly with only one ionic sublattice, whereas the perpendicular and tilted orientations form comparatively few interfacial contacts. On the CsCl termination, the $Cs^+$ sublattice additionally stabilizes an F-down configuration through eight F···Cs contacts, explaining why this orientation lies only 0.40 eV above the global minimum. The generality of this adsorption mechanism was further assessed by varying the PVDF chain length. Side-on adsorption calculations for oligomers containing two to six repeat units show that the adsorption energy increases from −4.392 to −5.004 eV with increasing chain length (Figure 7 (c), Table S3), following an approximately linear relationship:

$$E_{\mathrm{ads}} = -0.140\,n - 4.10\ \mathrm{eV}\ (R^2 = 0.943),$$

where $n$ is the number of $CH_2$–$CF_2$ repeat units. The incremental stabilization per additional repeat unit is small relative to the total binding, and the normalized adsorption energy E_ads/n decreases monotonically from −2.20 eV (n = 2) to −0.83 eV (n = 6). Stabilization is therefore dominated by a chain-length-independent contribution associated with establishing the initial flat-lying contact, with each further repeat unit contributing only a modest increment; the binding does not scale with contact area. The fitted intercept is a parameter of the regression over n = 2–6 and is not interpreted as a physical binding energy, since the $n \to 0$ limit corresponds to no adsorbed polymer. Across the series the slab reference energy is constant to within 0.6 meV (Table S3), confirming that the trend is not an artifact of the energy reference. Consistent with this interpretation, the minimum polymer–surface separation in side-on geometry remains 2.68 Å throughout the series, and neither surface reconstruction nor chemical bond formation is observed.

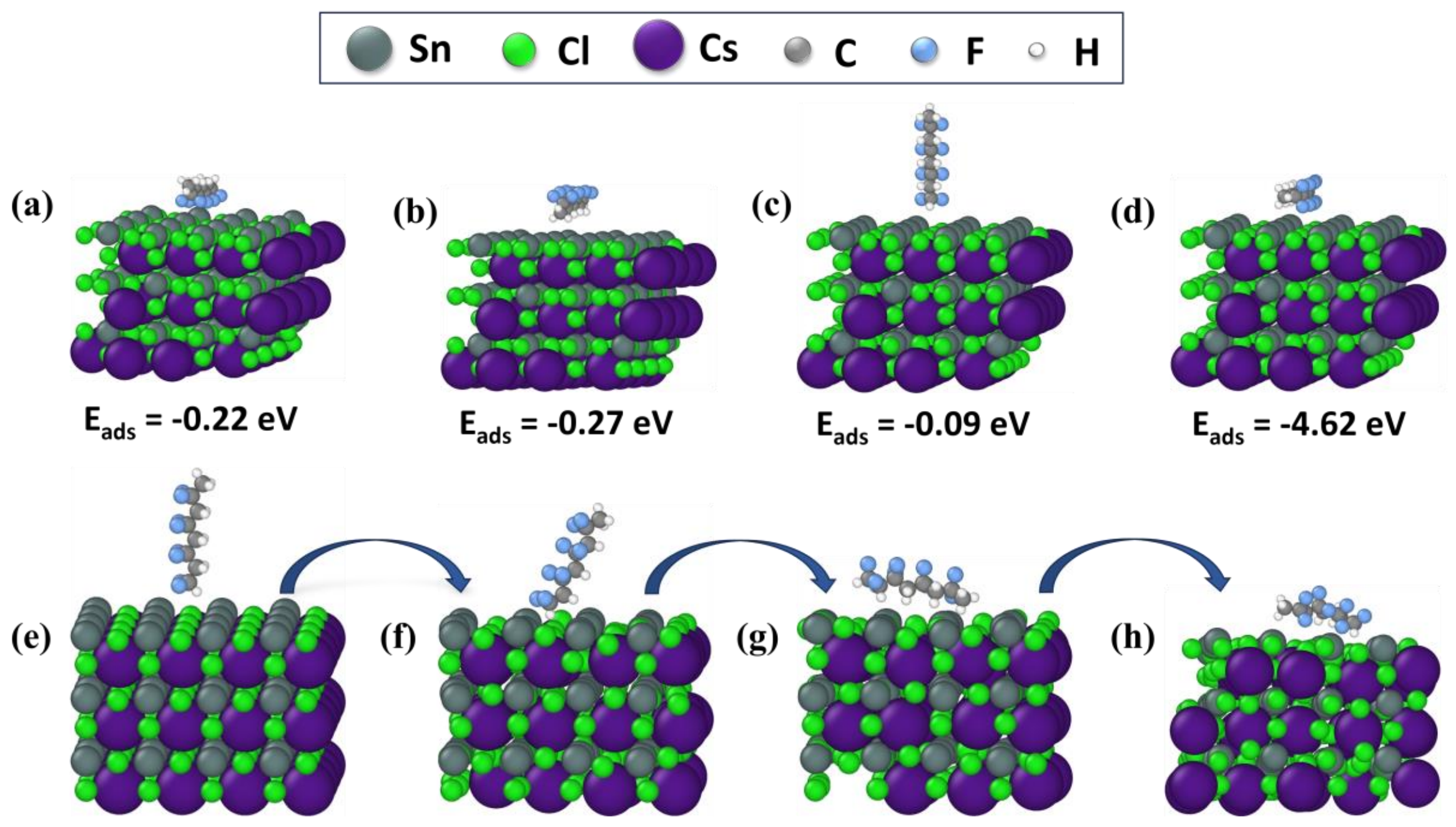

***Figure 6.*** *$CsSnCl_3$ (100) surface slab with different PVDF orientations on $SnCl_2$ termination. (a) F-down, i.e. F atoms face the slab. (b) H-down, i.e. H atoms face the slab. (c) Perpendicular orientation w.r.t slab. (d) F & H down, i.e. both the atoms face the slab. (e)-(h) shows the time evolution of the $CsSnCl_3$ (100) surface slab with PVDF in perpendicular orientation w.r.t slab on $SnCl_2$ termination at 300 K simulated via MD with time-interval of 100 fs. This reveals the thermodynamically unfavourable nature of PVDF perpendicular orientation as the chain readily reorients in the side-on configuration early in simulation and is maintained throughout the course of the 100 ps simulation.*

***Finite-temperature conformational stability of the adsorbed PVDF overlayer***

To examine the finite-temperature stability of the side-on adsorption geometry, molecular dynamics simulations were performed in the NPT ensemble at 300 K and 1 bar for 100 ps using the 4 × 4 supercell. Two independent trajectories were initialized from the relaxed side-on configuration and the relaxed perpendicular configuration (Figures 6(e–h) and S31–S32). In both simulations, the polymer converges to the same adsorbed state. The initially perpendicular chain spontaneously reorients into a quasi-planar configuration with a time-averaged tilt angle of 3.8° ± 2.9°, while the side-on configuration remains stable throughout the simulation with an average tilt angle of 4.9° ± 3.8°. These results demonstrate that the side-on geometry is not only the static energy minimum but also the preferred finite-temperature adsorption state. Throughout the trajectory, the adsorbed polymer retains an extended β-PVDF-like all-trans conformational character, with a mean end-to-end distance of 7.52 ± 0.56 Å compared with 8.9 Å for the fully extended chain, a *trans* dihedral fraction of 0.49, and a mean backbone dihedral angle of 123.3°. Thus, adsorption preserves approximately 84% of the maximum chain extension without inducing significant chain coiling or conformational collapse. Analysis of the equilibrated trajectories further reveals persistent vertical segregation of the polymer functional groups. Fluorine atoms remain at an average height of 3.59 Å above the perovskite surface, whereas hydrogen atoms occupy an average height of 2.77 Å, maintaining a separation of approximately 0.8 Å throughout the simulation. Consequently, the outer surface of the adsorbed overlayer remains fluorine-rich, while the methylene groups remain preferentially oriented toward the perovskite substrate. This stable fluorine-enriched interface provides the structural framework for the lithium insertion mechanism discussed in the following section.

### 2.6. Lithium Insertion Thermodynamics, Interfacial Coordination Geometry, and Transport Kinetics at the Passivated Interface

#### *Site Energetics on the Bare and Passivated Surfaces*

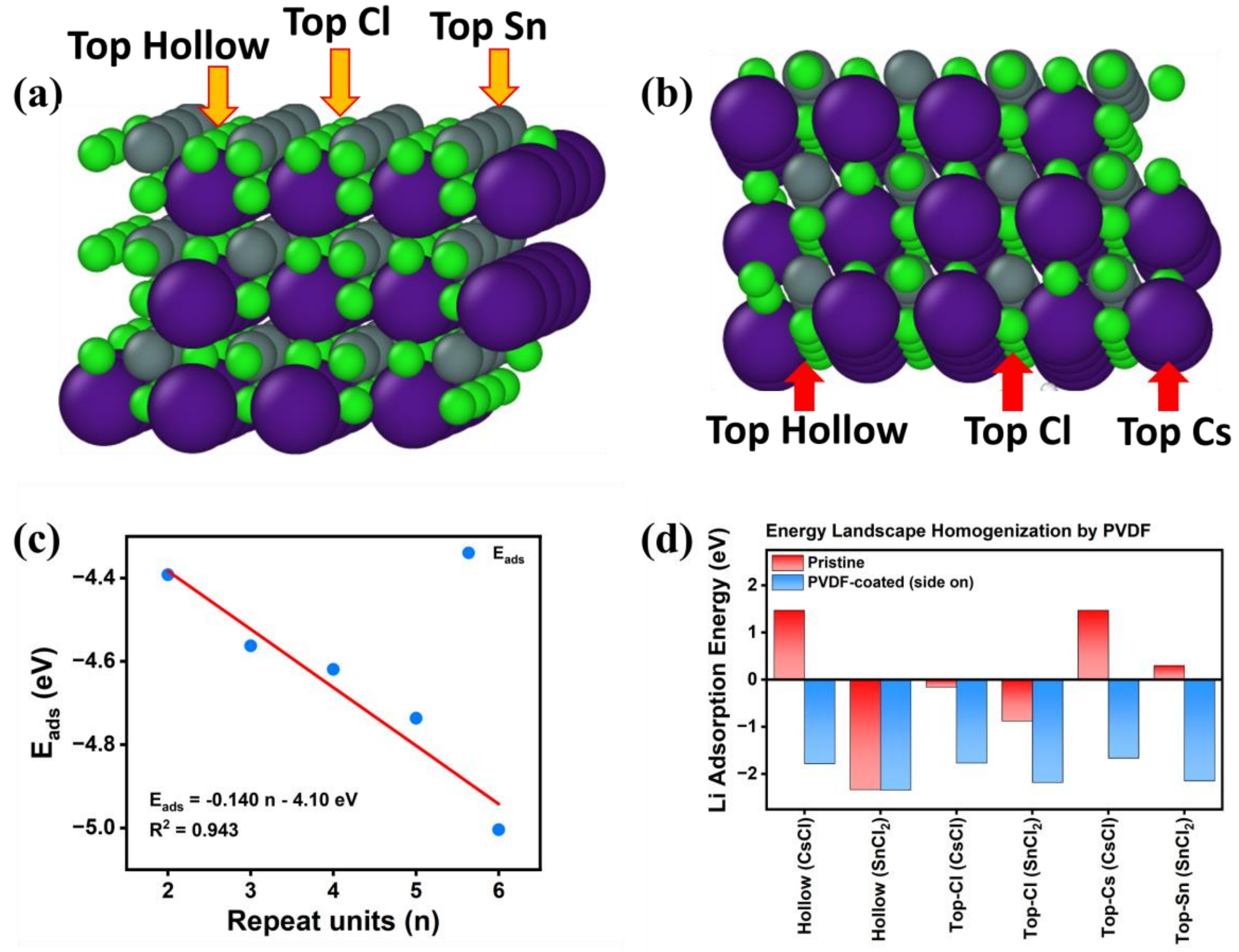

***Figure 7.*** *(a) Top-view representation of high-symmetry* $Li^+$ *adsorption sites on the* $SnCl_2$*-terminated perovskite (100) surface, including Hollow, Top-Cl, and Top-Sn sites. (b) Corresponding adsorption sites on the CsCl-terminated surface, including Hollow, Top-Cl, and Top-Cs sites. These sites define the high-symmetry adsorption configurations used for* $Li^+$ *adsorption energy calculations. (c) Adsorption energy of all-trans PVDF oligomers on the* $SnCl_2$*-terminated* $CsSnCl_3$*(100) surface as a function of chain length, for two to six* $CH_2$*-*$CF_2$ *repeat units in the side-on geometry. The linear dependence indicates an additive non-covalent interaction. The slope gives the adsorption energy per repeat unit, while the intercept collects the length-independent contributions of surface relaxation and the terminal groups. (d) Energy landscape homogenization by the PVDF binder. Comparison of* $Li^+$ *adsorption energies across high-symmetry surface sites for pristine (open bars) and PVDF-coated (filled bars) interfaces. The polymer layer reduces the large site-dependent binding-energy variation of the bare surface into a narrower and more uniform* $Li^+$ *binding-energy distribution.*

the energetic landscape for lithium insertion at the $CsSnCl_3$ interface. The relaxed surfaces expose several high-symmetry adsorption sites for $Li^+$, namely the Hollow site located between the surface cation and anion sublattices and atop sites directly above surface $Cl^-$ (Top-Cl), $Sn^{2+}$ (Top-Sn, $SnCl_2$ termination only), and $Cs^+$ (Top-Cs, CsCl termination only) (Figure 7(a,b)). To identify the equilibrium adsorption geometry at each site, $Li^+$ was initialized at four vertical separations (1.8, 2.2, 2.6, and 3.0 Å) above each lateral position and independently relaxed. The lowest-energy configuration was retained for analysis. This procedure avoids trapping the optimizer in metastable geometries within the confined region between the rigid perovskite surface and the adsorbed polymer layer. Adsorption energies are referenced to bulk bcc lithium ($\mu_{Li} = -2.3845$ eV atom$^{-1}$), such that negative values correspond to insertion sites more stable than metallic lithium. On the pristine $SnCl_2$ termination, lithium adsorption is strongly site dependent (Figure 7(d)). The Hollow site is the global minimum with an adsorption energy of −2.33 eV, followed by Top-Cl (−0.87 eV), whereas Top-Sn is unfavorable (+0.30 eV), consistent with electrostatic repulsion from the exposed $Sn^{2+}$ cation. The adsorption landscape therefore spans 2.63 eV across the three sites. The CsCl termination exhibits a qualitatively different behavior. Only the Top-Cl site is thermodynamically favorable (−0.16 eV). A lithium ion initialized at the Hollow position relaxes directly onto the neighboring Top-Cs site, and both initial configurations converge to the same final structure (+1.47 eV). Consequently, the pristine CsCl surface contains only two distinct adsorption minima separated by 1.63 eV, reflecting the weaker electrostatic stabilization provided by the larger and more diffuse $Cs^+$ cation. Adsorption changes significantly after PVDF passivation. On the $SnCl_2$ termination, the Hollow site remains essentially unchanged (−2.337 eV), whereas the polymer strongly stabilizes the previously unfavorable Top-Sn and Top-Cl sites to −2.143 and −2.178 eV, respectively. The adsorption-energy spread is therefore compressed from 2.63 to only 0.194 eV, representing an approximately fourteen-fold reduction. An analogous effect is observed on the CsCl termination. Following adsorption of the side-on PVDF layer, the Hollow, Top-Cl and Top-Cs sites converge to nearly identical adsorption energies of −1.780, −1.766 and −1.660 eV, respectively, reducing the energetic spread from 1.63 to 0.120 eV. Thus, irrespective of surface termination, PVDF transforms a highly heterogeneous adsorption landscape into one composed of nearly degenerate lithium binding sites. The origin of this energetic leveling is revealed by the relaxed coordination geometries. On the $SnCl_2$ termination, lithium forms mixed Li–F/Li–Cl coordination at every adsorption site. The Li–F distances remain constant, varying only from 1.864 to 1.941 Å, whereas the accompanying Li–Cl distances range from 2.203 to 2.307 Å. Unlike the lattice-derived chloride coordination, which necessarily depends on the underlying adsorption site, the fluorine coordination supplied by PVDF remains essentially unchanged. The polymer therefore provides a nearly site-independent coordination environment that compensates for differences in the underlying inorganic surface, thereby homogenizing the adsorption energies. The geometric landscape is similarly leveled. On $SnCl_2$, the equilibrium lithium height varies by 2.08 Å across the pristine surface but by only 0.98 Å after polymer adsorption. On the CsCl termination the corresponding variation decreases from 1.52 to 1.02 Å, while the relaxed lateral positions of the three adsorption sites converge to within 0.96–1.84 Å of one another. Consequently, PVDF removes distinctions between adsorption sites in both energy and geometry, producing a nearly uniform interfacial lithium environment. Repeating the calculations with the perovskite slab held rigid separates the intrinsic lithium–interface interaction from substrate relaxation. Under these conditions the adsorption-energy spread decreases from 0.93 to 0.49 eV on $SnCl_2$ and from 0.74 to 0.38 eV on CsCl, indicating that substrate relaxation accounts for approximately half of the total energetic leveling, with the remainder arising directly from polymer-mediated coordination.

***Coordination and Dynamics at the Bare and Passivated Interfaces***

The adsorption calculations identify the equilibrium insertion sites and their relative thermodynamic stability. We next examined how lithium explores this modified energy landscape under finite-temperature conditions using 0.9 ns molecular dynamics simulations in the NPT ensemble at 300 K and 1 bar on both pristine and PVDF-passivated interfaces. On the pristine surface, lithium resides 2.23 ± 0.30 Å above the surface plane and is coordinated exclusively by lattice chloride ions (Figures 8(a–d) and S33).

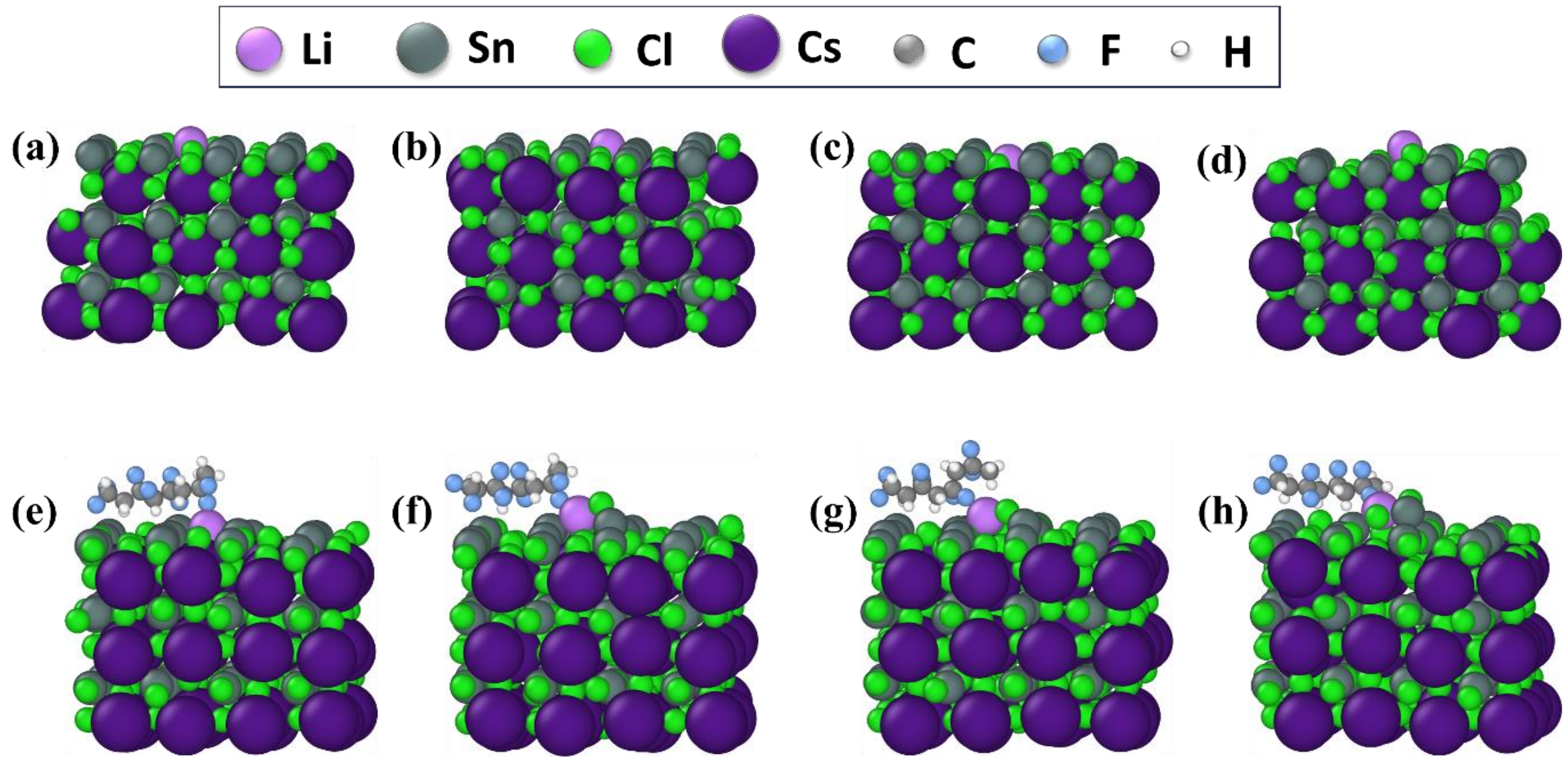

***Figure 8.*** *Molecular dynamics trajectories of a single Li adatom on the (a–d) pristine and (e–h) PVDF-passivated* $CsSnCl_3$ *surface over a total simulation time of 0.8 ns. The displayed trajectories correspond to the final 200 ps of the simulation, sampled at 500 fs intervals. On the pristine surface, Li remains strongly bound to surface Cl sites and exhibits greater lateral diffusion. In contrast, PVDF passivation lifts the Li ion away from the inorganic surface through Li–F interactions while suppressing its lateral mobility.*

Following PVDF adsorption, the equilibrium position shifts to 2.75 ± 0.44 Å above the surface, placing lithium between the inorganic surface and the fluorinated polymer backbone (Figures 8(e–h) and S34). The broader height distribution indicates that the polymer lifts lithium from the deep chloride potential well without replacing it by an equally confining interaction. This structural change is accompanied by a corresponding evolution of the coordination environment. The average Li–Cl coordination number decreases only slightly from 2.91 to 2.65, while a new Li–F coordination number of 0.96 emerges. Rather than replacing chloride coordination, PVDF introduces an additional coordination partner, producing a mixed inorganic–organic first coordination shell. The preference for fluorine coordination becomes more apparent after correcting for atomic abundance. Because fluorine atoms constitute only a small fraction of all possible coordinating atoms, coordination frequencies were normalized by elemental density. The resulting radial distribution functions (Figure 9(a)) show a sharp Li–F peak at 1.97 Å and a Li–Cl peak at 2.37 Å. After normalization, the probability of a Li–F contact exceeds that of Li–Cl by approximately a factor of 24, demonstrating that fluorine is the intrinsically preferred coordination partner despite its much lower concentration. The coordination shell remains highly dynamic throughout the trajectory. Individual Li–F contacts persist for an average of 1.59 ps, compared with 0.72 ps for Li–Cl contacts at the passivated interface and 1.67 ps for Li–Cl contacts on the pristine surface (Figure 9(b)). During the simulation lithium exchanges repeatedly between fluorine-coordinated and uncoordinated states while maintaining an average Li–F coordination number close to unity. The polymer therefore stabilizes lithium through rapidly exchanging fluorine interactions rather than permanent Li–F complexes, producing a dynamic mixed coordination environment that continuously regulates lithium position within the interfacial region. The polymer itself remains structurally stable throughout this process. During the lithium-loaded trajectory the adsorbed chain retains its flat-lying configuration with an average tilt angle of 7.8° ± 7.5°. Although lithium slightly increases thermal fluctuations relative to the unloaded interface, neither desorption nor disruption of the fluorine-rich surface layer is observed.

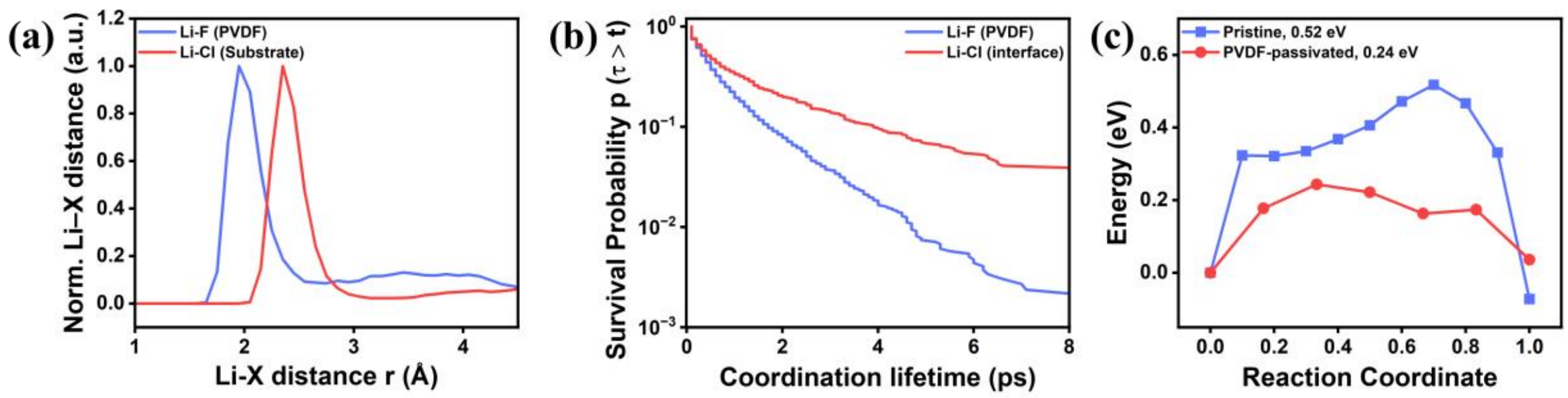

***Figure 9.Finite-temperature MD analysis of Li-ion interfacial dynamics at bare and PVDF-coated perovskite interfaces over a 0.9 ns trajectory at 300 K.*** *(a) Normalized Li-X distance distributions obtained from the PVDF-coated interface, showing the preferential interaction of Li with PVDF fluorine atoms, first peak at 1.97 Å, compared with chlorine atoms at the perovskite surface, first peak at 2.37 Å. Density-normalized, the per-atom probability of a Li-F contact exceeds that of a Li-Cl contact by a factor of 24. (b) Coordination residence survival probability, $P(\tau > t)$, for Li-F and Li-Cl coordination environments. The longer persistence of Li-F coordination, indicates enhanced dynamic stabilization of Li through interactions with the fluorinated polymer layer. (c) Computed minimum energy pathways for lateral $Li^+$ surface diffusion, Hollow site to Hollow site migration, on the pristine $SnCl_2(100)$ surface (black, 0.518 eV) and at the PVDF-passivated interface (red, 0.243 eV). The lower barrier under the polymer is the kinetic counterpart of the compressed adsorption landscape of Figure 7(c), while displacement of lithium out of the surface plane costs 1.08 eV.*

### *Migration Energetics*

The transport mechanism inferred from molecular dynamics was further examined using climbing-image nudged elastic band calculations for elementary lithium migration pathways. Initial and final configurations were independently relaxed before constructing each minimum-energy path. Lateral migration between neighboring adsorption sites is facilitated by polymer passivation rather than inhibited. On the pristine $SnCl_2$ surface, the migration barrier between adjacent Hollow sites is 0.518 eV, whereas beneath the PVDF overlayer the corresponding barrier decreases to 0.243 eV (Figure 9 (c)). This reduction follows directly from the thermodynamic leveling of the adsorption landscape: once adsorption sites become nearly degenerate, the energetic penalty for migration between them is correspondingly reduced. PVDF therefore promotes local redistribution of lithium within the interfacial plane. In contrast, migration away from the interface is energetically disfavored. Transfer of lithium from the surface plane into the polymer-covered interfacial region proceeds monotonically along the minimum-energy path without an intermediate saddle point and is endothermic by 1.08 eV. This quantity therefore represents a reaction energy rather than an activation barrier. By contrast, insertion into the perovskite subsurface is kinetically hindered, with a migration barrier of 1.267 eV beneath the PVDF overlayer, essentially unchanged from the pristine surface value of 1.251 eV. Taken together, these results reveal the mechanism by which PVDF regulates interfacial lithium. The polymer does not immobilize lithium; instead, it flattens the in-plane adsorption landscape, allowing efficient redistribution among nearly equivalent coordination sites while maintaining an energetic penalty for motion away from the interface. Lithium is therefore laterally mobile but vertically confined beneath the fluorine-rich polymer overlayer. This combination of thermodynamic homogenization and spatial confinement provides the atomistic basis for the controlled lithium accommodation described in the following sections.

### 2.7 Two-Dimensional Interfacial Capacity and its Dependence on Polymer Loading

Having established how PVDF modifies lithium adsorption thermodynamics, coordination, and transport at the interface, we next investigated how polymer loading governs the maximum two-dimensional lithium accommodation capacity. Polymer loading was quantified through the areal density of adsorbed PVDF ($\Gamma$), allowing direct comparison across different surface configurations.

***Two independent routes to polymer loading***

The effect of polymer loading was isolated using two independent structural routes. In the first route, the surface cell was enlarged while maintaining a single adsorbed PVDF chain, thereby decreasing the areal chain density. In the second route, the surface cell was fixed while the oligomer length was varied, increasing the polymer density within the same interfacial area. Comparing these two approaches enables the influence of polymer loading to be distinguished from effects associated with the particular supercell or chain-length representation. Along the first route, single four-repeat-unit PVDF chains were adsorbed on $4 \times 4$, $5 \times 5$ and $6 \times 6$ surface supercells, corresponding to chain densities of 0.198, 0.126 and 0.088 chains $nm^{-2}$, respectively. Along the second route, the surface cell remained fixed at $4 \times 4$ while hydrogen-capped oligomers containing two to six $–CH_2–CF_2–$ repeat units were considered, spanning areal monomer densities from 0.395 to 1.185 repeat units $nm^{-2}$. Both routes coincide at the parent system (four repeat units on the $4 \times 4$ surface), allowing the two approaches to be compared directly. Sequential deterministic lithium insertion calculations were then performed for every model, following the structural evolution of the interfacial lithium layer as a function of the total lithium population (Figure 10, 11). The adsorption energy of the first inserted lithium ion varies by less than 3 meV across the complete loading series. Thus, the intrinsic chemical affinity of an isolated binding site is essentially independent of polymer loading. Instead, the differences between systems arise from how the polymer redistributes the finite interfacial area available for lithium accommodation. To quantify the structural evolution of the growing lithium layer, the vertical thickness of the lithium adlayer, $\Delta Z_{Li}$, was monitored together with the fraction of lithium retained within the planar interfacial region. A value of $\Delta Z_{Li} \approx 2.5$ Å is adopted as a geometric indicator for the onset of out-of-plane lithium agglomeration, marking the transition from predominantly two-dimensional interfacial storage to three-dimensional accumulation.

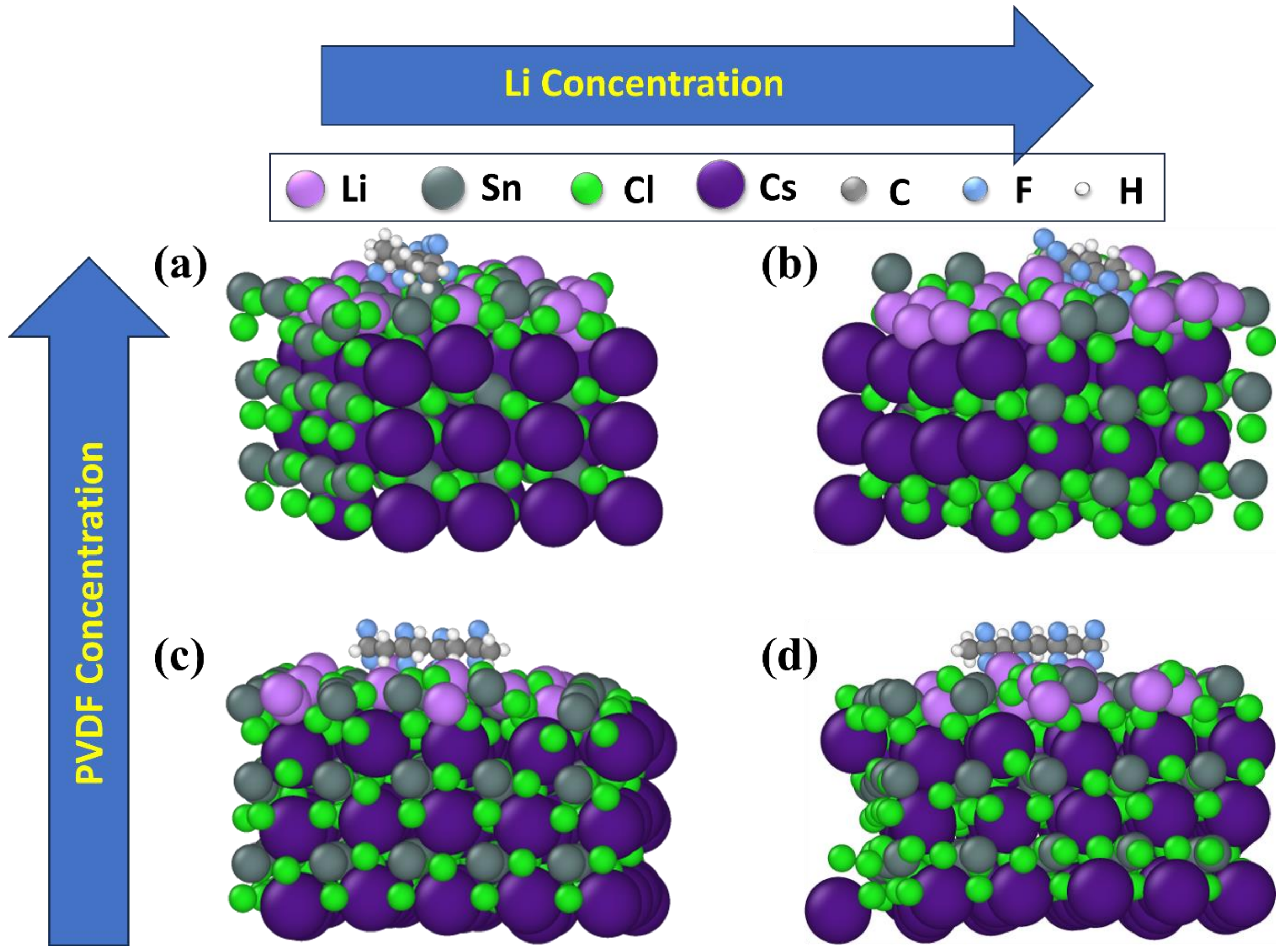

***Figure 10.*** *Structural visualization of the interfacial lithium adlayer as a function of lithium coverage. Panels (a) and (b) show systems with dense PVDF coverage and increasing lithium concentration, while panels (c) and (d) show systems with lower PVDF coverage and increasing lithium concentration. Complete lithium monolayer formation begins at a lower lithium concentration in the 5 × 5 system (dilute PVDF coverage, approximately 0.13 chains nm$^{-2}$) compared with the 4 × 4 system (dense PVDF coverage, approximately 0.20 chains nm$^{-2}$). This behavior demonstrates that the surface density of the polymer strongly influences the distribution and organization of lithium at the interface.*

***Effect of polymer surface density***

Along the density series, the two-dimensional accommodation capacity increases monotonically with polymer coverage. For the highest coverage (0.198 chains $nm^{-2}$), the interface remains essentially planar until an areal lithium density of approximately 4.03 Li $nm^{-2}$ ($N = 20$), beyond which rapid vertical growth begins (Figure 11(a)). At intermediate coverage (0.126 chains $nm^{-2}$), the transition occurs near 2.21 Li $nm^{-2}$, while the lowest coverage (0.088 chains $nm^{-2}$) reaches the transition at approximately 1.98 Li $nm^{-2}$. The denser polymer layer therefore delays the onset of out-of-plane lithium accumulation by providing a more continuous fluorine-rich electrostatic environment together with stronger geometric confinement. This interpretation is independently supported by the fraction of lithium retained within the interfacial plane (Figure 11(b)), which remains consistently higher for the higher-density interfaces over the entire loading range. When normalized per polymer chain, however, the transition occurs within a comparatively narrow window of approximately 18–22 lithium ions per chain (Figure 11(c)). Although the areal capacity changes by approximately a factor of two across the density series, the chain-level capacity remains nearly constant, indicating that each PVDF chain defines an approximately fixed electrostatically stabilized interfacial volume.

*Effect of oligomer length*

The chain-length series reveals a qualitatively different behaviour. At constant chain density, increasing polymer loading does not continuously improve accommodation capacity; instead, the planar storage limit initially increases before reaching a maximum. The accommodation threshold rises from approximately 2.38 Li nm⁻² for the shortest chains to 4.03 Li nm⁻² for the four-repeat-unit oligomer, before decreasing for longer chains (Figure 11(c), Table S4). The parent oligomer therefore represents an optimum polymer loading rather than simply an intermediate composition. This optimum is also reflected in the lithium adlayer morphology. For lithium loadings up to N = 20, the four-repeat-unit chain produces the thinnest lithium layer at three of the four sampled conditions (N = 12, 16, and 20), while at N = 14 the three-repeat-unit chain is marginally thinner (1.74 versus 1.89 Å). At the highest loading sampled (N = 24), the ordering reverses and the four-repeat-unit system exhibits the largest $\Delta Z_{Li}$ (Table S4), consistent with the interface exceeding its accommodation limit at this coverage. Shorter chains leave substantial regions of bare perovskite exposed, allowing lithium to sample the heterogeneous adsorption landscape identified in Section 2.7, whereas longer chains increasingly occupy interfacial area otherwise available for lithium accommodation. Both routes identify the parent loading as the optimum, demonstrating that the governing variable is the amount of polymer occupying the interface rather than chain length or supercell dimensions individually. The two approaches coincide at the parent system by construction and show consistent normalization of polymer coverage. The non-monotonic dependence is resolved by the chain-length series, which spans loadings on both sides of the optimum, whereas the density series samples only the lower-coverage regime and captures the corresponding rising branch.

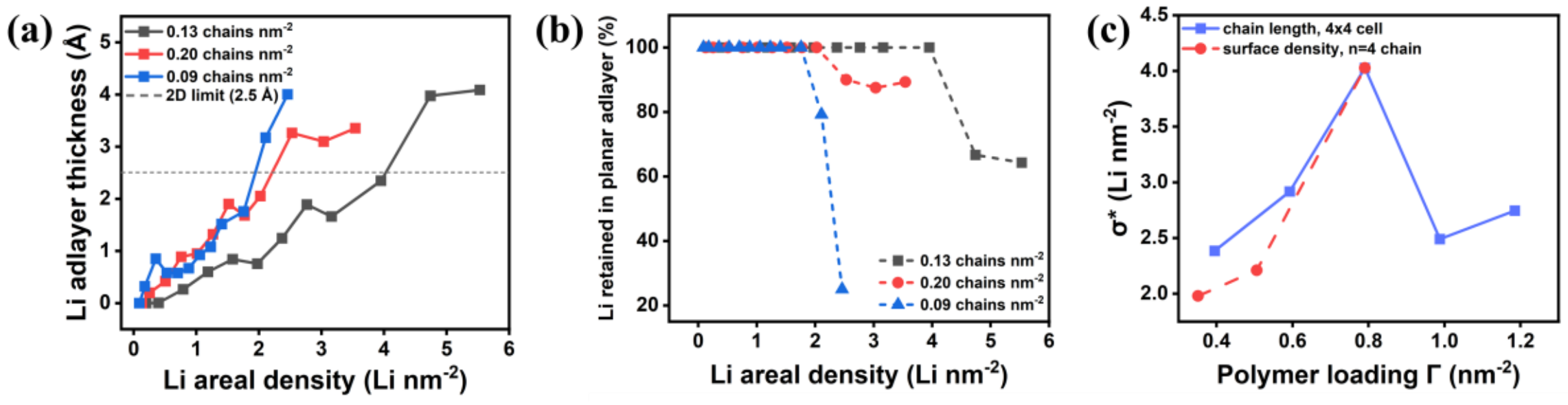

***Figure 11.*** *Interfacial lithium capacity and its dependence on polymer loading. (a) Adlayer thickness $\Delta Z_{Li}$(defined as the peak-to-peak difference in lithium atom heights, $max(h) - min(h)$) and (b) fraction of lithium retained within the planar interfacial region $f_{2D}$(calculated as the fraction of lithium atoms positioned within 2.2 Å of the lowest lithium atom, where $h - min(h) < 2.2$), both against Li areal density, for three PVDF surface densities of 0.198, 0.126 and 0.088 chains nm⁻² built in 4 × 4, 5 × 5 and 6 × 6 surface cells at fixed four-unit chain. The dashed line in (a) marks the 2.5 Å criterion at which the planar adlayer gives way to out-of-plane agglomeration. (c) Planar accommodation limit $\sigma^*$(determined via linear interpolation of the lithium areal density $\sigma$exactly where $\Delta Z_{Li}$crosses the 2.5 Å threshold), against polymer loading Γ along both routes: varying chain length at fixed cell (circles) and varying cell size at fixed chain (squares). The two coincide at the parent composition. $\sigma^*$rises to a maximum of 4.03 Li nm⁻² at Γ = 0.790 units nm⁻² and falls on either side, the structural counterpart of λ + θ = 1.*

***Chain-level capacity and fluorine-mediated interfacial confinement***

The parent hydrogen-capped PVDF oligomer used throughout the structural and dynamical analysis (four $CH_2$–$CF_2$ repeat units, $C_8H_{10}F_8$) contains eight fluorine atoms per polymer chain. Nevertheless, the sequential insertion calculations show that the corresponding interfacial region accommodates approximately twenty lithium ions before the onset of out-of-plane accumulation. Extending this analysis across the oligomer-length series further suggests that the storage capacity is not regulated by a fixed Li:F stoichiometric ratio (Table S5). Instead, the accommodation limit is determined by the polymer-defined interfacial environment established by the combined effects of fluorine-mediated coordination, adsorption-site homogenization, and geometric confinement. The capacity exceeding the number of available fluorine atoms excludes a static one-to-one Li–F coordination mechanism. Rather, PVDF acts as a collective interfacial stabilization layer in which fluorine atoms provide preferential but dynamically exchanging coordination sites, while the entire polymer overlayer contributes to confinement of the lithium population within the two-dimensional interfacial region. This interpretation is consistent with the molecular dynamics results, which reveal transient Li–F coordination together with continuous exchange between fluorine- and chloride-coordinated environments.

**2.8 Physical Origin of the Interfacial Balance Rule**

The coverage-dependent accommodation capacity established above provides the structural basis for the experimentally observed interfacial balance rule:

$$\lambda + \theta = 1$$

equivalently expressed experimentally as:

$$100 \times [Li^{+}]_{opt} + PVDF_{wt\%} \approx 25.$$

Understanding this balance requires considering three coupled consequences of increasing polymer loading. First, PVDF introduces fluorine-rich coordination environments that preferentially stabilize lithium ions. Second, it homogenizes the adsorption landscape by compressing the site-to-site adsorption-energy variation from 2.63 eV on the pristine surface to only 0.19 eV after passivation, allowing lithium to occupy numerous energetically similar configurations. Third, because the projected perovskite surface area is finite, increasing polymer loading inevitably reduces the remaining exposed interfacial area through which lithium can be accommodated. Polymer loading therefore cannot increase the available accommodation volume indefinitely. Beyond an optimum, the gain in coordination quality is offset by the progressive occupation of the finite interface by the polymer itself. The calculations distinguish between two related quantities. Sequential insertion calculations determine the maximum amount of lithium that can be accommodated within the two-dimensional interfacial region before out-of-plane accumulation begins. Electrochemical measurements determine the bulk lithium concentration required to populate this interfacial reservoir. These quantities are connected through the capture efficiency established by the polymer. Interfaces that bind lithium more strongly and more uniformly require a lower bulk lithium concentration to achieve the same interfacial occupation. At low polymer loading, incomplete passivation produces a heterogeneous adsorption landscape and a relatively small planar accommodation capacity. Consequently, higher lithium concentrations are required to populate the interface, yet out-of-plane lithium aggregation begins at comparatively low surface loading. Increasing polymer loading initially improves both the chemical stabilization and the geometric accommodation capacity, lowering the bulk lithium concentration required for efficient interfacial occupation. Beyond the optimum, however, further increases in polymer loading reduce the free interfacial area available for lithium accommodation. Although the local coordination environment remains favorable, the finite two-dimensional interfacial reservoir becomes progressively smaller, causing the accommodation capacity to decrease. The experimentally observed optimization rule therefore reflects the balance between lithium supply and the finite accommodation capacity of a polymer-modified interface of constant projected area. This atomistic picture is directly supported by high-resolution X-ray photoelectron spectroscopy. At the experimentally optimal composition, the F 1s spectra (Figure 4(c)) exhibit the maximum intensity of the Li–F component near 685.5 eV, consistent with the computational prediction that fluorine-mediated lithium coordination reaches its highest population under these conditions. As lithium loading increases beyond the optimum, the available two-dimensional accommodation sites become saturated. Additional lithium ions are forced into out-of-plane aggregation, becoming progressively separated from the fluorine-rich interfacial environment. Consequently, the fraction of lithium participating in Li–F coordination decreases, producing the experimentally observed reduction in the Li–F spectral intensity. The Sn 3d spectra (Figure 4c) provide an important complementary constraint. Across the entire composition range, no $Sn^{4+}$ features are observed, demonstrating that the perovskite lattice remains chemically intact. The deterioration in electrochemical performance away from the optimum therefore does not arise from bulk degradation of the perovskite but from exceeding the finite two-dimensional accommodation capacity established by the adsorbed polymer overlayer. Taken together, the adsorption calculations, sequential insertion analysis, finite-temperature molecular dynamics, migration energetics, and X-ray photoelectron spectroscopy provide a mechanism for the optimization relationship. The optimum results from balancing lithium availability against the finite two-dimensional accommodation capacity defined by the PVDF-modified interface, beyond which additional lithium is diverted into out-of-plane aggregation rather than reversible interfacial storage.

## 3. Conclusions

This study establishes the PVDF binder as an active thermodynamic and kinetic regulator of interfacial charge storage in halide perovskite electrodes, beyond its conventional structural role. The interfacial scaling relationship is quantified by the Interfacial Balance Constant ($\xi_{Int} = 25 \pm 2.5$), which in normalized form gives the Interfacial Balance Rule ($\lambda + \theta = 1$). This relationship defines the capacitance optimum in both rigid inorganic $CsSnCl_3$ and hybrid $MASnCl_3$ systems, demonstrating that lithium supply and polymer loading occupy complementary fractions of a finite interfacial accommodation capacity. Capacitive reversibility is therefore regulated by the two-dimensional binder–electrolyte interface rather than solely by the bulk perovskite lattice. Atomistic modeling with a pre-trained MACE machine-learned interatomic potential reveals that PVDF adopts a side-on adsorption geometry, simultaneously interacting with cationic and anionic surface sublattices. This configuration introduces fluorine-mediated coordination, homogenizes lithium adsorption energetics, and lowers the in-plane migration barrier, so that lithium remains laterally mobile within the interfacial plane while migration away from the interface is energetically unfavorable. The resulting accommodation mechanism arises from dynamically exchanging Li–F and Li–Cl coordination rather than static ion binding. Independent variation of polymer coverage through surface density and chain length demonstrates that increasing PVDF participation initially enhances planar lithium accommodation, whereas excessive polymer occupation reduces available interfacial space and lowers accommodation capacity. This provides the atomistic basis for the experimentally observed balance relationship: increasing binder participation reduces the lithium concentration required to achieve the optimized interfacial state until excessive coverage limits further accommodation. XPS confirms preservation of the $Sn^{2+}$ oxidation state across non-optimal electrolyte compositions, excluding irreversible perovskite degradation as the origin of performance variation. Instead, performance loss arises from exceeding or underutilizing the finite two-dimensional lithium accommodation capacity, beyond which lithium is diverted toward out-of-plane aggregation. Although the simulations represent an idealized defect-free surface, they provide a mechanistic interpretation of the interfacial processes governing the experimental trends. This work establishes binder loading as an active design parameter for perovskite-based and related mixed ionic–electronic interfaces, providing a foundation for predictive design and AI-driven screening of functional binder architectures for next-generation energy-storage materials.

## 4. Methods

### 4.1. Materials

Cesium chloride (CsCl, 99%), methylammonium chloride (MACl), stannous chloride dihydrate ($SnCl_2·2H_2O$, 99%), poly(vinylidene fluoride) (PVDF), N-methyl-2-pyrrolidone (NMP), Super P conductive carbon, and lithium bis(trifluoromethanesulfonyl)imide (LiTFSI) were obtained from TCI Chemicals and Sigma-Aldrich. Oleylamine (99%), toluene (99%) was purchased from SRL Chemicals. All chemicals and reagents were used strictly as received without further purification.

### 4.2. Synthesis of Inorganic $CsSnCl_3$ Nanocrystals

The primary lead-free all-inorganic $CsSnCl_3$ nanocrystals (NCs) were synthesized utilizing, ligand-assisted reprecipitation (LARP) method. In a representative synthesis, 0.10 mM of CsCl, 0.10 mM of $SnCl_2·2H_2O$, and 0.30 mM of oleylamine (acting as the capping agent and structural director) were sequentially dispersed in 10 mL of anhydrous toluene. The mixture was kept under continuous magnetic stirring at 40 °C for 2 h to ensure complete dissolution and precursor complexation. The fundamental chemical reaction governing the formation of the inorganic perovskite lattice is described as follows:

$$CsCl + SnCl_2 \rightarrow CsSnCl_3$$

Following the thermal stirring phase, the solution was subjected to probe ultrasonication for 1 hr to induce rapid, uniform nucleation and constrain the crystal size to the nanometer regime. The resulting colloidal precipitate was isolated via centrifugation at 7000 rpm for 10 min. To remove any unreacted precursor species or excess oleylamine, the collected nanocrystals were thoroughly washed multiple times with toluene. Finally, the phase-pure $CsSnCl_3$ nanocrystals were dried in a hot air oven at 50 °C for 24h prior to electrode formulation.

### 4.3. Synthesis of Organic-Inorganic $MASnCl_3$ Nanocrystals

To evaluate the transferability of the proposed interfacial scaling rule, hybrid organic–inorganic $MASnCl_3$ nanocrystals were synthesized using the same protocol employed for $CsSnCl_3$ nanocrystals to maintain comparable morphological and structural characteristics. In the synthesis, 0.10 mM of MACl was used as the A-site precursor, while the quantities of $SnCl_2 \cdot 2H_2O$ (0.10 mM), oleylamine (0.30 mM), and toluene (10 mL) were kept identical. The reaction mixture was stirred at 40 °C for 2 h according to the following reaction:

$$CH_3NH_3Cl + SnCl_2 \rightarrow CH_3NH_3SnCl_3$$

The subsequent probe ultrasonication, centrifugal isolation, toluene washing cycles, and drying parameters (50 °C for 24 h) were executed in exact parallel to the steps described for the $CsSnCl_3$ nanocrystals in Section 4.2.

### 4.4 Computational Details

All atomistic simulations were performed using the MACE-MH-1 machine-learned interatomic potential with the matpes-r2scan head implemented through the Atomic Simulation Environment (ASE). The potential is pre-trained on meta-GGA $r^2SCAN$ reference data and was independently validated for the $CsSnCl_3$ perovskite, PVDF polymer, and lithium-related structures (Table S1). Reference energies, including clean-slab energies, isolated adsorbate energies, and the bulk lithium chemical potential, were evaluated in double precision. Large-scale interfacial calculations and molecular dynamics trajectories were performed in single precision after confirming that energy differences for the 906-atom interface differed by less than 0.04 eV between the two precision settings.

The $CsSnCl_3$(100) surface was modeled using an asymmetric slab constructed from the relaxed cubic bulk structure. Cleavage along the (100) direction generates inequivalent $SnCl_2$- and CsCl-terminated surfaces. A 4 × 4 surface supercell containing 22 atomic layers and a 20 Å vacuum region was employed. The central slab layers were fixed to suppress coupling between the two terminations while the remaining atoms were fully relaxed. Surface energies of the two terminations were obtained using the Simultaneous Equations Method for asymmetric slabs, and adsorption energies were referenced to the corresponding relaxed clean surface.

PVDF was represented by an all-trans hydrogen-capped oligomer containing four $CH_2$–$CF_2$ repeat units ($C_8H_{10}F_8$), selected to reproduce the alternating fluorine coordination environment while remaining compatible with the lateral surface periodicity. Adsorption structures were obtained by relaxing multiple initial orientations and surface separations. Adsorption energies were calculated as:

$$E_{ads} = E_{slab+adsorbate} - E_{slab} - E_{adsorbate}$$

where negative values represent favorable binding. For lithium adsorption, the isolated lithium reference was replaced by the bulk bcc lithium chemical potential, allowing comparison of interfacial sites relative to lithium metal.

Structural relaxations were performed using FIRE followed by LBFGS optimization until the maximum residual force was below 0.05 eV $Å^{-1}$. Migration pathways were calculated using climbing-image nudged elastic band (CI-NEB) calculations. Molecular dynamics simulations were carried out at 300 K, 1 bar in NPT ensemble using Langevin dynamics with a 1 fs timestep. Lithium coordination, diffusion behavior, and vertical confinement were analyzed from the resulting trajectories.

The interfacial lithium accommodation capacity was determined by deterministic sequential loading. Lithium atoms were introduced into the polymer-modified interface using a deterministic sequential void-filling algorithm on a 40 × 40 fractional grid spanning 0.05 to 0.95 of the cell, laid over the interfacial midplane. The spacing is significantly finer than the 1.8 Å Li-Li separation floor and far below the length scale over which the Li-Li Coulomb interaction varies appreciably in all the studied supercells. The relaxed lithium configurations, rather than the initial loading geometries, were analyzed to determine the maximum planar accommodation density and the onset of the transition from confined two-dimensional accommodation to out-of-plane aggregation.

## ACKNOWLEDGEMENT

M.B. acknowledges the Science and Engineering Research Board (SERB), INDIA, under the award no. CRG/2021/001744 dated 07/03/2022 for partial support for this research work. A.K. acknowledges the Council of Scientific & Industrial Research (CSIR), India, for PhD fellowship. A.K.P. acknowledges the Ministry of Education for their PhD fellowship.

## Conflicts of interest

The authors declare no competing financial interests.

**TOC Graphic**

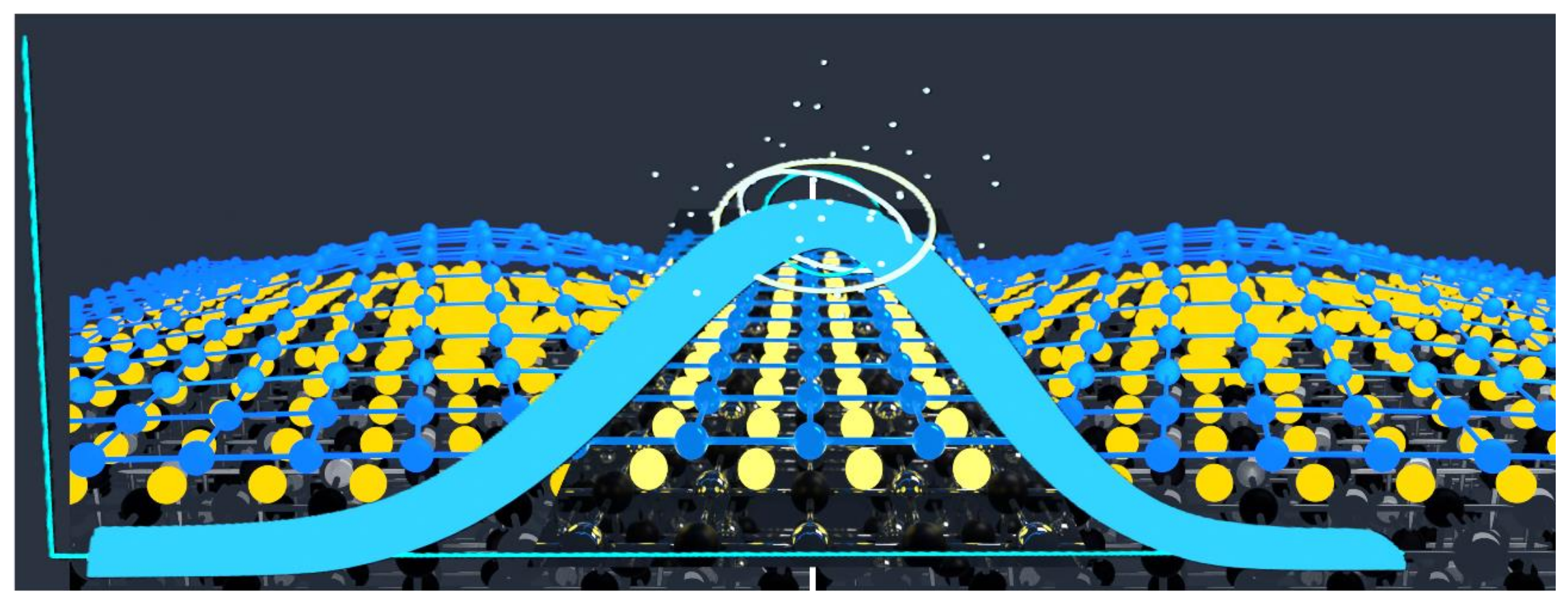

# Supplementary Information

**An Interfacial Balance Rule Governs Binder–Electrolyte Coupling in Lead-Free Perovskite Energy Storage**

Arun Kumar[†, 1], Ayush Kumar Pandey[†, 1], Ankur Yadav[†], Vishnu Saraswat^, Shiladitya Sengupta[†], Abhishek Tewari[#, $, *] and Monojit Bag[†, ‡, *]

[†] *Department of Physics, Indian Institute of Technology Roorkee, Roorkee 247667, Uttarakhand, India*

‡*Centre for Nanotechnology, Indian Institute of Technology Roorkee, Roorkee 247667, Uttarakhand, India*

^*Department of Electronics and Communication Engineering, SR University, Warangal 506371, Telangana, India*

[#] *Department of Metallurgical and Materials Engineering, Indian Institute of Technology Roorkee, Roorkee 247667, Uttarakhand, India*

[$] *Mehta Family School of Data Science and Artificial Intelligence, Indian Institute of Technology Roorkee, Roorkee 247667, Uttarakhand, India*

[1] *Equal contribution*

*Corresponding author: (Monojit Bag): monojit.bag@ph.iitr.ac.in,

(Abhishek Tewari): abhishek@mt.iitr.ac.in

## S1. Calculations

**Areal ($C_A$)** and **specific ($C_S$) capacitance** calculated from GCD measurements using these equations.

$$C_A = \frac{I\int(Vdt)}{m \times V^2} \tag{S1}$$

$$C_s = \frac{I\int(Vdt)}{A \times V^2} \tag{S2}$$

Energy density and power density calculated as

$$E = \frac{1}{7.2} \times C_s(V)^2 \tag{S3}$$

$$P = \frac{E}{t} \tag{S4}$$

**Areal ($C_A$)** and **specific ($C_S$) capacitance** calculated from CV measurements using these equations.

$$C_A = \frac{\int(IdV)}{V \times s \times A} \tag{S5}$$

$$C_s = \frac{\int(IdV)}{V \times s \times m} \tag{S6}$$

where ∫ (IdV) is the active area under CV spectra, V is potential window (V), s is the scan rate (mV $s^{-1}$), A is the active electrode area, m is mass of the active material (g) and t is the galvanostatic discharging time.

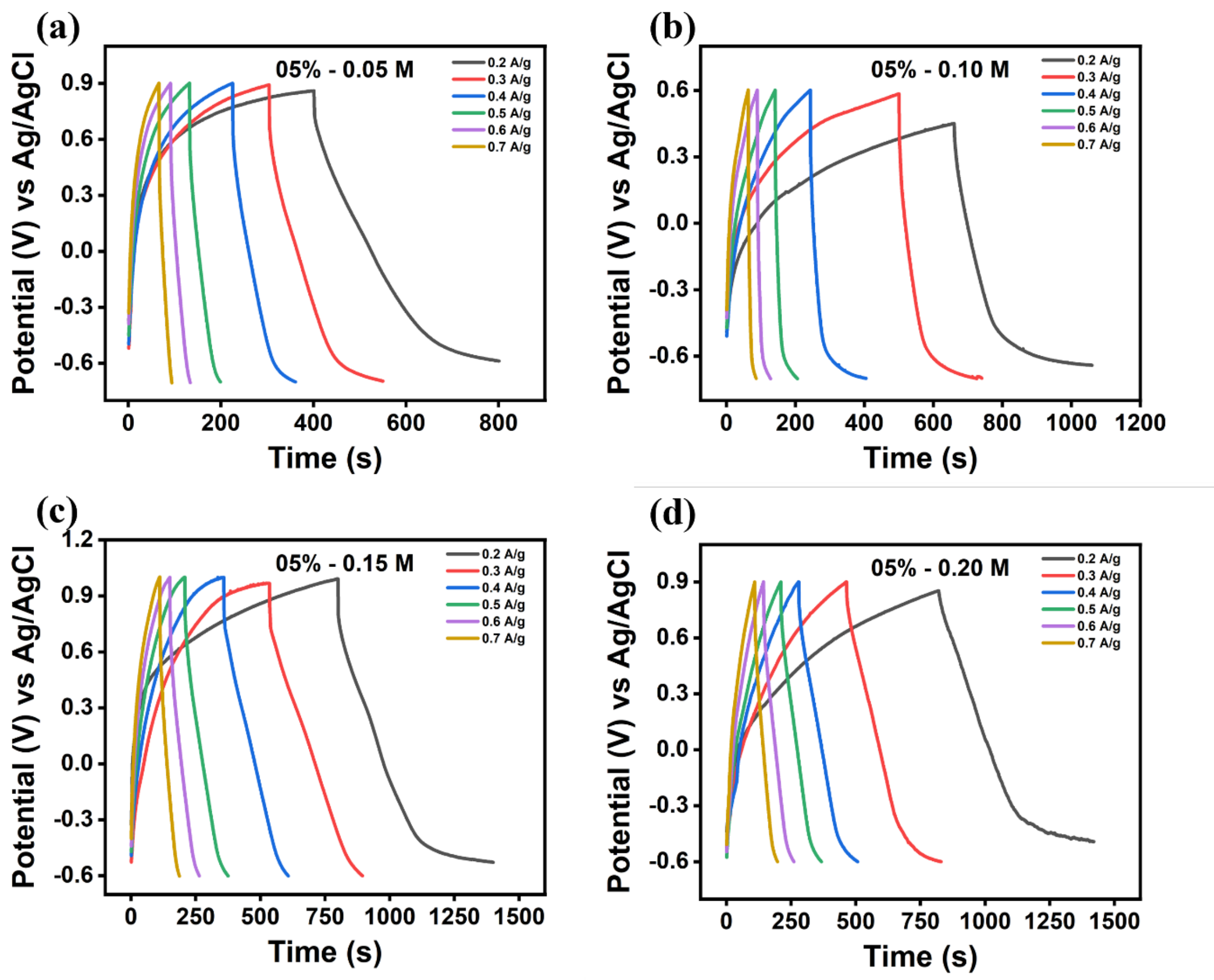

**Figure S1.** Galvanostatic charge-discharge (GCD) profiles of the $CsSnCl_3$ electrode containing 5 wt.% PVDF at various current densities (0.2-0.7 A/g) in LiTFSI electrolytes with concentrations of (a) 0.05 M, (b) 0.10 M, (c) 0.15 M, and (d) 0.20 M.

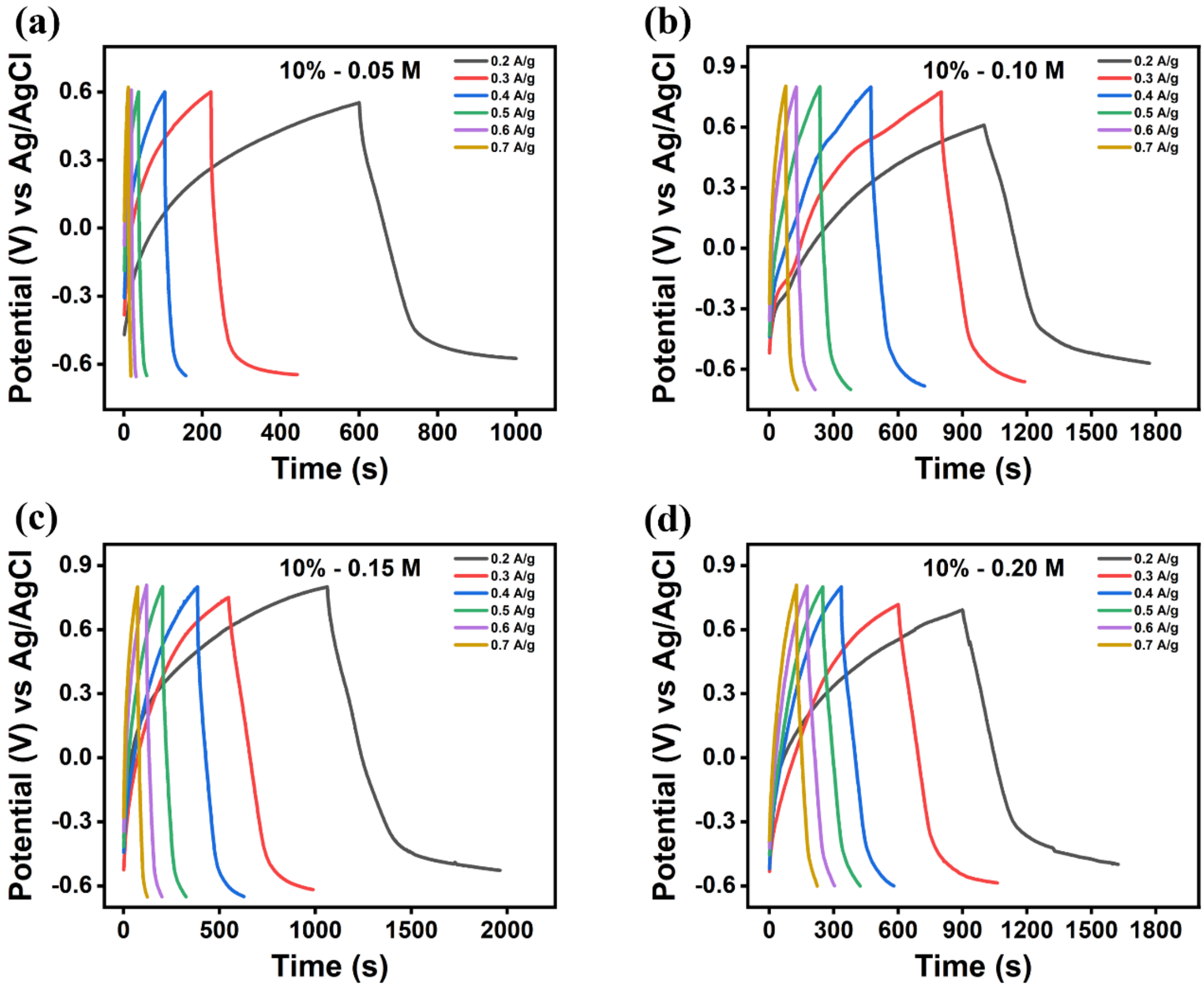

**Figure S2.** Galvanostatic charge-discharge (GCD) profiles of the $CsSnCl_3$ electrode containing 10 wt.% PVDF at various current densities (0.2-0.7 A/g) in LiTFSI electrolytes with concentrations of (a) 0.05 M, (b) 0.10 M, (c) 0.15 M, and (d) 0.20 M.

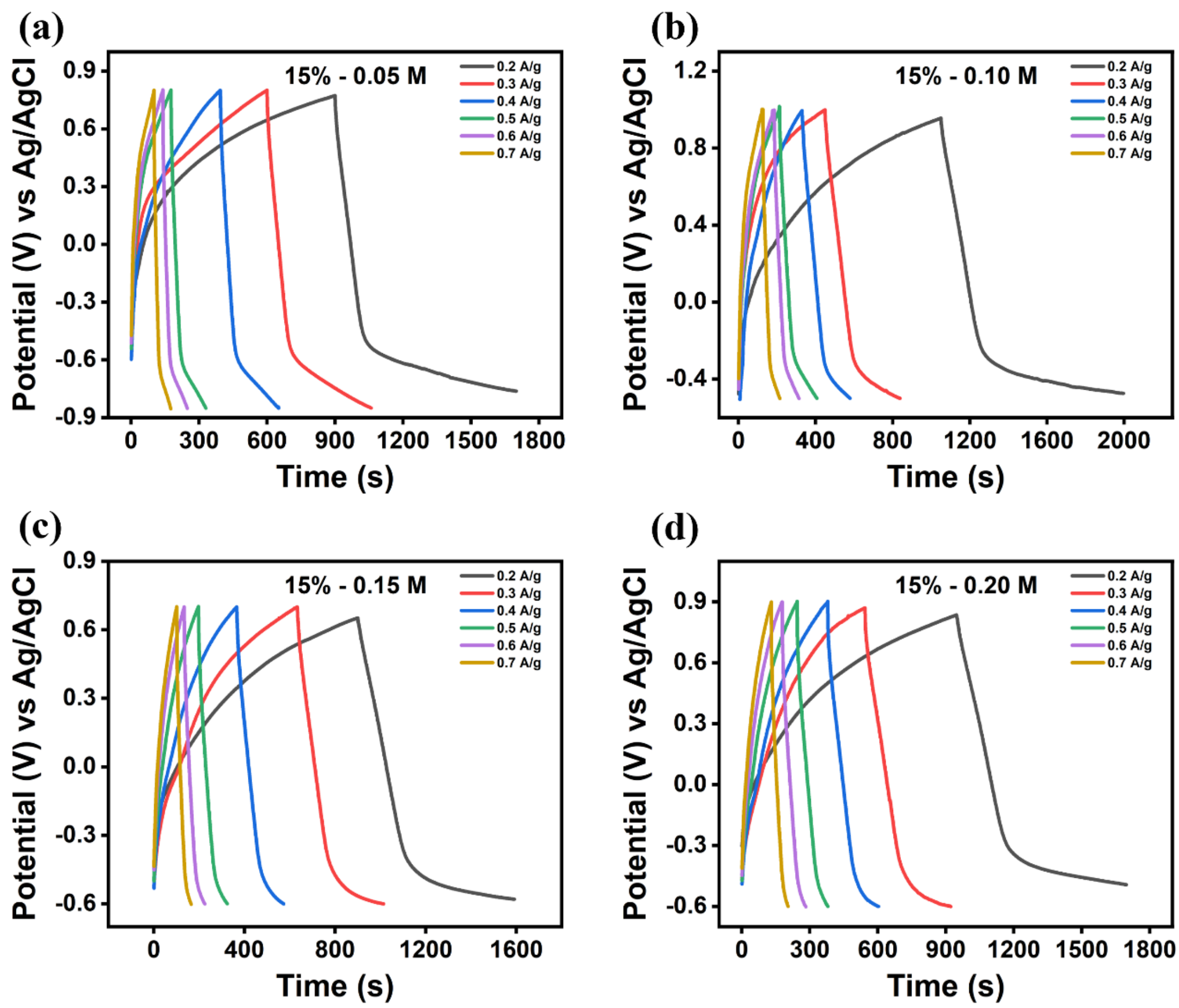

**Figure S3.** Galvanostatic charge-discharge (GCD) profiles of the $CsSnCl_3$ electrode containing 15 wt.% PVDF at various current densities (0.2-0.7 A/g) in LiTFSI electrolytes with concentrations of (a) 0.05 M, (b) 0.10 M, (c) 0.15 M, and (d) 0.20 M.

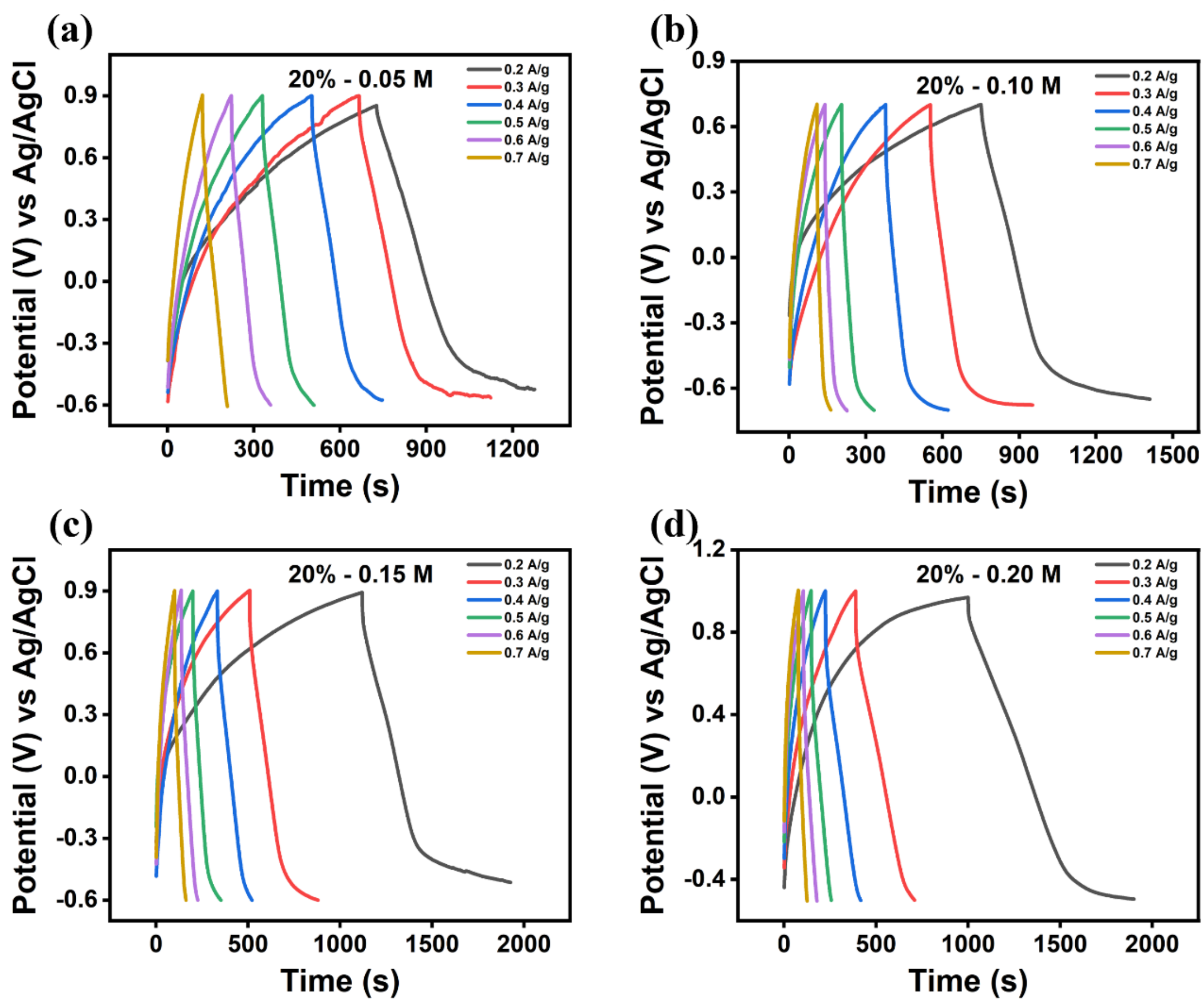

**Figure S4.** Galvanostatic charge-discharge (GCD) profiles of the $CsSnCl_3$ electrode containing 20 wt.% PVDF at various current densities (0.2-0.7 A/g) in LiTFSI electrolytes with concentrations of (a) 0.05 M, (b) 0.10 M, (c) 0.15 M, and (d) 0.20 M.

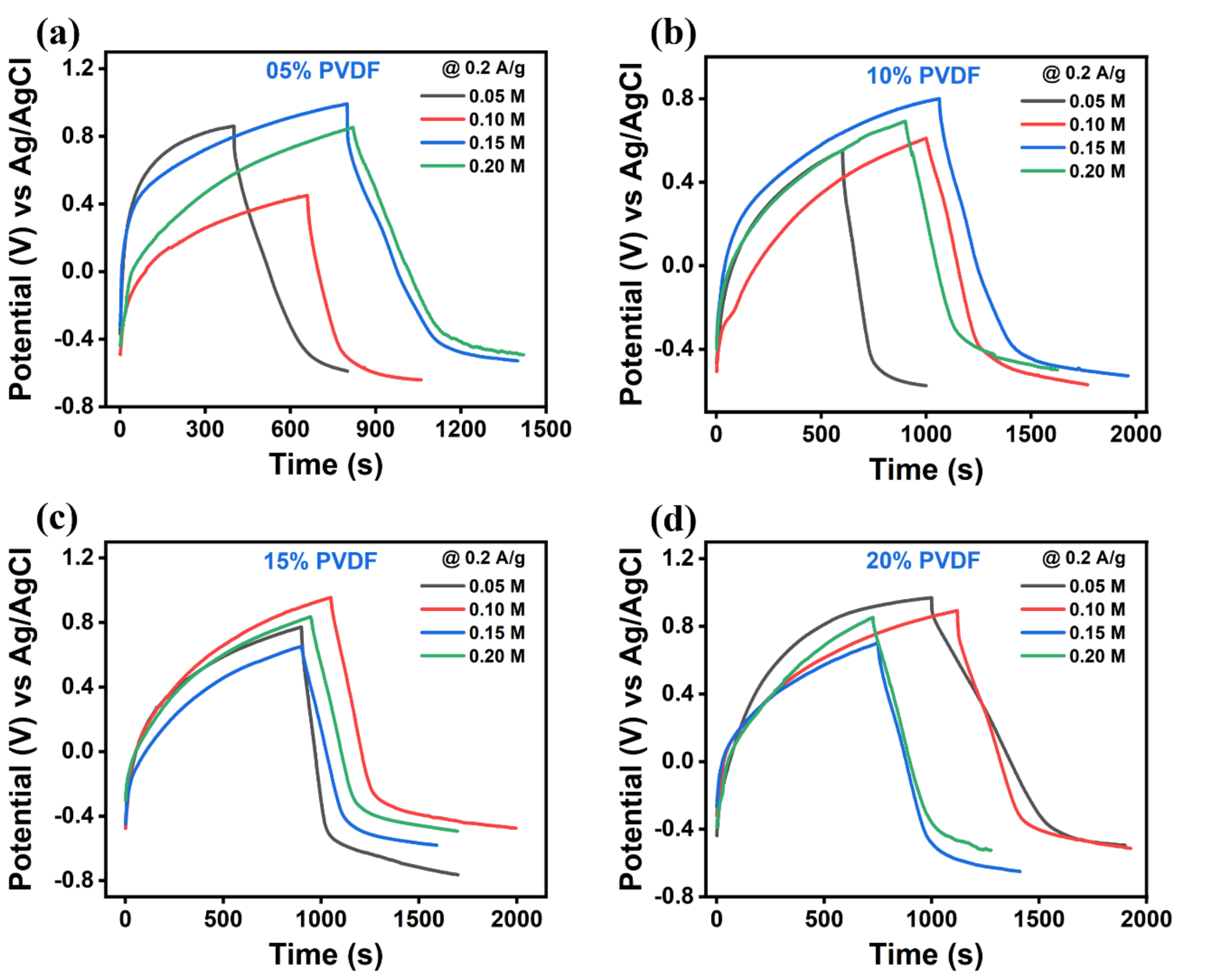

**Figure S5.** Comparison of galvanostatic charge-discharge (GCD) profiles at 0.2 A/g for $CsSnCl_3$ electrodes with different PVDF contents (5-20 wt.%) in LiTFSI electrolytes of varying concentrations (0.05-0.20 M). Panels (a-d) correspond to 5, 10, 15, and 20 wt.% PVDF, respectively.

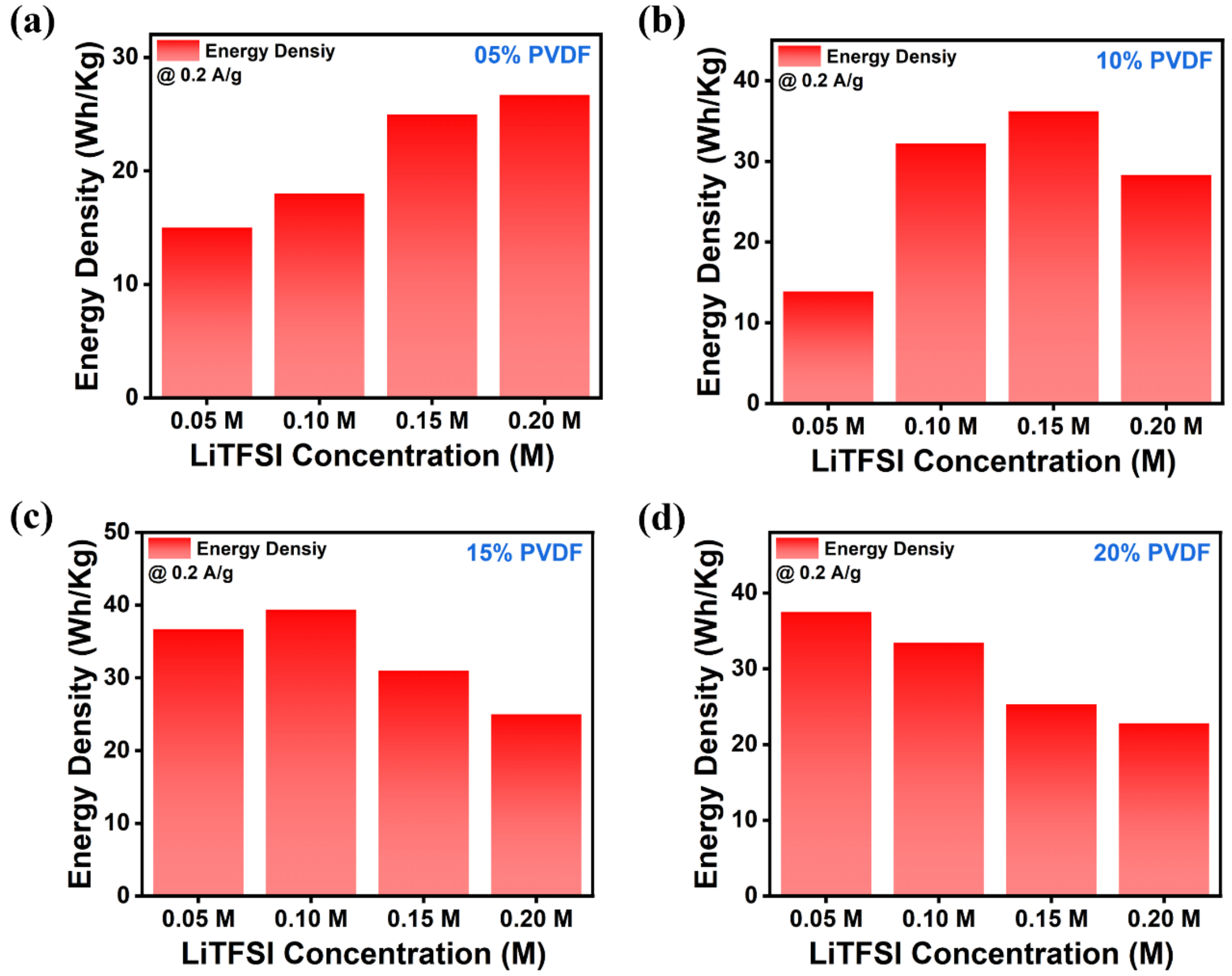

**Figure S6.** Comparison of energy density as a function of LiTFSI electrolyte concentration (0.05-0.20 M) for $CsSnCl_3$ electrodes containing different PVDF contents: (a) 5 wt.%, (b) 10 wt.%, (c) 15 wt.%, and (d) 20 wt.%. Energy density values were obtained from GCD measurements at 0.2 A/g.

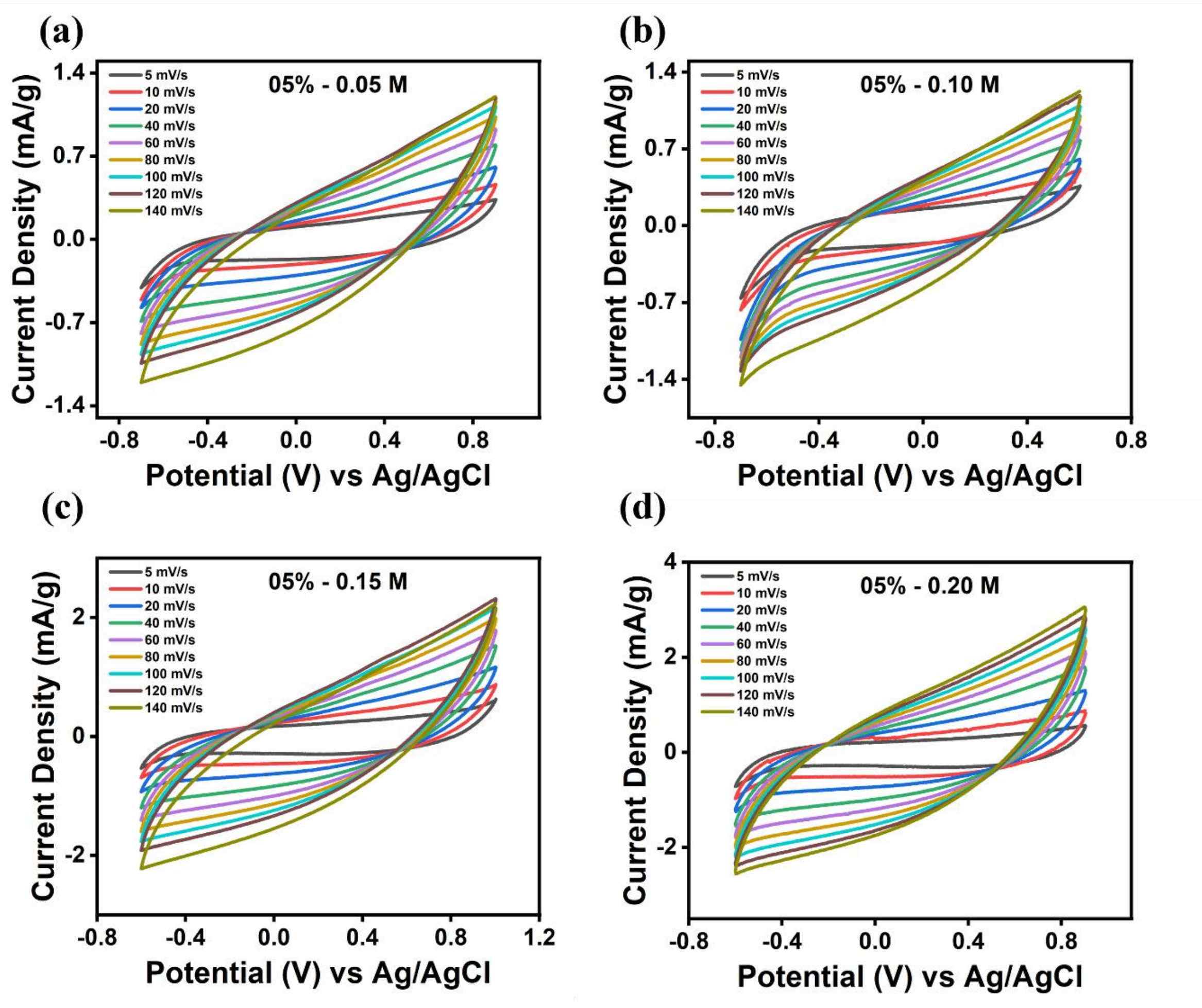

**Figure S7.** Comparison of cyclic voltammetry (CV) profiles at scan rates of 5-140 mV/s for the $CsSnCl_3$ electrode containing 5 wt.% PVDF in LiTFSI electrolytes of varying concentrations: (a) 0.05 M, (b) 0.10 M, (c) 0.15 M, and (d) 0.20 M.

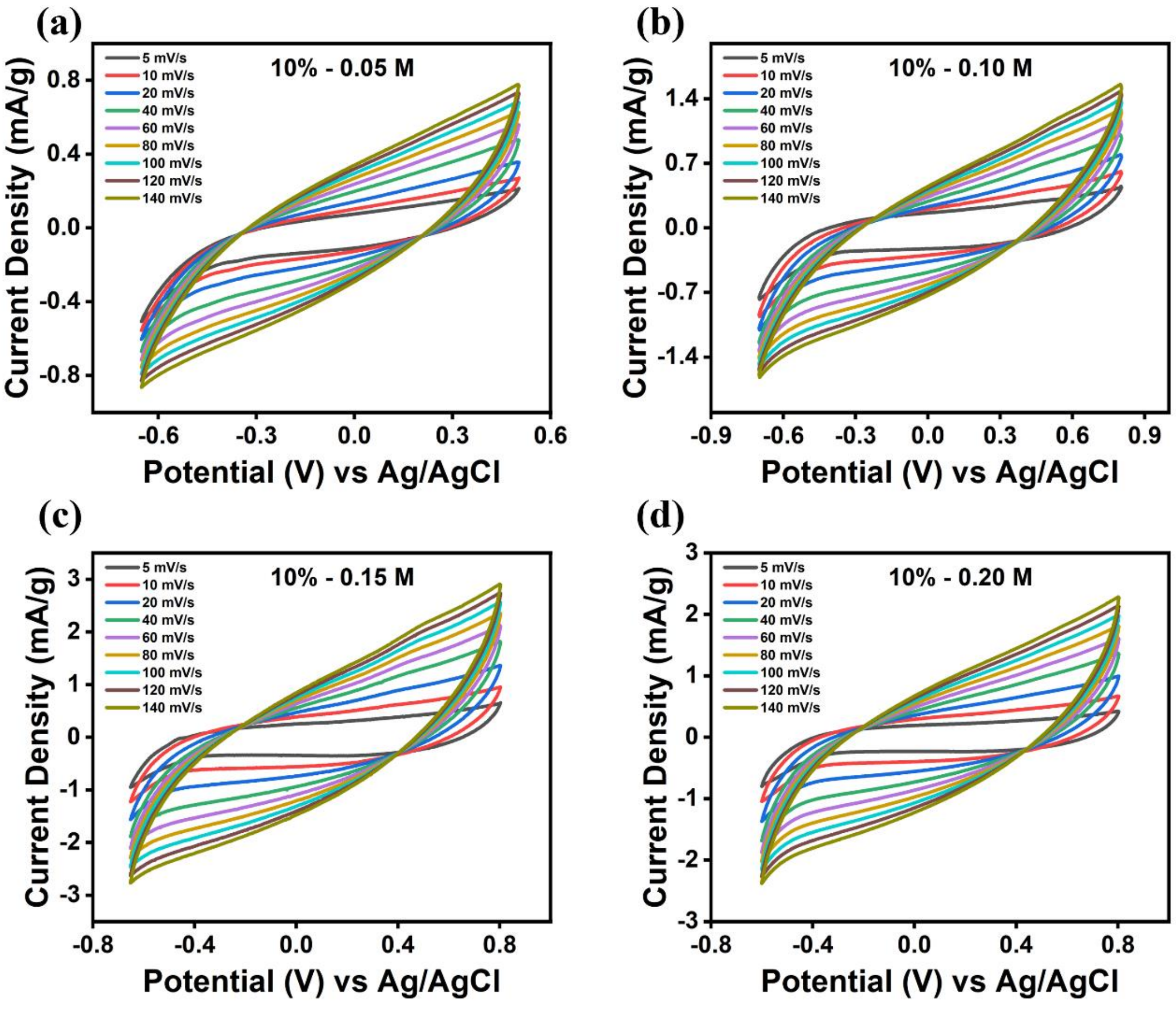

**Figure S8.** Comparison of cyclic voltammetry (CV) profiles at scan rates of 5-140 mV/s for the $CsSnCl_3$ electrode containing 10 wt.% PVDF in LiTFSI electrolytes of varying concentrations: (a) 0.05 M, (b) 0.10 M, (c) 0.15 M, and (d) 0.20 M.

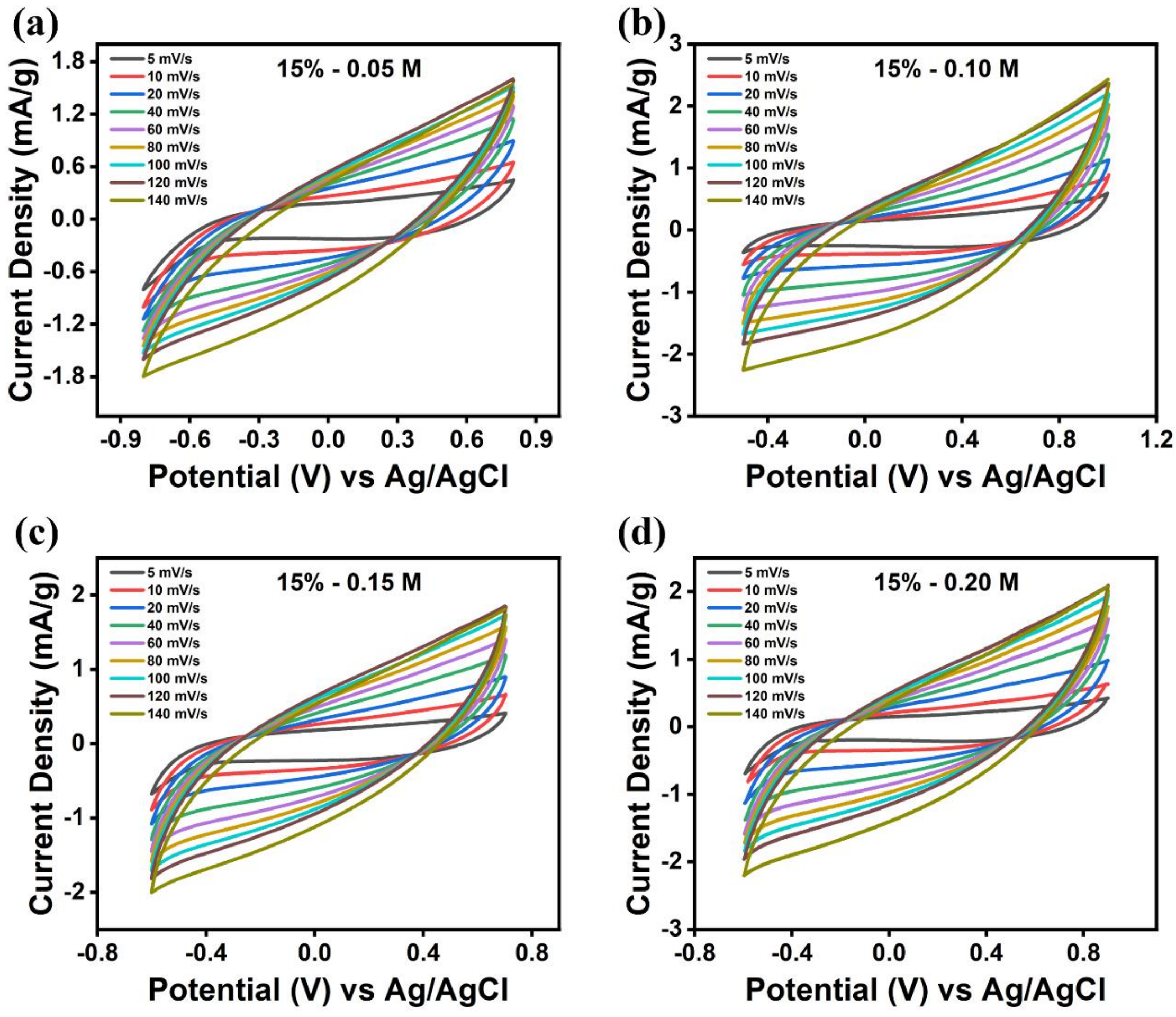

**Figure S9.** Comparison of cyclic voltammetry (CV) profiles at scan rates of 5-140 mV/s for the $CsSnCl_3$ electrode containing 15 wt.% PVDF in LiTFSI electrolytes of varying concentrations: (a) 0.05 M, (b) 0.10 M, (c) 0.15 M, and (d) 0.20 M.

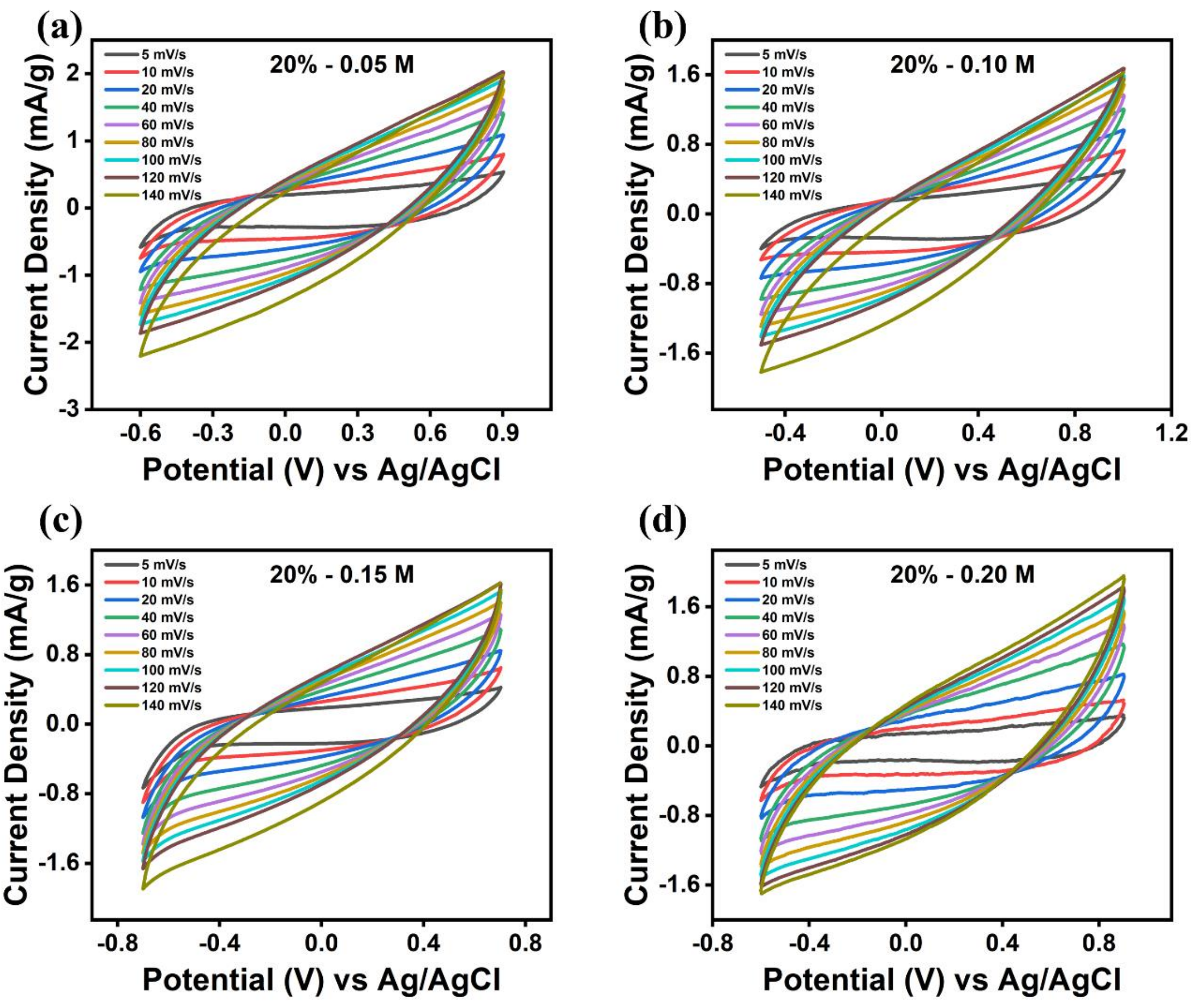

**Figure S10.** Comparison of cyclic voltammetry (CV) profiles at scan rates of 5-140 mV/s for the $CsSnCl_3$ electrode containing 20 wt.% PVDF in LiTFSI electrolytes of varying concentrations: (a) 0.05 M, (b) 0.10 M, (c) 0.15 M, and (d) 0.20 M.

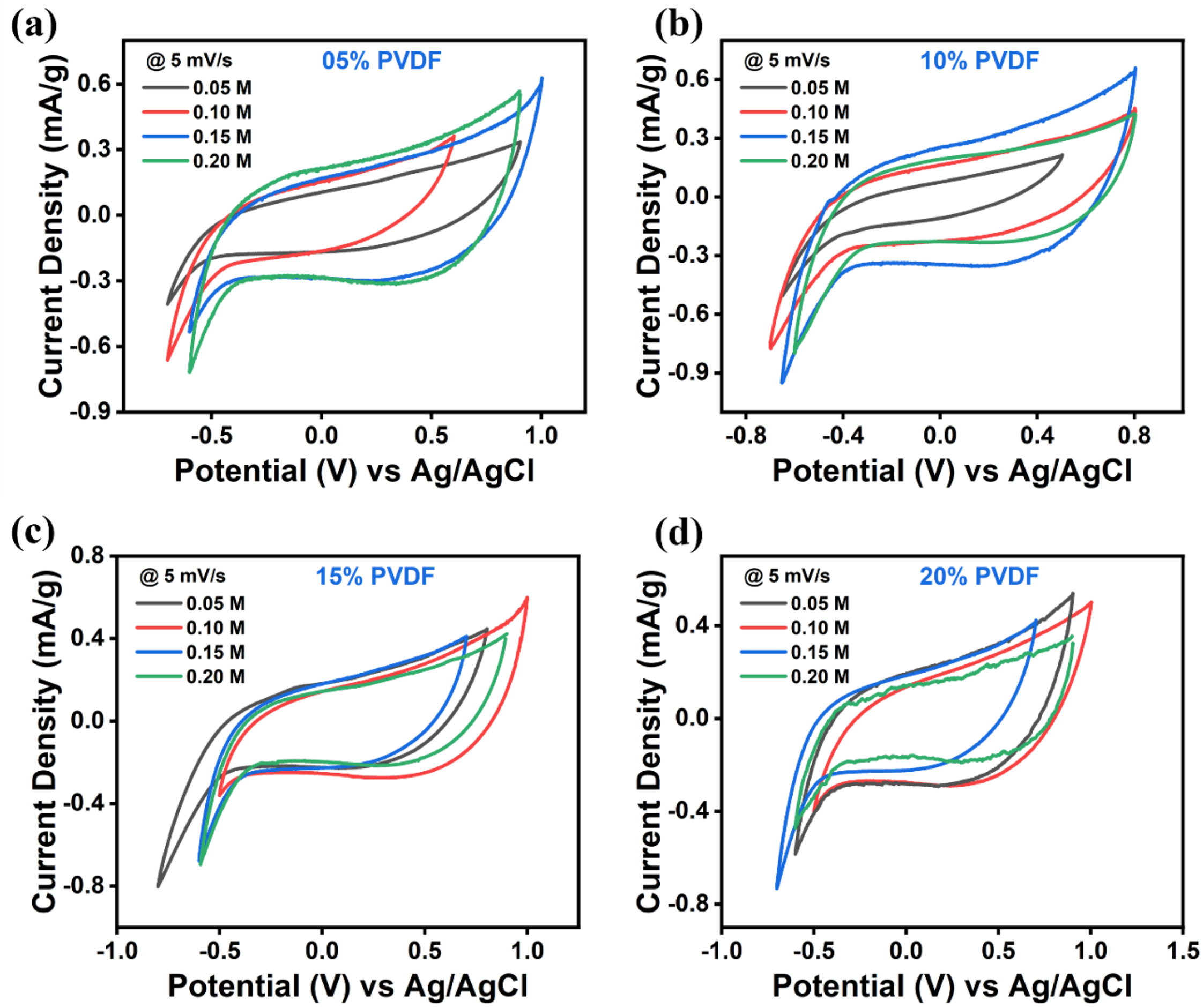

**Figure S11.** Comparison of cyclic voltammetry (CV) profiles at a scan rate of 5 mV/s for $CsSnCl_3$ electrodes with varying PVDF contents (5-20 wt.%) in LiTFSI electrolytes of different concentrations (0.05-0.20 M). Panels (a-d) correspond to 5, 10, 15, and 20 wt.% PVDF, respectively.

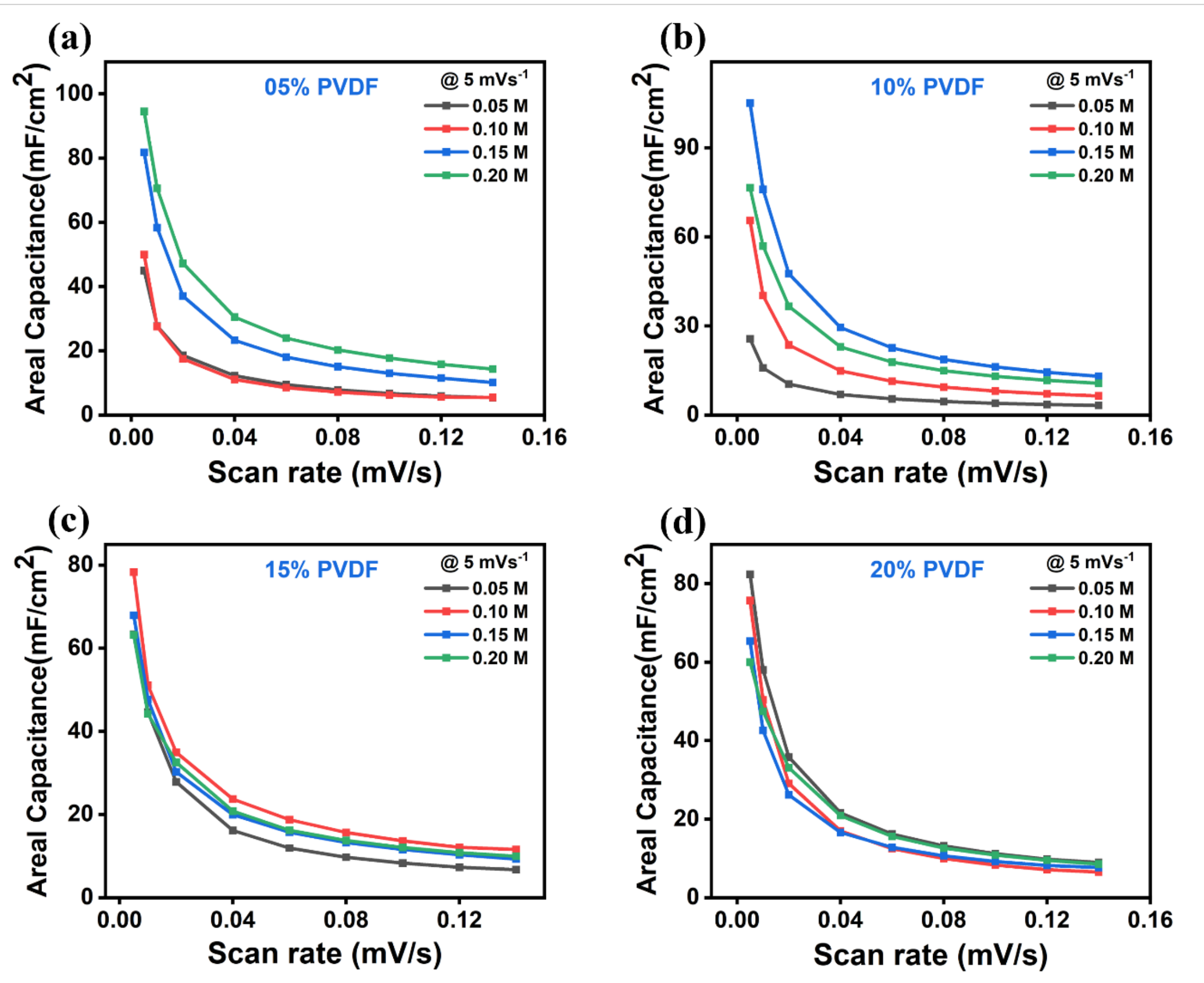

**Figure S12.** Comparison of areal capacitance as a function of scan rate for $CsSnCl_3$ electrodes with different PVDF contents (5-20 wt.%) in LiTFSI electrolytes of varying concentrations (0.05-0.20 M). Panels (a-d) correspond to 5, 10, 15, and 20 wt.% PVDF, respectively.

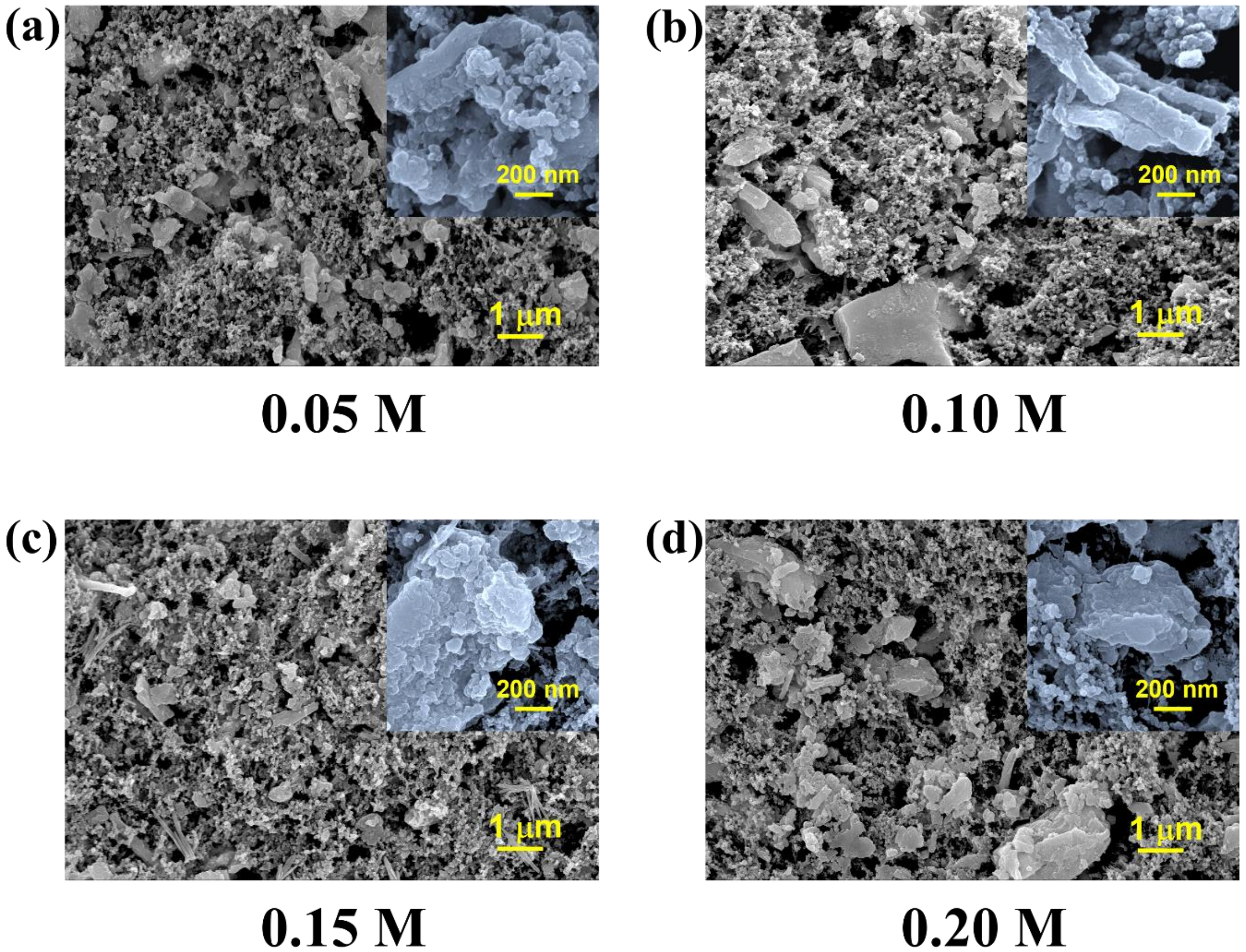

**Figure S13.** Scanning electron microscopy (SEM) images of the 15 wt.% PVDF $CsSnCl_3$ electrodes after electrochemical measurements performed in LiTFSI electrolytes of varying concentrations: (a) 0.05 M, (b) 0.10 M, (c) 0.15 M, and (d) 0.20 M.

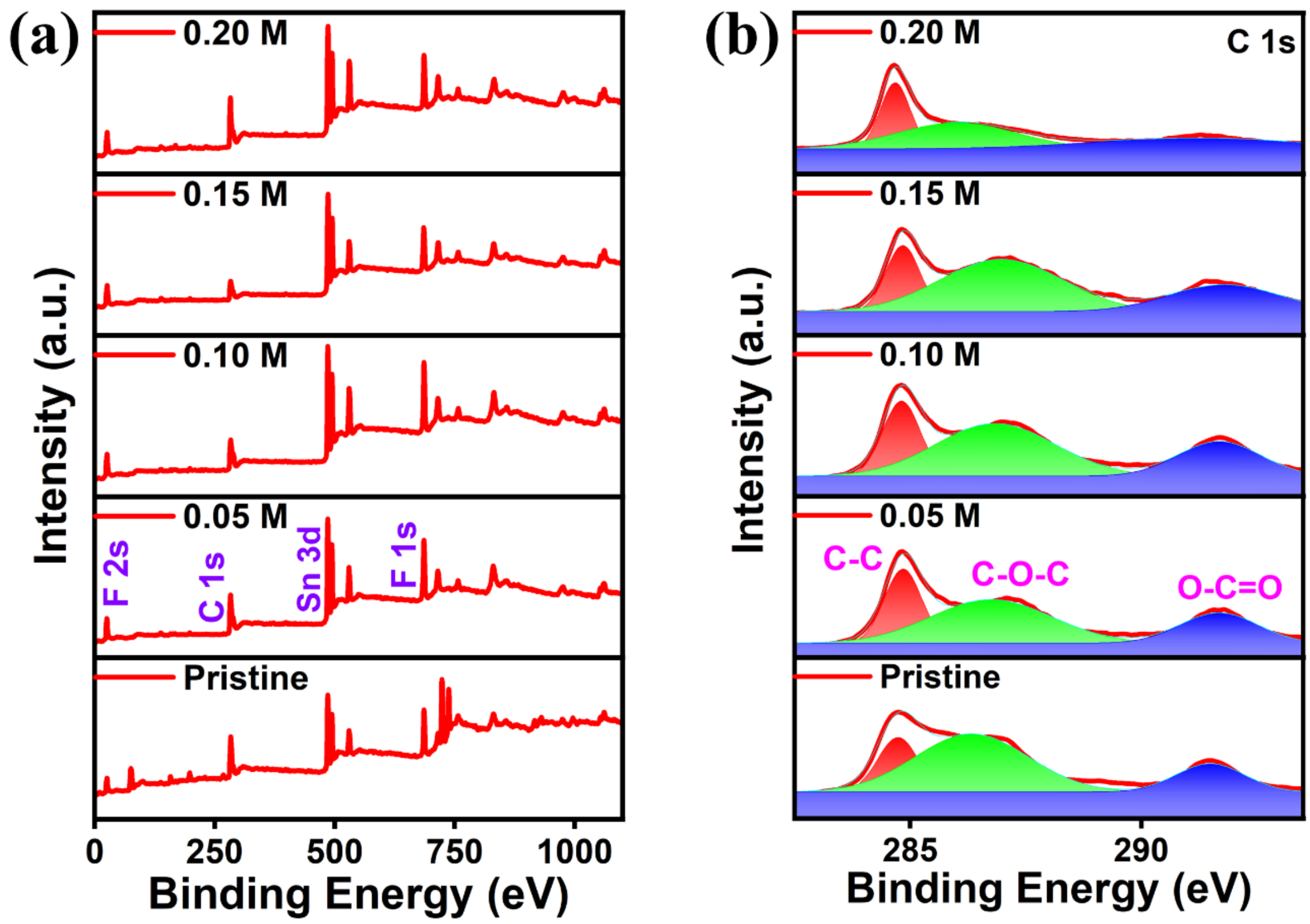

**Figure S14.** XPS characterization of the pristine and treated $CsSnCl_3$ samples. (a) Survey spectra of the pristine sample and samples treated with LiTFSI concentrations of 0.05, 0.10, 0.15, and 0.20 M, showing the presence of F, C, and Sn elements. (b) High-resolution C 1s spectra of the corresponding samples, with deconvoluted peaks assigned to C–C, C–O–C, and O–C=O bonding states.

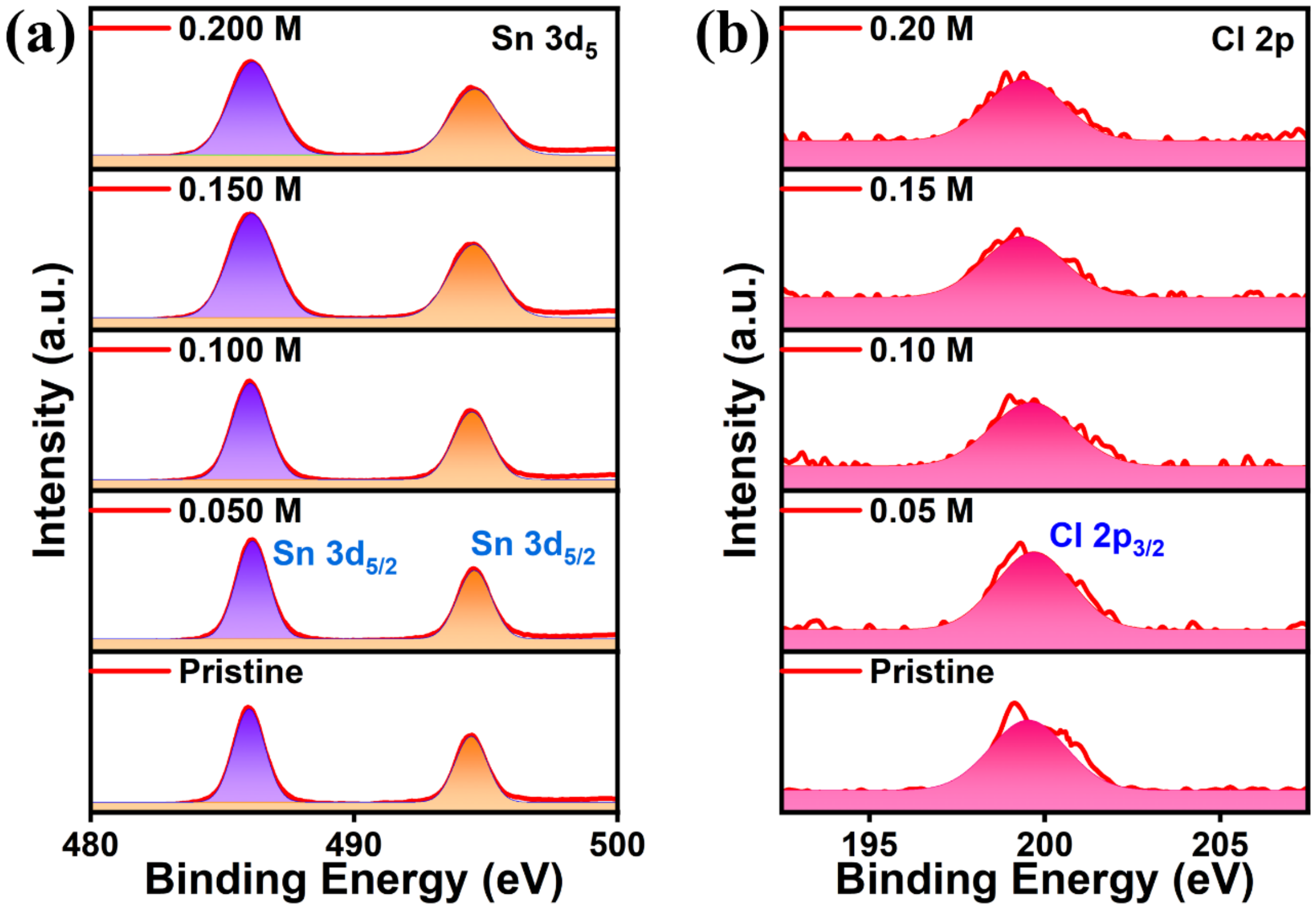

**Figure S15.** XPS characterization of the pristine and treated $CsSnCl_3$ samples. (a) High-resolution Sn 3d spectra of the pristine sample and samples treated with LiTFSI electrolyte concentrations of 0.05, 0.10, 0.15, and 0.20 M. (b) High-resolution Cl 2p spectra of the corresponding samples.

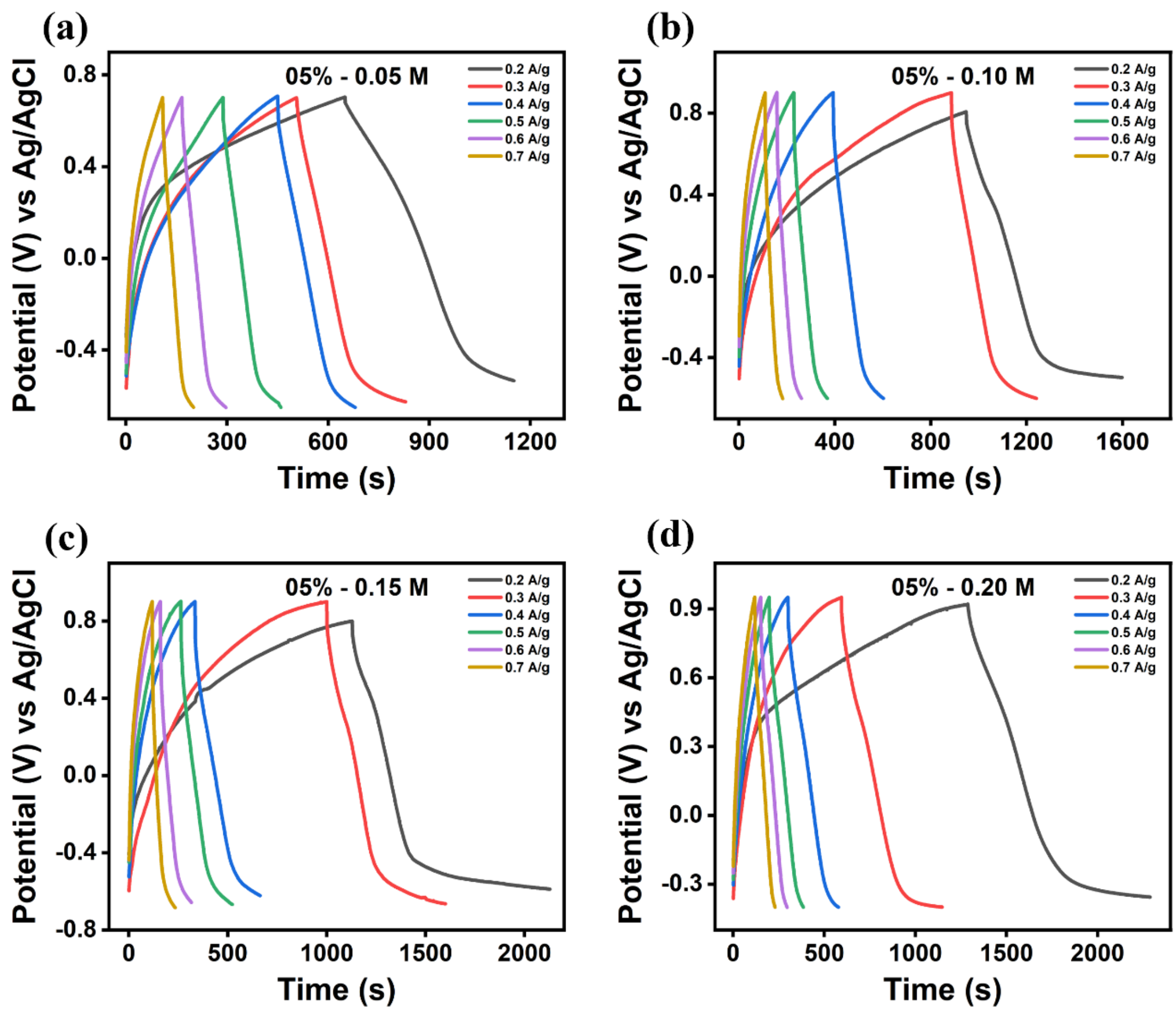

**Figure S16.** Galvanostatic charge-discharge (GCD) profiles of the $MASnCl_3$ electrode containing 5 wt.% PVDF at various current densities (0.2-0.7 A/g) in LiTFSI electrolytes with concentrations of (a) 0.05 M, (b) 0.10 M, (c) 0.15 M, and (d) 0.20 M.

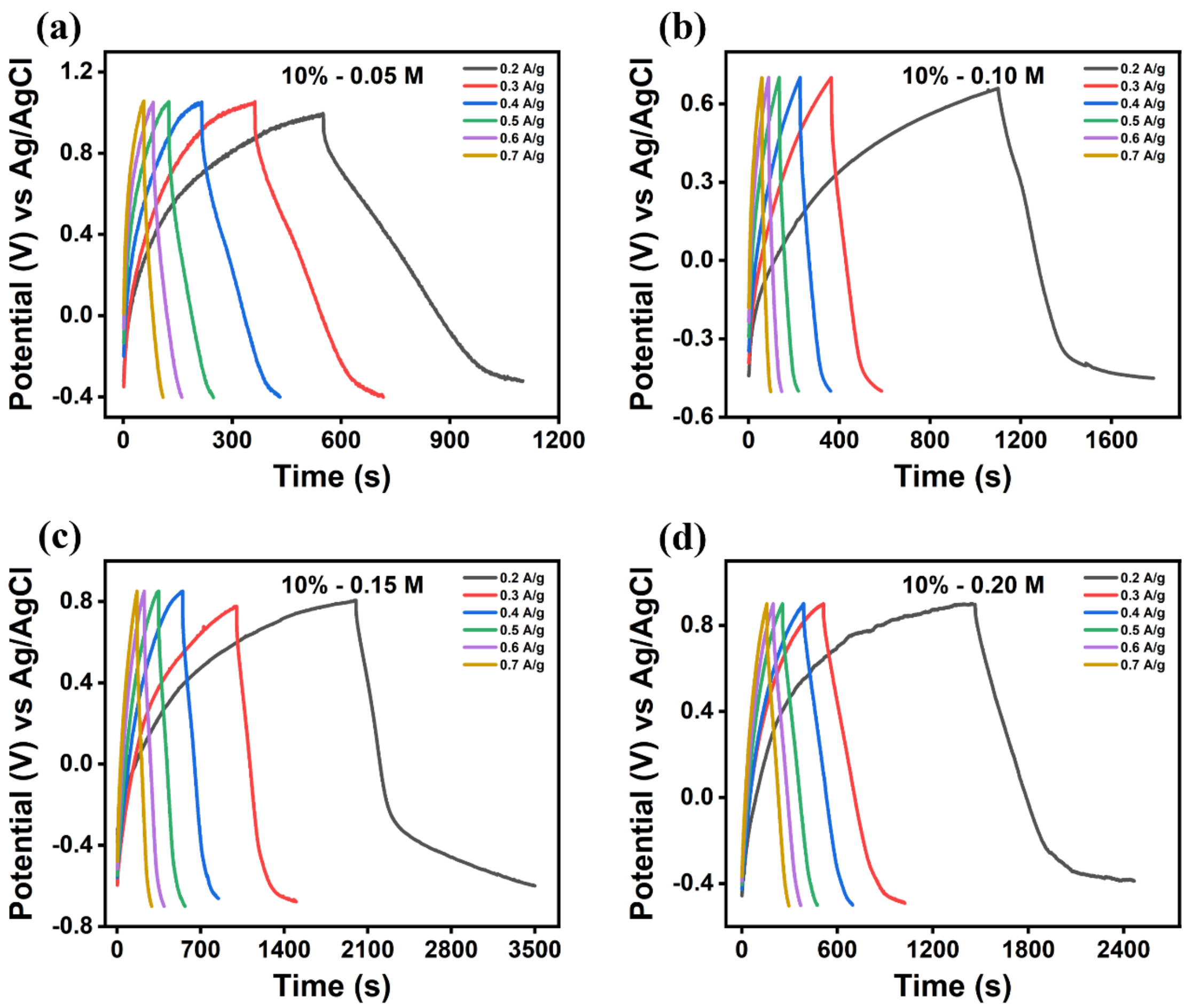

**Figure S17.** Galvanostatic charge-discharge (GCD) profiles of the $MASnCl_3$ electrode containing 10 wt.% PVDF at various current densities (0.2-0.7 A/g) in LiTFSI electrolytes with concentrations of (a) 0.05 M, (b) 0.10 M, (c) 0.15 M, and (d) 0.20 M.

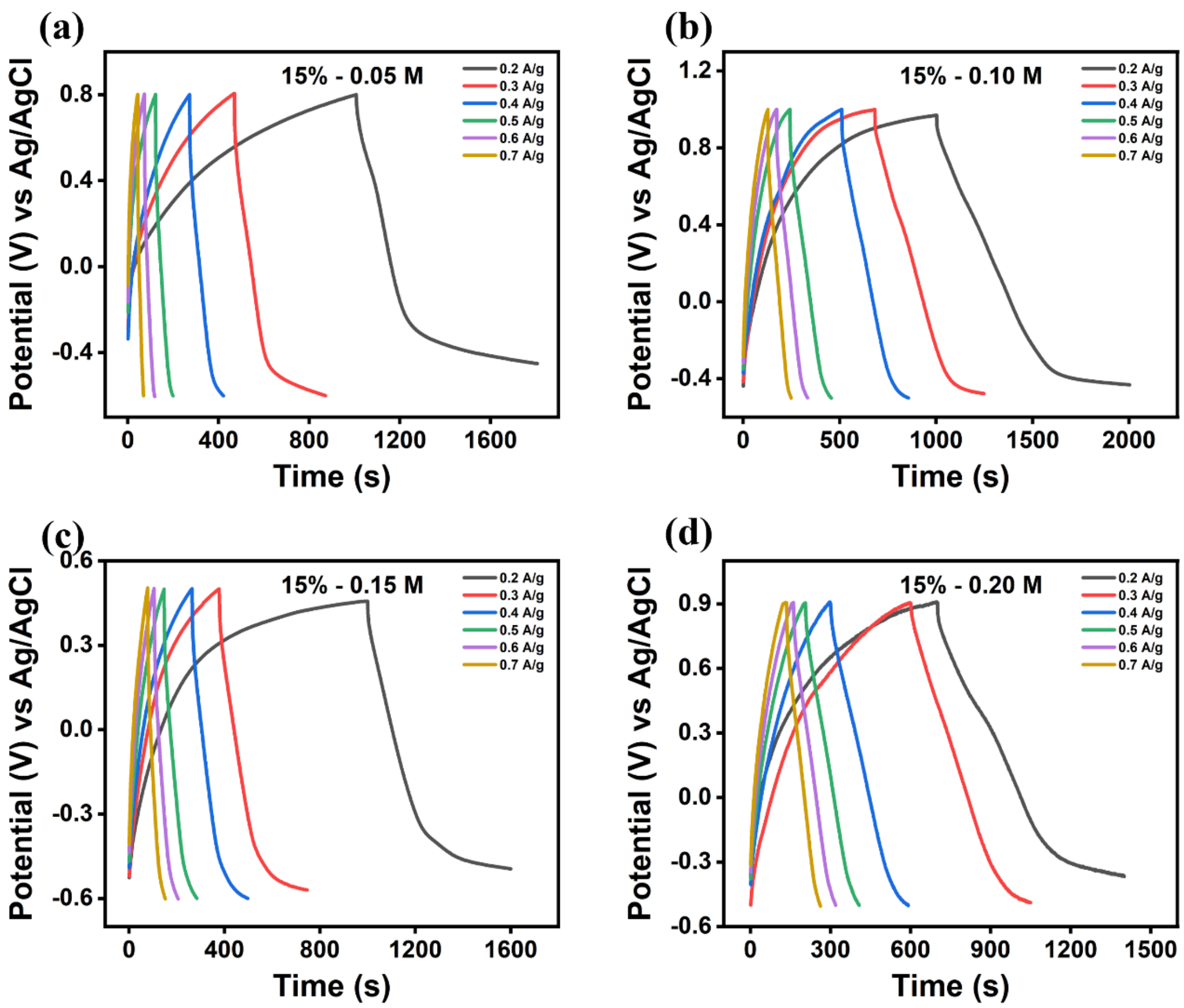

**Figure S18.** Galvanostatic charge-discharge (GCD) profiles of the $MASnCl_3$ electrode containing 15 wt.% PVDF at various current densities (0.2-0.7 A/g) in LiTFSI electrolytes with concentrations of (a) 0.05 M, (b) 0.10 M, (c) 0.15 M, and (d) 0.20 M.

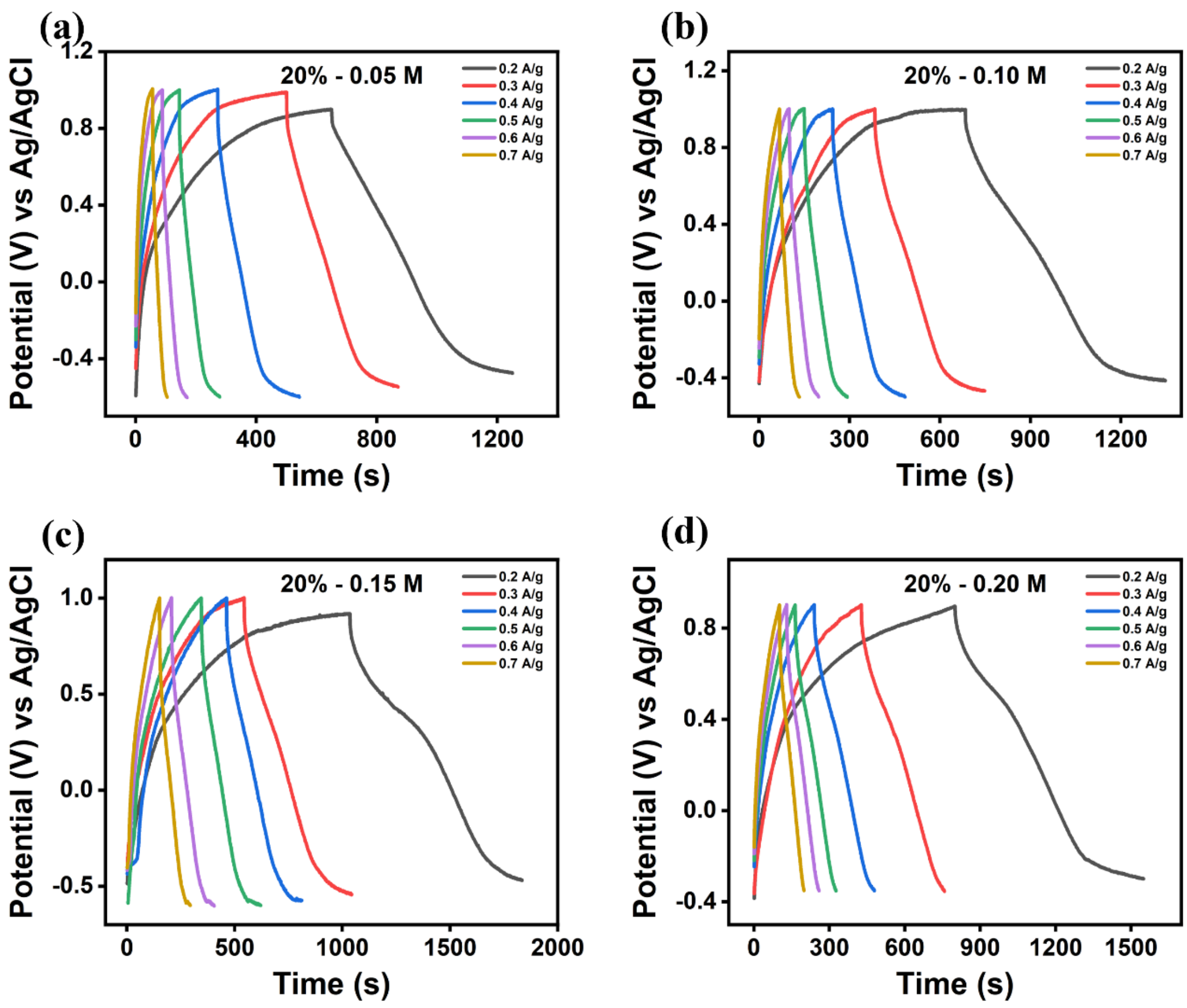

**Figure S19.** Galvanostatic charge-discharge (GCD) profiles of the $MASnCl_3$ electrode containing 20 wt.% PVDF at various current densities (0.2-0.7 A/g) in LiTFSI electrolytes with concentrations of (a) 0.05 M, (b) 0.10 M, (c) 0.15 M, and (d) 0.20 M.

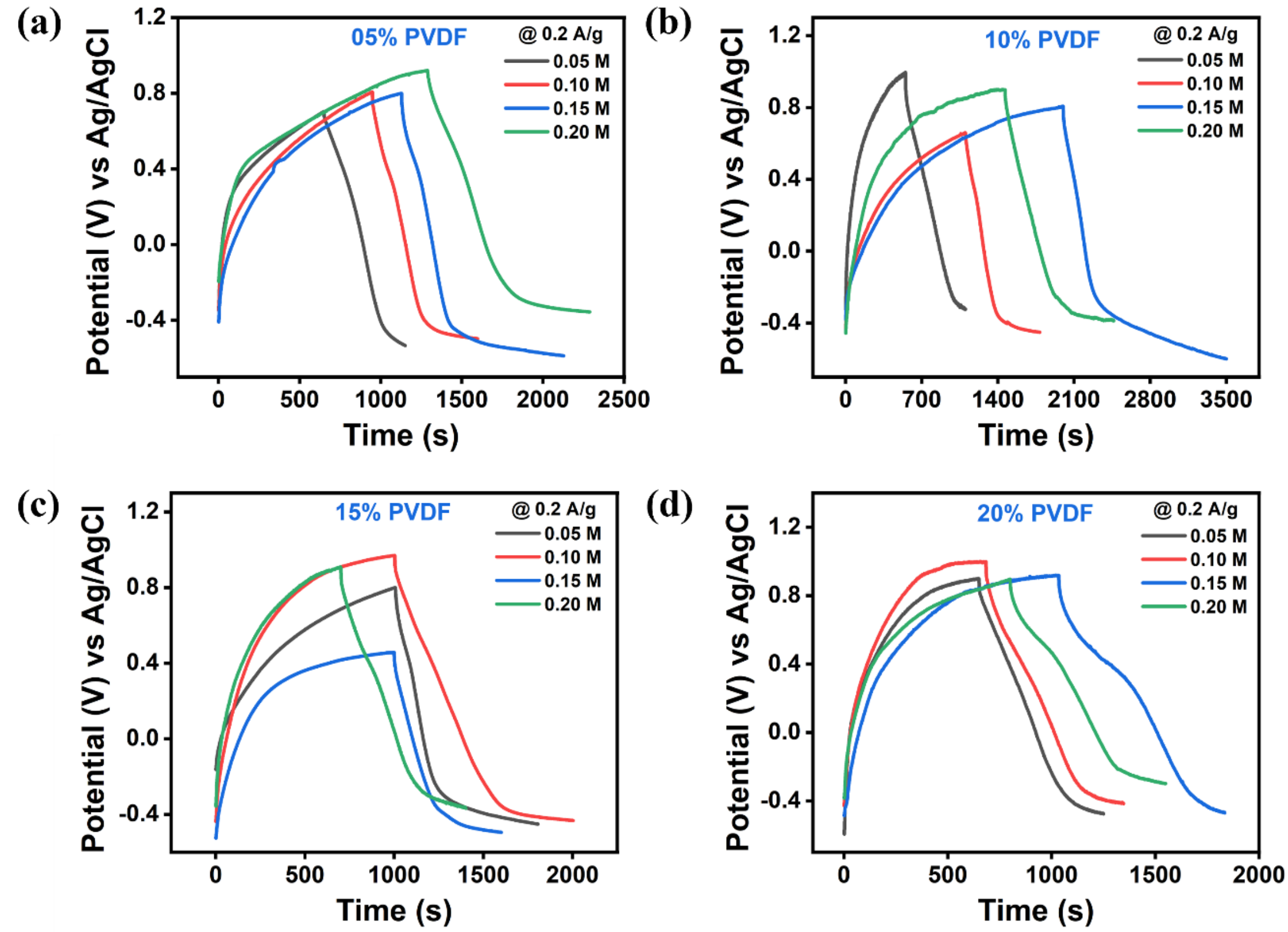

**Figure S20.** Comparison of galvanostatic charge-discharge (GCD) profiles at 0.2 A/g for $MASnCl_3$ electrodes with different PVDF contents (5-20 wt.%) in LiTFSI electrolytes of varying concentrations (0.05-0.20 M). Panels (a-d) correspond to 5, 10, 15, and 20 wt.% PVDF, respectively.

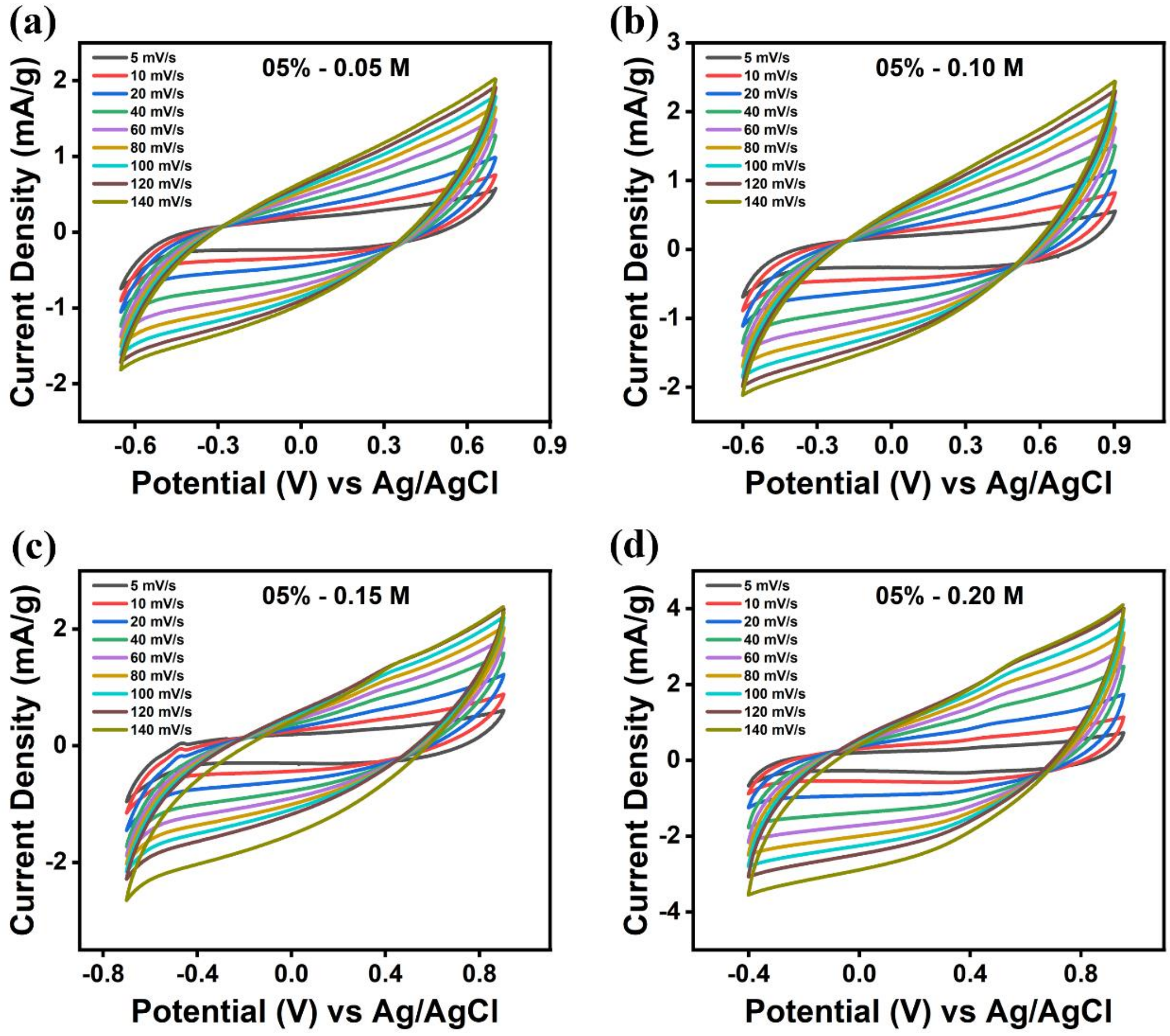

**Figure S21.** Comparison of cyclic voltammetry (CV) profiles at scan rates of 5-140 mV/s for the $MASnCl_3$ electrode containing 5 wt.% PVDF in LiTFSI electrolytes of varying concentrations: (a) 0.05 M, (b) 0.10 M, (c) 0.15 M, and (d) 0.20 M.

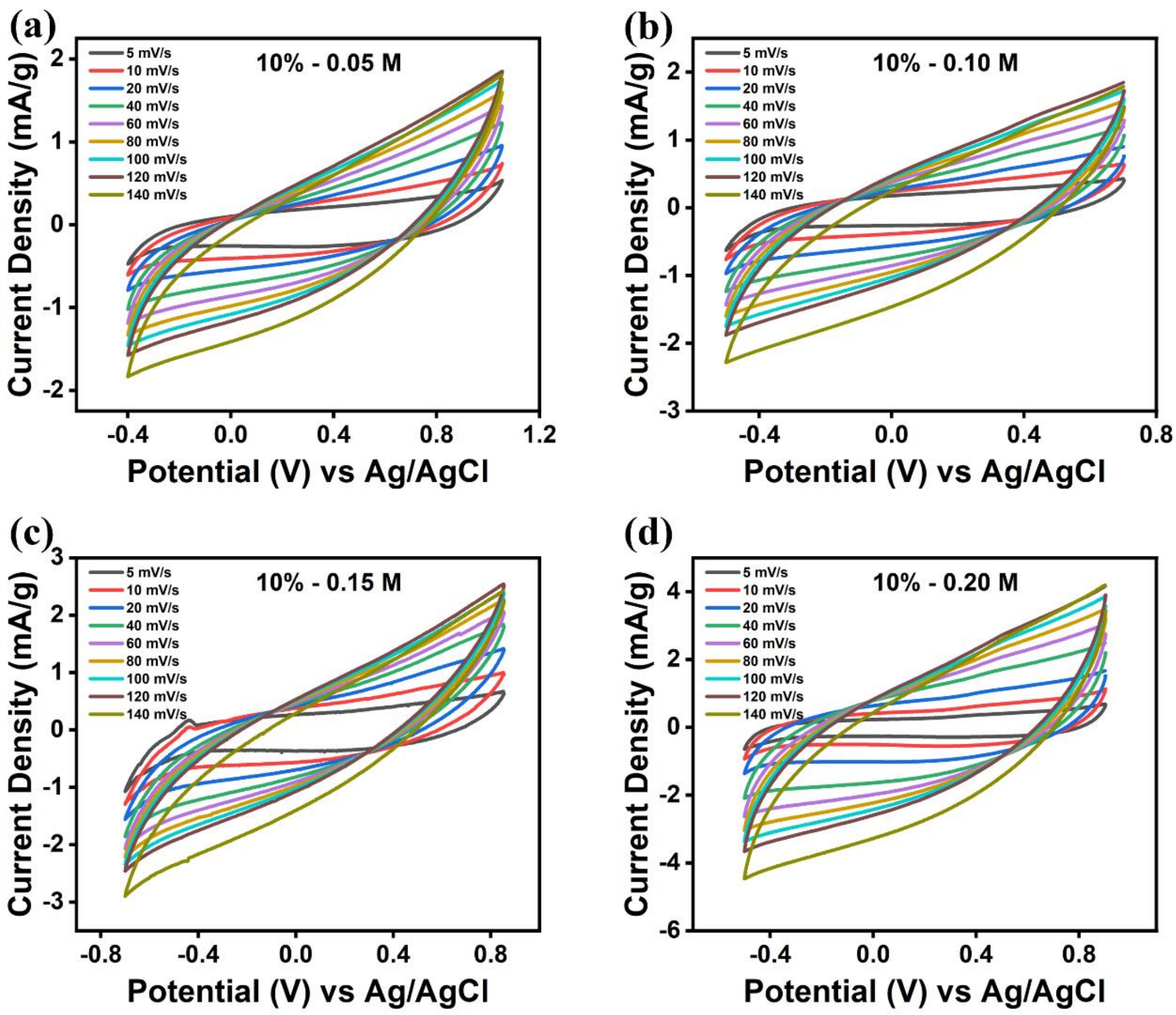

**Figure S22.** Comparison of cyclic voltammetry (CV) profiles at scan rates of 5-140 mV/s for the $MASnCl_3$ electrode containing 10 wt.% PVDF in LiTFSI electrolytes of varying concentrations: (a) 0.05 M, (b) 0.10 M, (c) 0.15 M, and (d) 0.20 M.

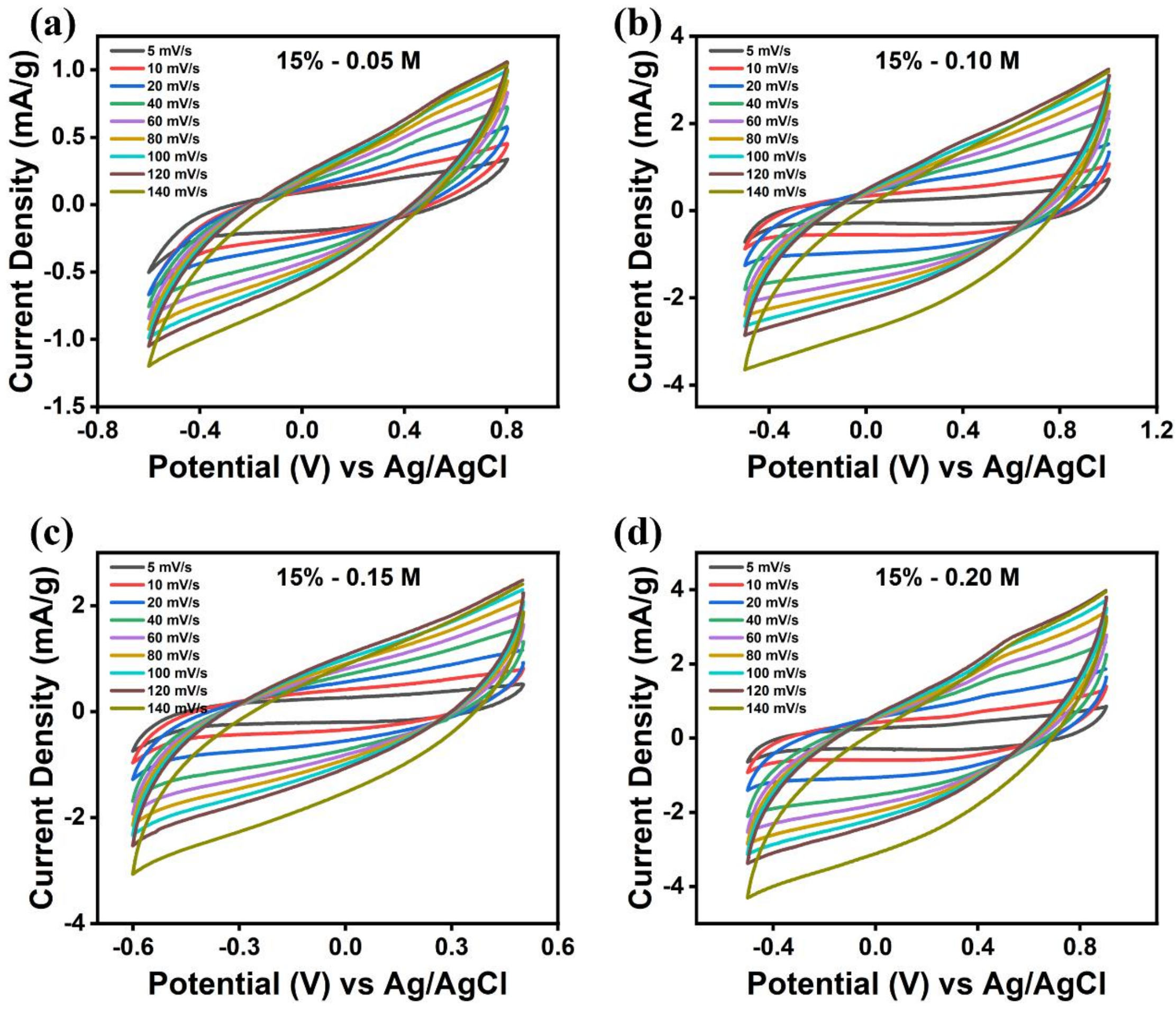

**Figure S23.** Comparison of cyclic voltammetry (CV) profiles at scan rates of 5-140 mV/s for the $MASnCl_3$ electrode containing 15 wt.% PVDF in LiTFSI electrolytes of varying concentrations: (a) 0.05 M, (b) 0.10 M, (c) 0.15 M, and (d) 0.20 M.

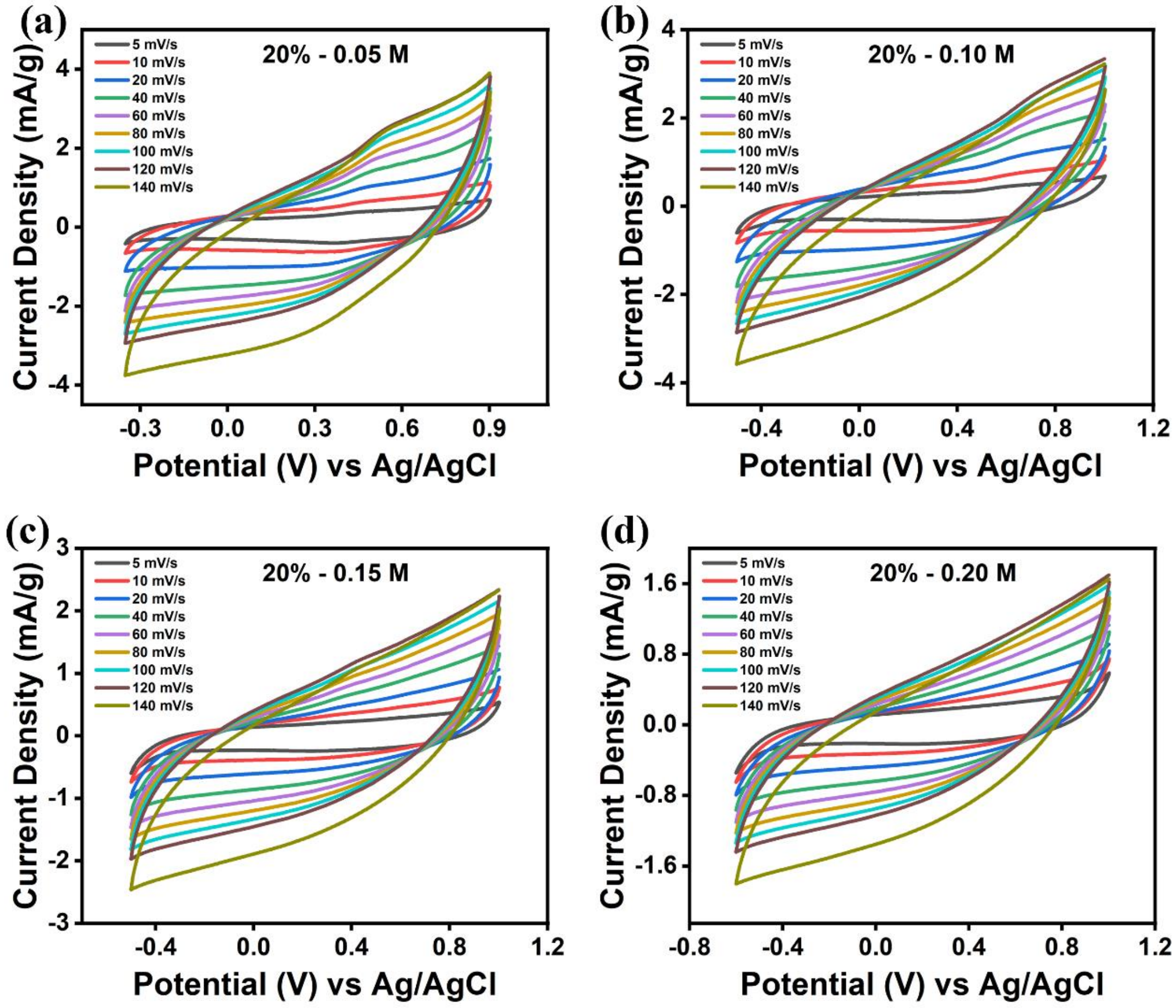

**Figure S24.** Comparison of cyclic voltammetry (CV) profiles at scan rates of 5-140 mV/s for the $MASnCl_3$ electrode containing 20 wt.% PVDF in LiTFSI electrolytes of varying concentrations: (a) 0.05 M, (b) 0.10 M, (c) 0.15 M, and (d) 0.20 M.

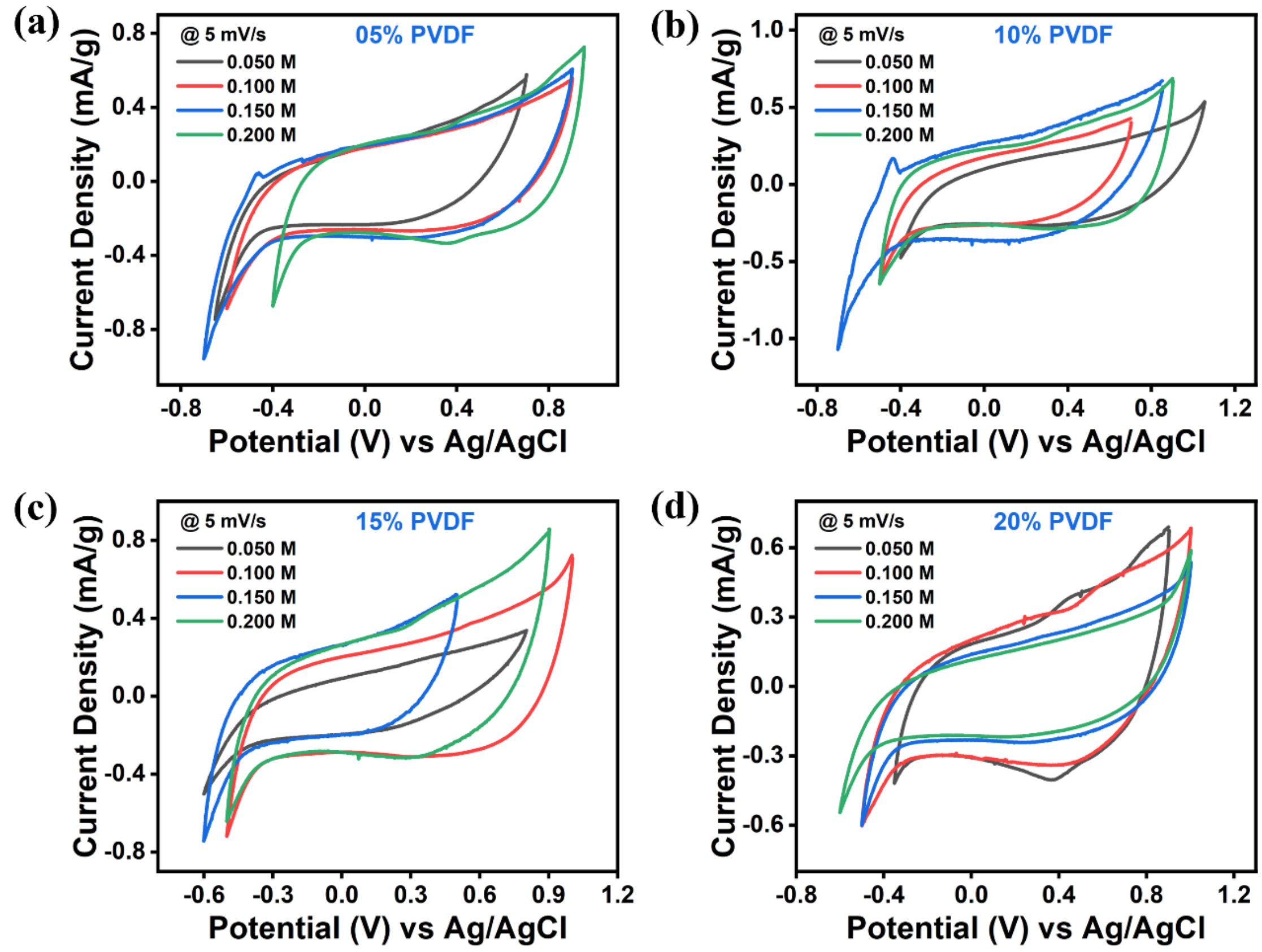

**Figure S25.** Comparison of cyclic voltammetry (CV) profiles at a scan rate of 5 mV/s for $MASnCl_3$ electrodes with varying PVDF contents (5-20 wt.%) in LiTFSI electrolytes of different concentrations (0.05-0.20 M). Panels (a-d) correspond to 5, 10, 15, and 20 wt.% PVDF, respectively.

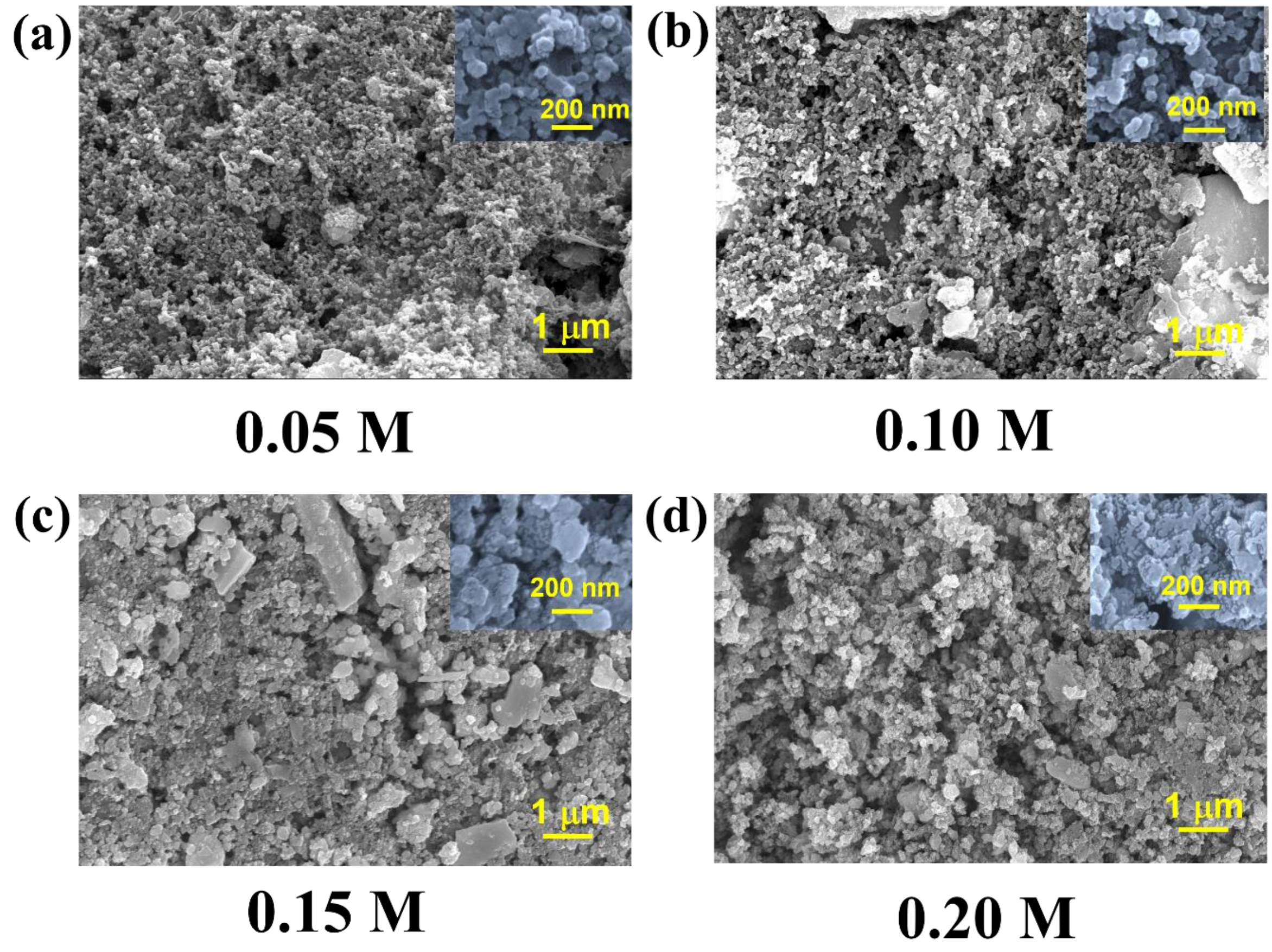

**Figure S26.** Scanning electron microscopy (SEM) images of the 15 wt.% PVDF $MASnCl_3$ electrodes after electrochemical measurements performed in LiTFSI electrolytes of varying concentrations: (a) 0.05 M, (b) 0.10 M, (c) 0.15 M, and (d) 0.20 M.

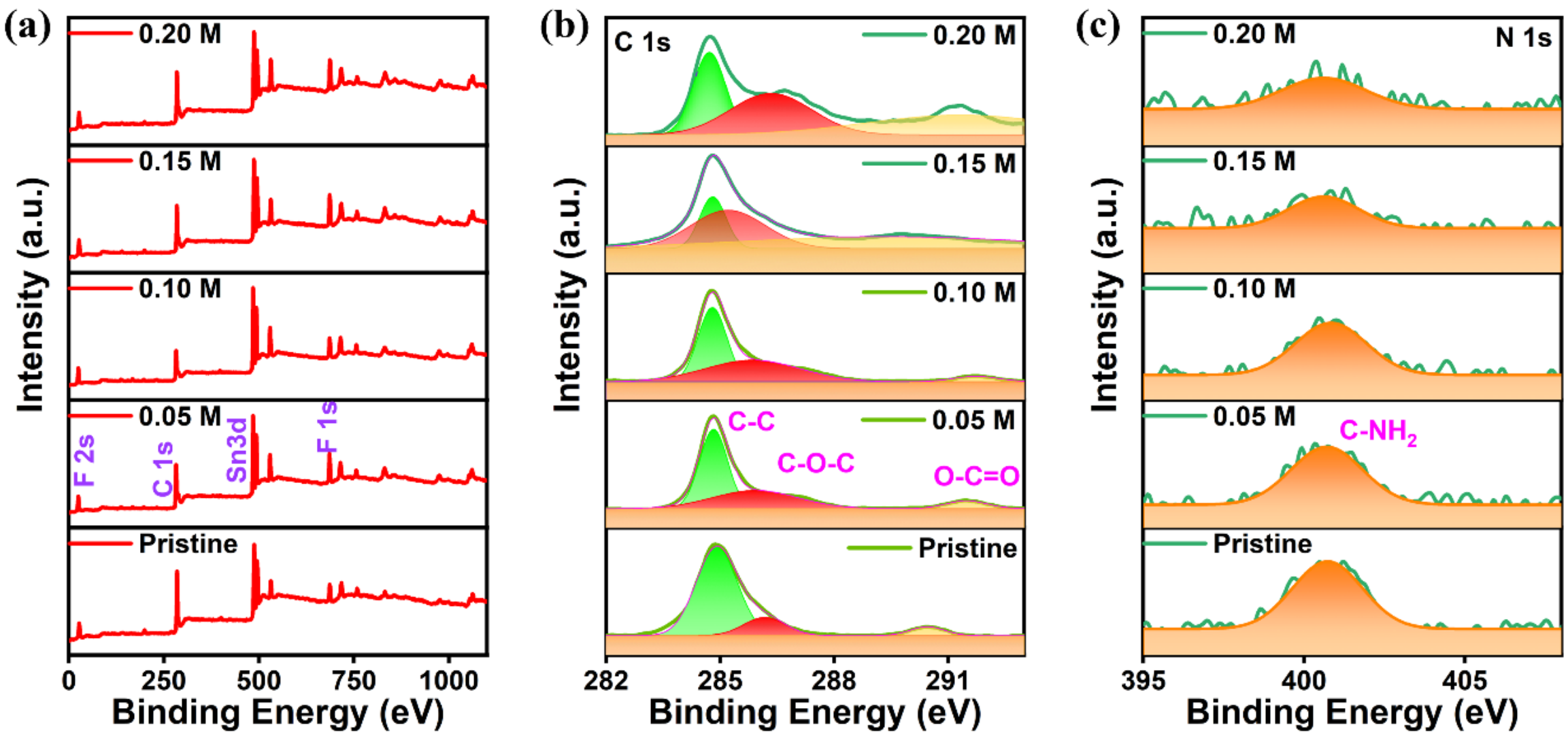

**Figure S27.** XPS characterization of the pristine and treated $MASnCl_3$ samples. (a) Survey spectra of the pristine sample and samples treated with LiTFSI concentrations of 0.05, 0.10, 0.15, and 0.20 M, showing the presence of F, C, and Sn elements. (b) High-resolution C 1s spectra of the corresponding samples, with deconvoluted peaks assigned to C–C, C–O–C, and O–C=O bonding states. (c) High-resolution N 1s spectra of the corresponding samples.

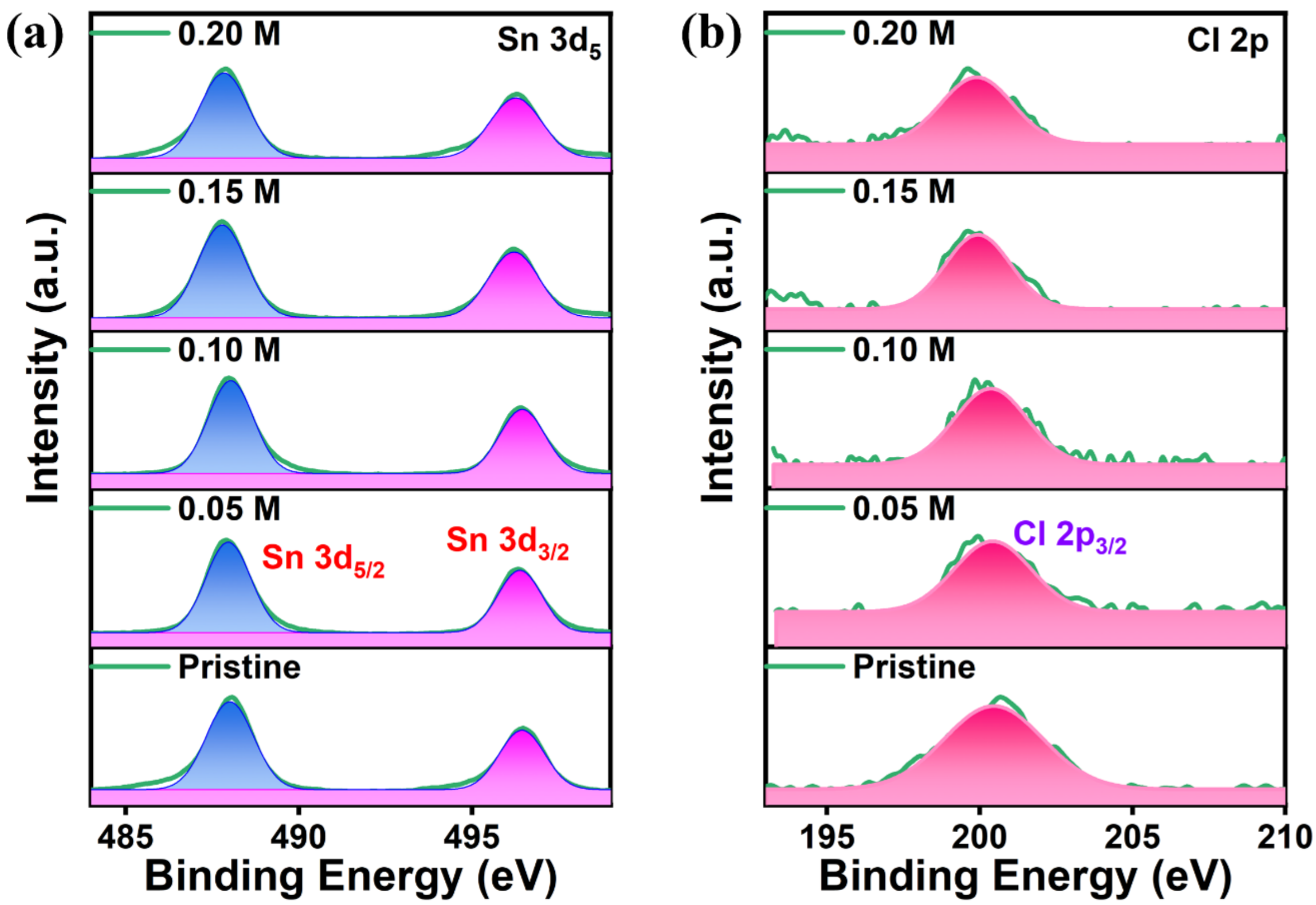

**Figure S28.** XPS characterization of the pristine and treated $MASnCl_3$ samples. (a) High-resolution Sn 3d spectra of the pristine sample and samples treated with LiTFSI electrolyte concentrations of 0.05, 0.10, 0.15, and 0.20 M. (b) High-resolution Cl 2p spectra of the corresponding samples.

**Table S1.** Structural and energetic properties computed with the MACE potential compared to reference values.

| System | Property | This work | Reference Value | Source | Deviation |
|---|---|---|---|---|---|
| $CsSnCl_3$, cubic α ($Pm\bar{3}m$) | Lattice constant $a$ | 5.577 Å | 5.60 / 5.63 Å[1] | Exp. / DFT-PBE | −0.4 / −0.9 % |
| $CsSnCl_3$ | Bulk modulus $B_0$ | 26.15 GPa | 22.653 GPa[2] | DFT-PBE | +15.4 % |
| $CsSnCl_3$(100)-$SnCl_2$ | Surface energy $\gamma$ | 0.058 J m$^{-2}$ | 0.055 J m$^{-2}$[3] | DFT-PBEsol | +5 % |
| $CsSnCl_3$(100)-CsCl | Surface energy $\gamma$ | 0.080 J m$^{-2}$ | 0.069 J m$^{-2}$[3] | DFT-PBEsol | +16 % |
| β-PVDF | Mean C–C | 1.515 Å | 1.54 Å[4] | Crystallographic | −1.6 % |
| β-PVDF | Mean C–F | 1.379 Å | 1.34 Å[4] | Crystallographic | +2.9 % |
| β-PVDF | Mean C–H | 1.093 Å | 1.09 Å[4] | Crystallographic | +0.3 % |
| β-PVDF | Mean C–C–C angle | 114.7° | 112°[4] | Crystallographic | +2.5 % |
| Li, bcc | Lattice constant $a$ | 3.497 Å | 3.51 Å[5] | Kittel | −0.4 % |
| Li, bcc | Cohesive energy | 1.436 eV | 1.63 eV[6] | Kittel | −11.9 % |
| Li, bcc | Chemical potential $\mu_{\mathrm{Li}}$ | −2.3845 eV atom$^{-1}$ | — | — | — |

**Table S2.** Adsorption energies and interfacial contacts for five PVDF orientations on $CsSnCl_3$ (100).

| Termination | Orientation | $E_{ads}$ (eV) | F···cation | H···anion | $d_{min}$ (Å) | Best displacement (Å) |
|---|---|---|---|---|---|---|
| $SnCl_2$ | Side-on | −4.615 | 4 | 11 | 2.68 | 0.0 |
| | H-down | −0.418 | 0 | 10 | 3.15 | +0.8 |
| | F-down | −0.247 | 8 | 0 | 3.32 | +0.4 |
| | Tilted 45° | −0.184 | 0 | 1 | 2.60 | +0.8 |
| | Perpendicular | −0.150 | 0 | 3 | 3.09 | +0.8 |
| CsCl | Side-on | −4.612 | 4 | 5 | 2.76 | 0.0 |
| | F-down | −4.213 | 8 | 0 | 3.25 | 0.0 |
| | Perpendicular | −3.692 | 2 | 2 | 3.00 | 0.0 |
| | H-down | −0.112 | 0 | 6 | 3.07 | +0.8 |
| | Tilted 45° | −0.084 | 0 | 2 | 3.56 | +0.8 |

**Table S3.** Adsorption energy of side-on all-*trans* PVDF oligomers on $SnCl_2$ (100) as a function of chain length $n$.

| $n$ | F atoms | $E_{total}$ (eV) | $E_{mol}$ (eV) | $E_{ads}$ (eV) |
|---|---|---|---|---|
| 2 | 4 | −13449.8409 | −83.5631 | −4.392 |
| 3 | 6 | −13488.0096 | −121.5612 | −4.563 |
| 4 | 8 | −13526.0640 | −159.5590 | −4.619 |
| 5 | 10 | −13564.1779 | −197.5555 | −4.737 |
| 6 | 12 | −13602.4403 | −235.5503 | −5.004 |

**Table S4.** $\Delta Z_{\mathrm{Li}}$ (Å) as a function of lithium count $N$ for different oligomer lengths $n$. Values in parentheses are $w_{90-10}$ (Å).

| $N$ | $n = 2$ | $n = 3$ | $n = 4$ | $n = 5$ | $n = 6$ |
|---|---|---|---|---|---|
| 12 | 2.50 | 1.47 | 1.24 | 2.48 | 2.11 |
| 14 | 2.57 | 1.74 | 1.89 | 2.55 | 2.52 |
| 16 | 2.51 | 3.71 | 1.66 | 4.25 | 2.98 |
| 20 | 3.17 | 3.82 | 2.34 | 3.33 | 2.66 |
| 24 | 3.81 | 3.01 | 3.97 | 3.02 | 2.34 |

**Table S5.** Planar accommodation limit $\sigma^*$ (Li nm$^{-2}$) at which $\Delta Z_{\mathrm{Li}}$ first exceeds 2.5 Å.

| Route | System | Γ (units nm$^{-2}$) | Chains nm$^{-2}$ | $\sigma^*$ (Li nm$^{-2}$) | Li per chain |
|---|---|---|---|---|---|
| Surface density | 6 × 6 | 0.351 | 0.088 | 1.98 | 22.5 |
| Surface density | 5 × 5 | 0.506 | 0.126 | 2.21 | 17.5 |
| Chain length | $n = 2$ | 0.395 | 0.198 | 2.38 | 12.1 |
| Chain length | $n = 3$ | 0.593 | 0.198 | 2.92 | 14.8 |
| Both | 4 × 4, $n = 4$ | 0.790 | 0.198 | 4.03 | 20.4 |
| Chain length | $n = 5$ | 0.988 | 0.198 | 2.49 | 12.6 |
| Chain length | $n = 6$ | 1.185 | 0.198 | 2.75 | 13.9 |

**Table S6.** Spearman correlation $\rho$ between geometric descriptors and mean Li–Li coordination number.

| Descriptor | $\rho$ |
|---|---|
| Standard deviation of lithium height | 0.818 |
| $\Delta Z_{\mathrm{Li}} = \max z_{\mathrm{Li}} - \min z_{\mathrm{Li}}$ | 0.794 |
| $1 - f_{\mathrm{2D}}$ (planar fraction, 2.2 Å window) | 0.766 |
| Fraction of Li atoms above 3 Å | 0.653 |

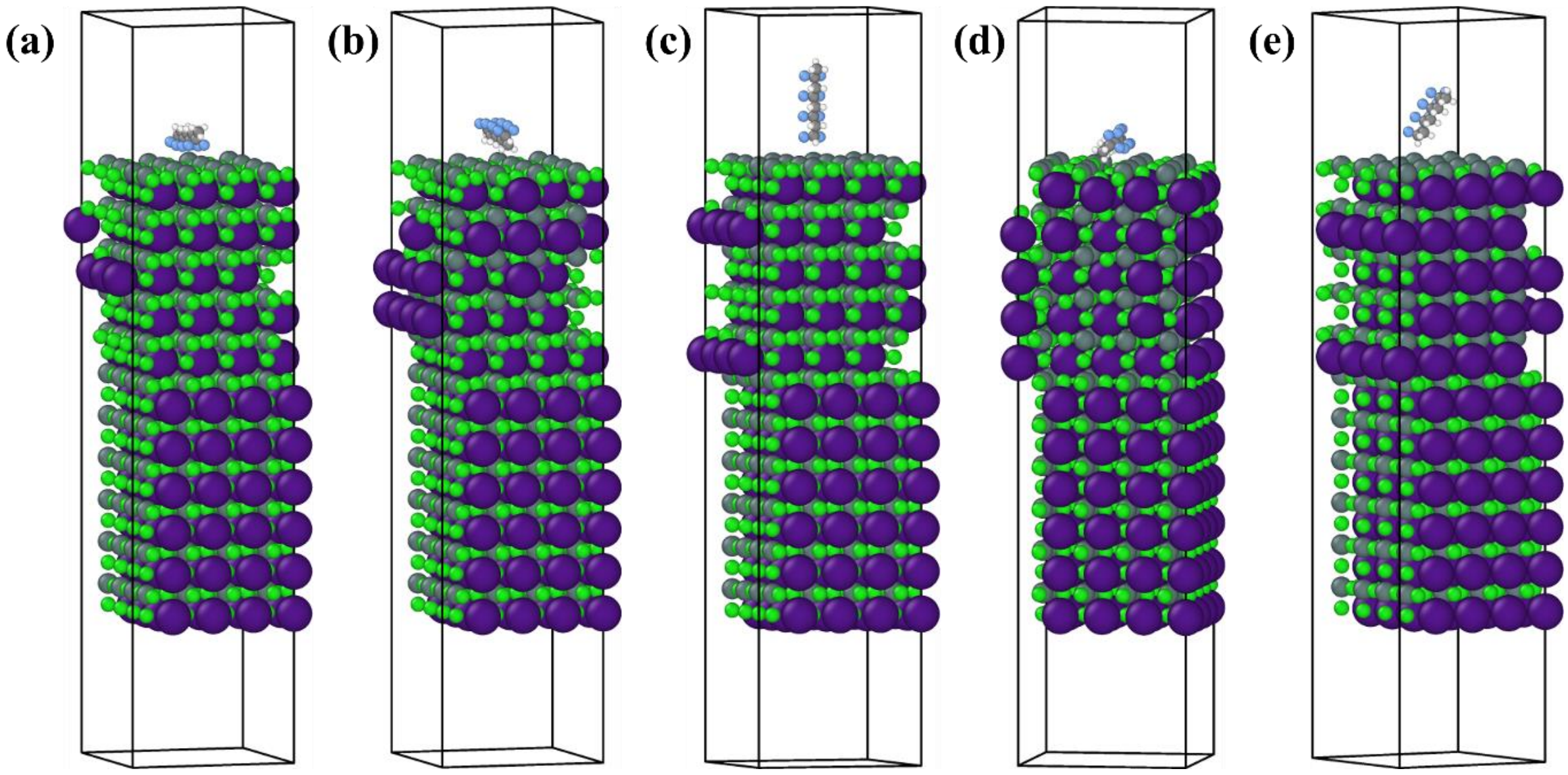

**Figure S29.** $CsSnCl_3$ (100) surface slab with different PVDF orientations on $SnCl_2$ termination. (a) F-down, i.e. F atoms face the slab. (b) H-down, i.e. H atoms face the slab. (c) Perpendicular orientation w.r.t slab. (d) F & H down, i.e. both the atoms face the slab. (e) Chain tilted at 45° w.r.t the slab.

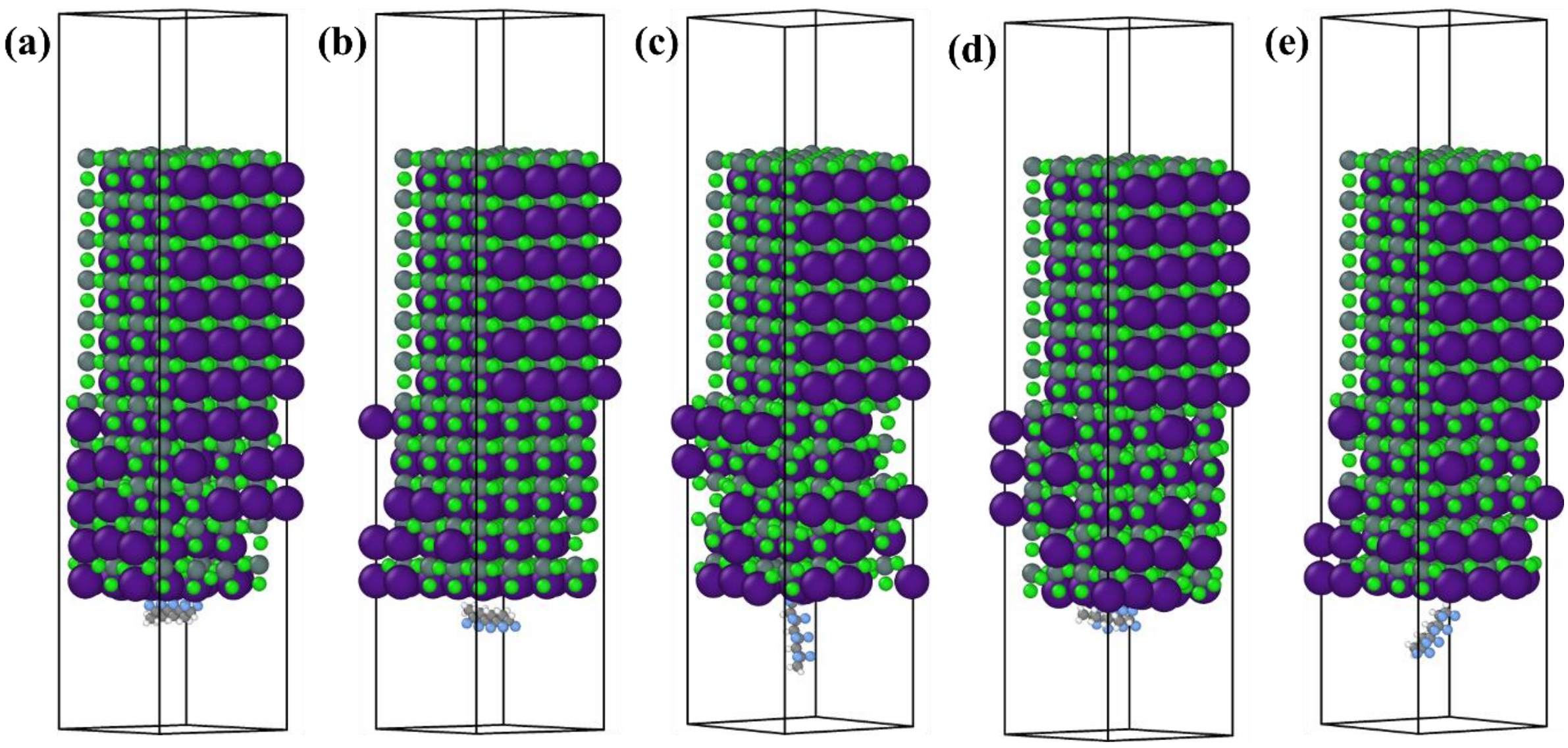

**Figure S30.** $CsSnCl_3$ (100) surface slab with different PVDF orientations on CsCl termination. (a) F-down, i.e. F atoms face the slab. (b) H-down, i.e. H atoms face the slab. (c) Perpendicular orientation w.r.t slab. (d) Side-on (F & H down), i.e. both the atoms face the slab. (e) Chain tilted at 45° w.r.t the slab.

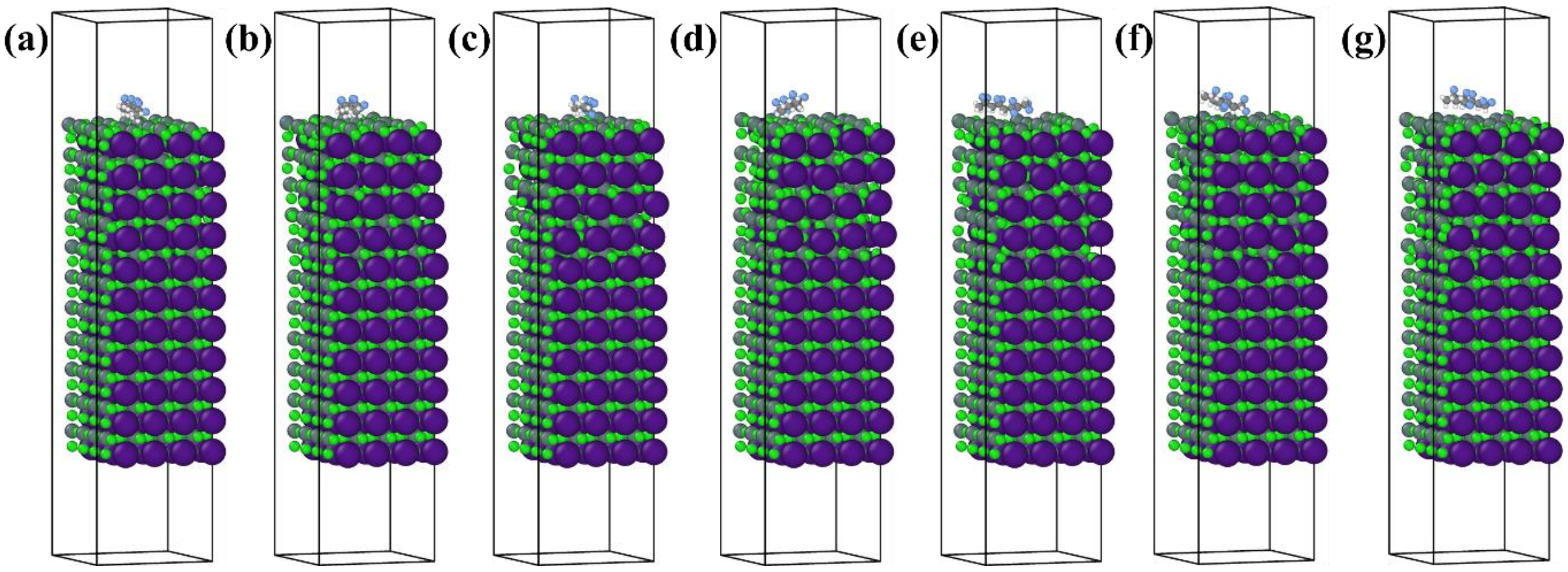

**Figure S31.** Time evolution of the $CsSnCl_3$ (100) surface slab with PVDF in side-on orientation on $SnCl_2$ termination at 300 K simulated via MD. (a)-(g) shows the trajectories with time-interval of 100 fs. This reveals the thermodynamically favourable nature of PVDF side-on orientation as it is maintained throughout the course of 100 ps simulation.

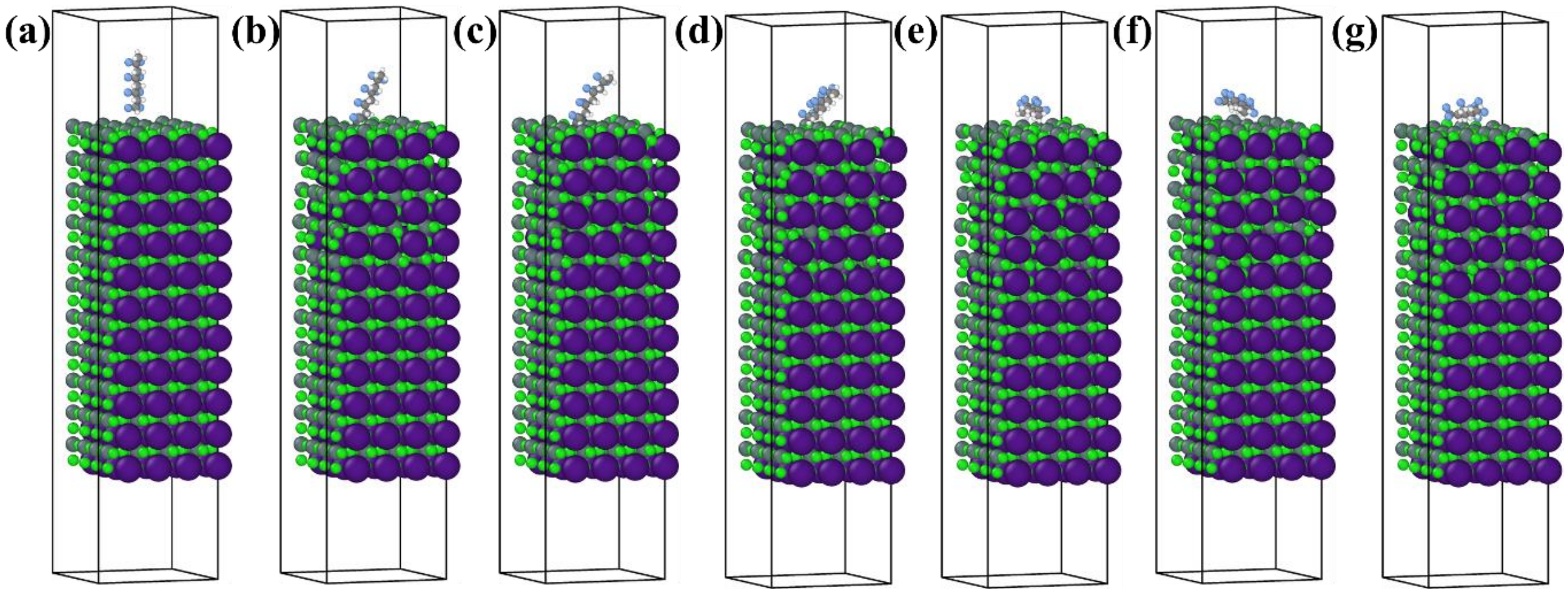

**Figure S32.** Time evolution of the $CsSnCl_3$ (100) surface slab with PVDF in perpendicular orientation w.r.t slab on $SnCl_2$ termination at 300 K simulated via MD. (a)-(g) shows the trajectories with time-interval of 40 fs. This reveals the thermodynamically unfavourable nature of PVDF perpendicular orientation as the chain readily re-orients in the side-on configuration early in simulation and is maintained throughout the course of the 100 ps simulation.

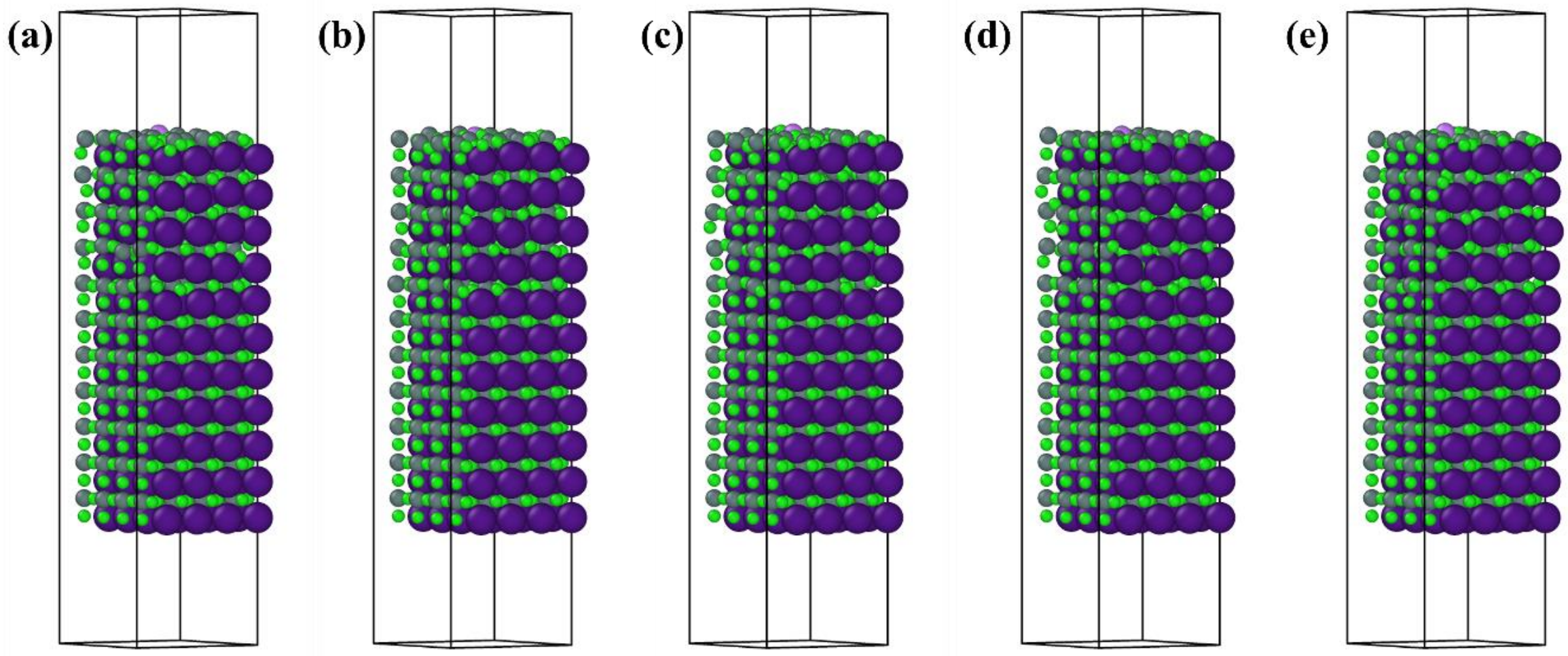

**Figure S33. Molecular dynamics (MD) trajectories of a single Li adatom on the pristine $CsSnCl_3$ surface.** The panels (a)-(e) shows the last 2000 trajectories in the 0.8 ns simulation with time-interval of 500 fs. On the bare surface, the Li atom is deeply bound to the perovskite lattice, maintaining an average Z-gap of 2.23 ± 0.30 Å. The adatom exhibits highly persistent coordination with surface chloride ions (average Li-Cl coordination = 2.91; bond lifetime = 4.03 frames). This unpassivated environment allows for relatively facile lateral diffusion across the surface, yielding a lateral mean squared displacement (MSD) of 0.13 $Å^2$.

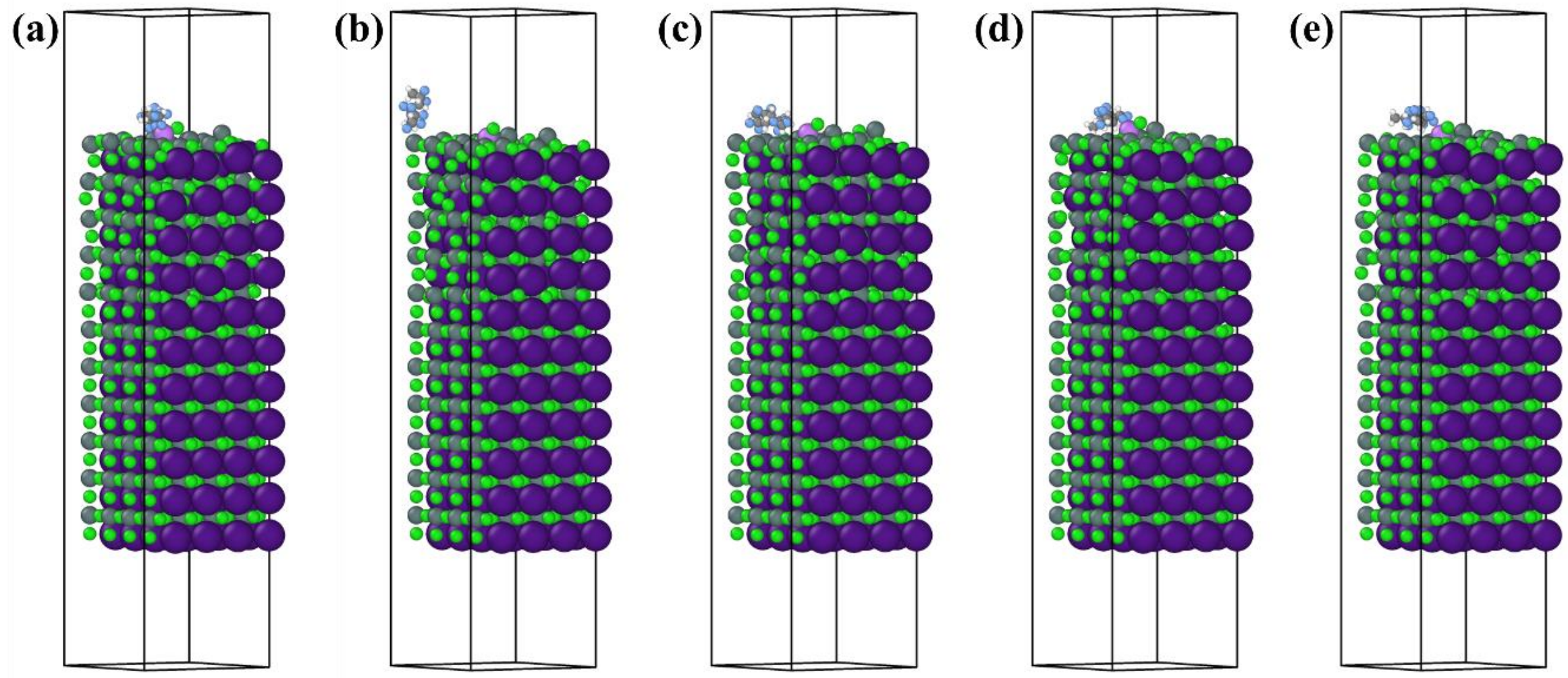

**Figure S34. Molecular dynamics (MD) trajectories of a single Li adatom at the PVDF-passivated $CsSnCl_3$ interface.** The panels (a)-(e) shows the last 2000 trajectories in the 0.8 ns simulation with time-interval of 500 fs. The PVDF binder adopts a highly stable, planar conformation against the surface (tilt angle = 7.8° ± 7.5°). The electronegative fluorine backbone actively solvates the Li ion (Li-F coordination = 0.96), physically lifting it away from the inorganic lattice (Z-gap = 2.75 ± 0.44 Å) and reducing the persistence of Li-Cl bonds to 1.81 frames.